\documentclass{ws-ijmpe}

\newcommand{\beq}{\begin{eqnarray}}
\newcommand{\eeq}{\end{eqnarray}}
\newcommand{\be}{\begin{equation}}
\newcommand{\ee}{\end{equation}}

\newcommand{{\SD}}{\rm SD}

\newcommand{{\Mc}}{\mathcal{M}}

\newcommand{\vex}{{\bm x}}

\newcommand{\ver}{{\bm r}}
\newcommand{\vesig}{{\bm \sigma}}

\newcommand{\vep}{{\bm p}}

\newcommand{\vez}{{\bm z}}

\newcommand{\veL}{{\bm L}}
\newcommand{\veR}{{\bm R}}

\newcommand{\ven}{{\bm n}}
\newcommand{\veu}{{\bm u}}

\newcommand{\veH}{{\bm H}}
\newcommand{\veE}{{\bm E}}

\newcommand{\veal}{{\bm \alpha}}

\newcommand{\llan}{\langle\langle}
\newcommand{\rran}{\rangle\rangle}
\newcommand{\lan}{\langle}
\newcommand{\ran}{\rangle}
\newcommand{\Tr}{\mbox{Tr}}
\newcommand{\ds}{\displaystyle}

\begin{document}

\catchline{}{}{}{}{}

\title{DECONFINEMENT AND QUARK--GLUON PLASMA}

\author{\footnotesize A.V. NEFEDIEV, YU.A. SIMONOV, M.A. TRUSOV}

\address{Institute of Theoretical and Experimental Physics, Russia 117218, Moscow, Bolshaya Cheremushkinskaya str. 25}

\maketitle

\begin{history}
\end{history}

\begin{abstract}
The theory of confinement and deconfinement is discussed as based
on the properties of the QCD vacuum. The latter are described by
field correlators of colour-electric and colour-magnetic fields in
the vacuum, which can be calculated analytically and on the
lattice. As a result one obtains a self-consistent theory of the
confined region in the $(\mu,T)$ plane with realistic hadron
properties. At the boundary of the confining region, the
colour-electric confining correlator vanishes, and the remaining
correlators describe strong nonperturbative dynamics in the
deconfined region with (weakly) bound states. Resulting equation
of state for $\mu=0$, $p(T)$, $\frac{\varepsilon-3P}{T}$ are in
good agreement with lattice data. Phase transition occurs due to
evaporation of a part of the colour-electric gluon condensate, and
the resulting critical temperatures $T_c(\mu)$ for different $n_f$
are in good correspondence with available data.
\end{abstract}

\newpage

\tableofcontents

\markboth{A.V. Nefediev, Yu.A. Simonov, M.A. Trusov}{Deconfinement
and quark-gluon plasma}

\newpage

\section{Introduction}

The fundamental problems of confinement and
deconfinement have recently become a hot topic because of a
possible observation of deconfined phase --- the Quark-Gluon Plasma (QGP)
at RHIC \cite{1}. The theory of the QGP was mainly centered on
perturbative ideas with an inclusion of some resummations and
nonperturbative effects \cite{2,3}. However, the experimental data
\cite{1} display a strong interacting QGP, with some properties
more similar to a liquid, rather than to a weakly interacting
plasma, with strong  collective effects and large energy loss of
fast constituents. These effects call for a nonperturbative treatment of QGP,
and hence for a nonperturbative theory of confined and deconfined phases of
QCD.

For the confined phase the corresponding methods have been
suggested in \cite{4} and formulated as the Field Correlator
Method (FCM) --- see \cite{4*} for a review. In the FCM at $T=0$,
the vacuum configurations support mostly two scalar quadratic in
field correlators, $D(x)$ and $D_1(x)$. The first correlator is
purely nonperturbative and ensures confinement, while the second
one contains both perturbative and  nonperturbative parts. As a
result one obtains linear confining potential from the $D(x)$ plus
perturbative and nonperturbative corrections coming from the
$D_1(x)$. The spectra of all possible bound states (mesons,
baryons, hybrids and glueballs) have been calculated in FCM, with
current masses of quarks, the string tension $\sigma$, and the
strong coupling constant $\alpha_s(\Lambda_Q)$ used as input, in
good agreement with the experiment (see \cite{5*,6*} for a review)
and lattice data \cite{7*}. One can conclude therefore that the
FCM describes well the confining region at $T=\mu=0$.

For $T>0$ several complications occur (see \cite{6}$^-$\cite{5}
for the details). First of all, colour-electric and
colour-magnetic field correlators are no longer identical to each
other and one has to do with five correlators instead of two:
$D^E(x)$, $D_1^E(x)$, $D^H(x)$, $D_1^H(x)$, and $ D_1^{EH}(x)$.
Correspondingly, the colour-electric and colour-magnetic
condensates are now different, as well as the string tensions,
which are simply related to the correlators, \be
\sigma^{E(H)}=\frac12\int D^{E(H)}(x)d^2 x. \label{sigmas} \ee It
is important to notice that only the electric tension $\sigma^E$
is responsible for confinement, while all others ensure spin-orbit
forces. Thus it was conjectured in \cite{6} that deconfinement
occurs if the correlator $D^E(x)$ vanishes at $T=T_c(\mu)$.
Indeed, this conjecture was confirmed later on the lattice, while
all other correlators were measured to stay intact \cite{7*}.
Finally, it was argued in \cite{6} (see also \cite{8*}) that,
since the vacuum energy density is proportional to the gluonic
condensate, one can derive the condition of vanishing of the
$D^E(0)$ directly from the second thermodynamic law. Consequently,
general features of the phase transitions and the very value of
the critical temperature $T_c(\mu)$ can be calculated in terms of
the difference of the gluonic condensates in the confining and
deconfining phases \cite{6,7}.

We now come to the central point of the paper: namely what is the
dynamics of the QGP and its Equation of State (EoS)? As will be
shown below, the basic role in the EoS is played by two effects:
the colour-electric (nonconfining) forces due to the correlator
$D_1^E(x)$ and colour-magnetic forces due to the correlator
$D^H(x)$. The former one is dominant for $T_c\leq T \leq 1.5 T_c$
and creates an effective mass (energy) of isolated quarks and
gluons, thus leading to typical curves for the pressure $p(T)$
similar to those observed in lattice calculations. Furthermore,
the same $D_1^E(x)$ generates potential $V_1(r)$ which is able to
bind quarks and gluons \cite{6*,6}, and provides strong
correlations in white $3q$ systems. Colour-magnetic forces which
stem from the $D^H(x)$ also can bind quarks at large distances and
provide a large ratio $\lan {\rm potential~energy}\ran/\lan {\rm
kinetic~ energy}\ran\gtrsim 10$ specific for liquids \cite{Nef}.
Moreover, $D^H(x)$ (and thence $\sigma^H$) grow with the
temperature, and finally become the basic interaction in the
dimensionally reduced $3D$ QCD, with the transition temperature
around $1.5\div 2T_c$.

The paper is organised as follows. In Section~\ref{Ess} we
introduce the FCM at $T=0$. In particular, we give necessary
essentials of the method, discuss colour-electric and
colour-magnetic interactions, explain the QCD string approach to
hadrons, and consider spin-dependent interactions. In
Section~\ref{T0} we generalise the FCM for nonzero temperatures:
we introduce the winding measure of integration, define the Single
Line Approximation (SLA), and derive the EoS of the QGP in the
framework of the SLA. In Section~\ref{BSLA} we discuss
interactions between quarks and gluons above the $T_c$ and
investigate bound states of quarks and gluons due to these
interactions. In Section~\ref{Disc} we generalise the results of
Section~\ref{T0} for nonzero chemical potentials $\mu>0$. In
Section~\ref{Concl} we compare the results discussed in this
review with other approaches found in the literature and discuss
various solved and unsolved problems.

\section{FCM at $T=0$}\label{Ess}

\subsection{Essentials of the method}

To describe the dynamics of quarks and gluons in both confining and deconfined phases of QCD we
start from the gauge-covariant Green's function of a single quark (or gluon) in the field of
other quarks and gluons plus vacuum fields and use the Fock--Feynman--Schwinger Representation (FFSR)
(in Euclidean space-time) \cite{FFSR}:
\be
S(x,y|A)=\langle x|(m+\hat D)^{-1}|y\rangle=(m-\hat D)\int^\infty_0 ds(Dz)_{xy} e^{-K}\Phi_\sigma(x,y,s)
\label{1}
\ee
where $K$ is the kinetic energy term,
\be
K=m^2s+\frac14\int^s_0 d\tau \left(\frac{dz_\mu(\tau)}{d\tau}\right)^2,
\label{2}
\ee
with $m$ being the pole mass of the quark. The parallel transporter along the trajectory $z_\mu(\tau)$ of the quark
propagating from point $x$ to point $y$ is given by
\be
\Phi(x,y)=P_A\exp{\left[ig\int_{y}^{x}dz_{\mu}A_{\mu}\right]},
\label{PA}
\ee
while
\be
\Phi_\sigma(x,y,s)=\Phi(x,y)P_F\exp \left[g\int^s_0 \sigma_{\mu\nu} F_{\mu\nu}(z(\tau))d\tau\right],
\label{Ps}
\ee
where
\be
\sigma_{\mu\nu}=\frac{1}{4i}(\gamma_\mu\gamma_\nu-\gamma_\nu\gamma_\mu),\quad
\sigma_{\mu\nu}F_{\mu\nu}\equiv\sigma F=\left(\begin{array}{ll} \vesig \veH& \vesig\veE\\\vesig \veE& \vesig \veH\end{array}\right),
\label{3}
\ee
and $\vesig$ are the usual $2\times 2$ Pauli matrices.
The symbols $P_A$ and $P_F$ in (\ref{PA}) and (\ref{Ps}) stand for the ordering operators for the matrices $A_\mu$ and $F_{\mu\nu}$, respectively,
along the quark path.

Similarly, for one gluon Green's function, one writes
\be
G_{\mu\nu}(x,y|A)=\langle x|(\tilde{D}^2_\lambda \delta_{\mu\nu}-2ig\tilde{F}_{\mu\nu})^{-1}|y\rangle
\label{11}
\ee
and, proceeding in the same lines as for quarks, one arrives at the gluon Green's function in the FFSR:
\be
G_{\mu\nu}(x,y|A)=\int^{\infty}_0 ds (Dz)_{xy} e^{-K_0}\tilde{\Phi}_{\mu\nu}(x,y,s),
\label{012}
\ee
where
\be
K_0=\frac14\int^\infty_0\left(\frac{dz_\mu}{d\tau}\right)^2d\tau,\quad
\tilde{\Phi}_{\mu\nu}(x,y,s)=\left[\tilde{\Phi}(x,y)P_F\exp\left(2g\int^s_0d\tau\tilde{F}(z(\tau))\right)\right]_{\mu\nu}.
\label{13}
\ee
Here and below the tilde sign denotes quantities in the adjoint representation, for example,
$\tilde{F}^{bc}_{\mu\nu}\equiv i F_{\mu\nu}^a f^{abc}$.

The single-quark(gluon) Green's functions (\ref{1}) and (\ref{012}) can be used now as building blocks to write the Green's function
of a hadron. For example, for the quark--antiquark meson one has:
\be
G_{q\bar q}(x,y|A)=\mbox{Tr}(S_q(x,y|A)\Gamma S^\dagger_{\bar{q}}(x,y|A)\Gamma^\dagger),
\ee
where $\Gamma$ is a Dirac $4\times 4$ matrix which provides the correct quantum numbers of the meson
($\Gamma$=1, $\gamma_\mu$, $\gamma_5$, $\gamma_\mu\gamma_5$, $\ldots$) and the trace is taken in both Dirac and colour indices.
The next important step is building the physical Green's function of the meson,
\be
G_{q\bar q}(x,y)=\lan G_{q\bar q} (x,y|A)\ran_A,
\label{7}
\ee
where the averaging over gluonic fields is done with the usual Euclidean weight containing all gauge-fixing and ghost terms.
The exact form of this weight is inessential for what follows.

The actual average one needs to evaluate in order to proceed reads:
\be
\lan W_{\sigma_1\sigma_2}(C)\ran=\lan\Phi(C)\Phi_{\sigma_1}(C,s_1)\Phi_{\sigma_2}(C,s_2)\ran_A,
\label{6}
\ee
where the closed contour $C$ runs along the trajectories of the quark, $z_{1\mu}(\tau_1)$, and that of the antiquark, $z_{2\mu}(\tau_2)$.
Since the orderings $P_A$ and $P_F$ in $\Phi_{\sigma_1}$, and $\Phi_{\sigma_2}$, are universal, then $W_{\sigma_1\sigma_2}(C)$ is just a
Wegner--Wilson loop with the insertion of the operators $(\sigma_1 F)$ and $(\sigma_2 F)$ at proper places along the contour $C$.

To proceed it is convenient to rewrite (\ref{6}) in the form
\be
\lan W_{\sigma\sigma'}(C)\ran=\left\lan\exp ig\int_S d\pi_{\mu\nu}(z) F_{\mu\nu}(z)\right\ran,
\label{Wssp}
\ee
where the non-Abelian Stokes theorem \cite{10} was used and the integral is taken over the surface $S$ bounded by the contour $C$.
Notice that the differential
\be
d\pi_{\mu\nu}(z)=ds_{\mu\nu}(z)-i\sigma_{\mu\nu}^{(1)} d\tau_1+i\sigma_{\mu\nu}^{(2)}d\tau_2
\label{9}
\ee
contains not only the surface element $ds_{\mu\nu}$ but it also incorporates the spin variables. This is the most economical way to
include into consideration spin-dependent interactions between quarks ($\tau_1$ and $\tau_2$ are the proper-time variables for the quark
and the antiquark, respectively). Furthermore, if a cluster expansion theorem is applied to the right-hand side of (\ref{Wssp}),
then the result reads:
\be
\frac{1}{N_C}\lan W_{\sigma_1\sigma_2}(C)\ran=\exp\sum_{n=1}^\infty\frac{(ig)^n}{n!}\int d\pi(1)\ldots\int d\pi(n)\llan F(1)\ldots F(n)\rran,
\label{8}
\ee
where $d\pi(k)\equiv d\pi_{\mu_k\nu_k}(z)$ and $F(k)\equiv F_{\mu_k\nu_k}(u_k, x_0)\equiv\Phi(x_0, u_k)F_{\mu_k\nu_k}(u_k)\Phi(u_k, x_0)$.
Double brackets $\llan\ldots\rran$ denote irreducible (connected) averages, and the reference point $x_0$ is arbitrary.

Equation (\ref{8}) is exact and therefore its right-hand side does not depend on any particular choice of the surface $S$.
At this step one can make the first approximation by keeping only the lowest (quadratic, or Gaussian) field correlator
$\llan FF\rran$, while the surface is chosen to be the minimal area surface. As it was argued in \cite{11}, using
comparison with lattice data, this approximation (sometimes called the Gaussian approximation) has the accuracy of a few per cent.

As was mentioned before, the factors $(m-\hat D)$ in the Green's function (\ref{7}) need a special treatment in the
process of averaging over the gluonic fields --- it is shown in \cite{D} that one can use a simple replacement,
\be
m-\hat{D}\to m-i\hat{p},\quad p_\mu =\frac{1}{2}\left(\frac{dz_\mu}{d\tau}\right)_{\tau=s}.
\label{10}
\ee

Therefore, the first step is fulfilled, namely the derivation of
the physical Green's function of a quark--antiquark meson (one can
proceed along the same lines for baryons and hadrons with an
excited glue) in terms of the vacuum correlator $\llan FF\rran$.
In the next chapter the structure of this Gaussian correlator is
studied, in particular, its separation into colour-electric and
colour-magnetic parts.

\subsection{Colour-electric and colour-magnetic correlators}

The Gaussian correlator of background gluonic fields can be
parameterized via two scalar functions, $D$ and $D_1$ \cite{4}:
\begin{eqnarray}
\llan FF\rran&\equiv&D_{\mu\nu\lambda\rho}(z-z')=\frac{g^2}{N_c}\langle\langle \Tr F_{\mu\nu}(z)\Phi(z,z')F_{\lambda\rho}(z')\Phi^\dagger(z,z')\rangle\rangle
\label{HSO0}\\
&=&(\delta_{\mu\lambda}\delta_{\nu\rho}-\delta_{\mu\rho}\delta_{\nu\lambda})D(u)
+\frac12\left[\frac{\partial}{\partial u_\mu}(u_\lambda\delta_{\nu\rho}-u_\rho\delta_{\lambda\nu})+
\genfrac{(}{)}{0pt}{0}{\mu\leftrightarrow\nu}{\lambda\leftrightarrow\rho}\right]D_1(u),\nonumber
\label{FF}
\end{eqnarray}
where $u=z-z'$. Notice that, in order to proceed from (\ref{8}) to (\ref{HSO0}) one needs to show that, for the correlators in
(\ref{8}) taken at close points, $|z-z'|\lesssim T_g$\footnote{The gluonic correlation length $T_g$ defines the distance at which the
background gluonic fields are correlated or, more specifically, the correlator $D(u)$ defined in (\ref{HSO0}) decreases in all directions of
the Euclidean space, and the length $T_g$ governs this decrease. The correlation length $T_g$ extracted from the lattice data is quite small,
$T_g\lesssim 0.1$ fm \cite{Komas,BNS2}.}, the parallel transporters $\Phi$ passing through the point $x_0$ can
be reduced to the straight line between the points $z$ and $z'$.
To this end, notice that, for a generic configuration with $|z-z'|\ll|z-x_0|,|z'-x_0|\sim R$ ($R$ is the radius of the Wilson loop,
of order of the hadron size), one can write $\Phi(z-x_0)\Phi(x_0-z')\approx\Phi(z-z')+O(T_g^2/R^2)$.

For future convenience let us also write the correlator (\ref{HSO0}) in components:
\beq
\frac{g^2}{N_c}\lan\lan {\rm Tr} E_i(z)\Phi E_j(z')\Phi^\dagger\ran\ran&=&\delta_{ij}\left[D^E(u)+D_1^E(u)+u^2_4
\frac{\partial D_1^E}{\partial u^2}\right]+u_iu_j\frac{\partial D_1^E}{\partial u^2},\nonumber\\
\frac{g^2}{N_c}\lan\lan {\rm Tr} H_i(z)\Phi H_j(z')\Phi^\dagger\ran\ran&=&\delta_{ij}\left[D^H(u)+D_1^H(u)+
\veu^2\frac{\partial D_1^H}{\partial\veu^2}\right]-
u_iu_j\frac{\partial D_1^H}{\partial u^2},\label{Hs0}\\
\frac{g^2}{N_c}\lan\lan {\rm Tr} H_i(z)\Phi E_j(z')\Phi^\dagger\ran\ran&=&\varepsilon_{ijk}u_4u_k
\frac{\partial D_1^{EH}}{\partial u^2}.\nonumber
\label{HE0}
\eeq

For the sake of brevity we omit, in this chapter, the spin-dependent terms in the Wilson loop (\ref{8}) and consider only nonperturbative
contributions to the Gaussian correlator $\llan FF\rran$ responsible for confinement and given by the correlator $D$.
This correlator enters the Wilson loop (\ref{8}) multiplied by the surface elements $ds_{\mu\nu}ds_{\lambda\rho}$ and,
in what follows, we distinguish between the colour-electric and colour-magnetic contributions in (\ref{FF}). The former
are accompanied by the structure $ds_{0i}ds_{0i}$ and enters (\ref{8}) multiplied by electric correlator $D^E$,
whereas, for the latter, these are $ds_{jk}ds_{jk}$ and $D^H$, respectively.
The electric and magnetic string tensions are defined then according to (\ref{sigmas}).

We now proceed from the proper-time variables to the laboratory times by performing the following change of
variables:
\be
d\tau_1=\frac{dz_{10}}{2\mu_1},\quad d\tau_2=\frac{dz_{20}}{2\mu_2}
\label{taus}
\ee
and synchronise the quarks in the laboratory frame, putting $z_{10}=z_{20}=t$. The new variables $\mu_{1,2}$ which appeared in (\ref{taus})
are known as the auxiliary (or einbein) fields \cite{ein,DKS,ein3}. The physical meaning and the role played by the einbeins will be
discussed in detail below. Now we can write the surface element through the string profile function $w_\mu(t,\beta)$ as
\be
ds_{\mu\nu}(z)=\varepsilon^{ab}\partial_a w_\mu(t,\beta)\partial_b w_\nu(t,\beta)dt d\beta,\quad \{a,b\}=\{t,\beta\},
\ee
where $0\leqslant t\leqslant {t_{\rm max}}$, $0\leqslant\beta\leqslant 1$, and we adopt
the straight--line ansatz for the minimal string profile, so that the latter is defined by the trajectories of the
quarks \cite{DKS}:
\be
w_\mu(t,\beta)=\beta z_{1\mu}(t)+(1-\beta)z_{2\mu}(t).
\ee
For further convenience let us introduce two vectors:
\be
{\bm r}={\bm z}_1-{\bm z}_2,\quad{\bm \rho}=[({\bm z}_1-{\bm z}_2)\times(\beta\dot{{\bm z}}_1+(1-\beta)\dot{{\bm z}}_2)]\equiv r{\bm \omega},
\ee
where ${\bm\omega}$ is the angular rotation vector. This allows one to write the differentials in a compact form,
\be
ds_{0i}(z)ds_{0i}(z')={\bm r}(t){\bm r}(t')dtdt'd\beta d\beta',\quad
ds_{jk}(z)ds_{jk}(z')=2{\bm \rho}(t){\bm \rho}(t')dtdt'd\beta d\beta'.
\ee

Presenting the averaged Wilson loop (\ref{8}) (without spins) as
$$
\frac{1}{N_C}\langle Tr W(C)\rangle=e^{-J},\quad J=J^E+J^H,
$$
one can write for the electric and magnetic contributions separately:
\begin{eqnarray}
J^E=\int_0^{t_{\rm max}}dt\;dt'\int_0^1d\beta\;d\beta'\;{\bm r}(t){\bm r}(t')D^E(z-z'),\label{JE0}\\
J^H=\int_0^{t_{\rm max}}dt\;dt'\int_0^1d\beta\;d\beta'\;{\bm \rho}(t,\beta){\bm \rho}(t',\beta')D^H(z-z')\label{JH0}.
\end{eqnarray}

The correlation functions $D^{E,H}$ decrease in all directions of
the Euclidean space--time with the correlation length $T_g$ which
is measured on the lattice to be rather small, $T_g\approx 0.2\div
0.3$~fm \cite{7*} or even smaller, $T_g\lesssim 0.1$~fm
\cite{Komas,BNS2}. Therefore, only close points $z$ and $z'$ are
correlated, so that one can neglect the difference between ${\bm r}(t)$ and ${\bm r}(t')$, 
${\bm \rho}(t,\beta)$ and ${\bm \rho}(t',\beta')$ in (\ref{JE0}), (\ref{JH0}) and also write: 
\be
(z-z')^2=(z(t,\beta)-z(t',\beta'))^2=g^{ab}\xi_a\xi_b,\quad\xi_a=t-t',\quad\xi_b=\beta-\beta'.
\ee 
The induced metric tensor is $g^{ab}=g^a\delta^{ab}$,
$g^1g^2=det\;g=r^2+\rho^2=r^2(1+\omega^2)$. Now, after an
appropriate change of variables and introducing the string
tensions, according to (\ref{sigmas}), one readily finds:
\begin{eqnarray}
J^E=\sigma_E \int_0^{t_{\rm max}}dt\int_0^1d\beta\frac{r^2}{\sqrt{r^2+\rho^2}}=\sigma_E r\int_0^{t_{\rm max}}
dt\int_0^1d\beta\frac{1}{\sqrt{1+\omega^2}},\label{JE}\\
J^H=\sigma_H \int_0^{t_{\rm max}}dt\int_0^1d\beta\frac{\rho^2}{\sqrt{r^2+\rho^2}}=\sigma_H r\int_0^{t_{\rm max}}
dt\int_0^1d\beta\frac{\omega^2}{\sqrt{1+\omega^2}}.\label{JH}
\end{eqnarray}

For $\sigma_E=\sigma_H=\sigma$ the sum of $J^E$ and $J^H$
reproduces the well-known action of the Nambu--Goto string (see, for example, \cite{ein}$^-$\cite{ein3}),
\be
J=J^E+J^H=\sigma r\int_0^{t_{\rm max}}dt\int_0^1d\beta\sqrt{1+\omega^2}.
\ee

Notice that, for $\omega\ll 1$, $J_E=\sigma_E r\times t_{\rm max}+O(\omega^2)$ and $J_H=O(\omega^2)$, that is confinement is of a purely colour-electric
nature while the colour-magnetic contribution is entirely due to the rotation of the system.

\subsection{QCD string approach}

With the form of the colour-electric and colour-magnetic spin-independent interactions (\ref{JE}) and (\ref{JH}) in hand we are in a
position to build a quantum-mechanical model of hadron consisting of quarks (gluons) connected by straight-line Nambu-Goto strings.
Below we consider, as a paradigmatic case, a quark--antiquark meson.

We start from the kinetic energies of the quarks,
\be
K_i=m_i^2s_i+\frac14\int^{s_i}_0d\tau_i\left(\frac{dz_{i\mu}(\tau_i)}{d\tau_i}\right)^2=
\int_0^{t_{\rm max}} dt \left[\frac{m_i^2}{2\mu_i}+\frac{\mu_i}{2}+\frac{\mu_i\dot{\vez}_i^2}{2}\right],\quad i=1,2,
\label{Ki}
\ee
where (\ref{taus}) was used, and path integrals in $\mu_i$ appear through the substitution
$ds_i Dz_{i0}\to D\mu_i$. One can see therefore that
the einbeins should be treated on equal footings with other coordinates. However, since time derivatives of the einbeins, $\dot{\mu}_i$,
do not enter (\ref{Ki}), they can be eliminated with the help of the well-known integral:
\be
\int D\mu\exp\left[-\int_0^{t_{\rm max}}\left(\frac{a\mu}{2}+\frac{b}{2\mu}\right)dt\right]
\sim\exp\left[-\int_0^{t_{\rm max}}\sqrt{ab}\;dt\right].
\label{rr}
\ee

Then, with the help of (\ref{1}), (\ref{JE}), (\ref{JH}), and (\ref{Ki}), one can extract the Lagrangian of the quark--antiquark
system in the form (notice that hereinafter in this chapter we work in Minkowski space-time):
$$
L=-m_1\sqrt{1-\dot{{\bm z}}_1^2}-m_2\sqrt{1-\dot{{\bm z}}_2^2}
-\sigma_Er\int_0^1d\beta\frac{1}{\sqrt{1-[{\bm n}\times(\beta\dot{{\bm z}}_1+(1-\beta)\dot{{\bm z}}_2)]^2}}
$$
\be
+\sigma_Hr\int_0^1d\beta\frac{[{\bm n}\times(\beta\dot{{\bm z}}_1+(1-\beta)\dot{{\bm z}}_2)]^2}
{\sqrt{1-[{\bm n}\times(\beta\dot{{\bm z}}_1+(1-\beta)\dot{{\bm z}}_2)]^2}},\quad {\bm n}=\frac{{\bm r}}{r}.
\label{L1}
\ee

In fact, the einbein form of the kinetic energies is much more
convenient \cite{ein}$^-$\cite{ein3,ein2} since it does not
contain square roots (unbearable for path integrals) and one deals
with formally nonrelativistic kinetic terms, while the entire set
of relativistic corrections is summed by taking extrema in the
einbeins. Furthermore, extra (continuous) einbeins, $\nu(\beta)$
and $\eta(\beta)$, can be introduced in order to simplify the
string terms in (\ref{L1}), through the substitutions: \be
\int_0^1 d\beta\sqrt{AB}\to \int_0^1 d\beta
\left(\frac{A}{2\nu}+\frac{B\nu}{2}\right),\quad \int_0^1
d\beta\frac{A^2}{B}\to-\int_0^1 d\beta
\left(B\eta^2+2A\eta\right). \ee

Then, introducing the centre-of-mass position and the relative coordinate as
\be
\veR=\zeta_1\vex_1+(1-\zeta_1)\vex_2,\quad
\ver=\vex_1-\vex_2,\quad\zeta_1=\frac{\mu_1+\int_0^1\beta\nu d\beta}{\mu_1+\mu_2+ \int^1_0 \nu d\beta}, \quad\zeta_2=1-\zeta_1,
\label{Rr}
\ee
one can rewrite the Lagrangian in the form (the centre-of-mass motion is omitted for simplicity):
\be
L=-\sum_{i=1}^2\left[\frac{m_i^2}{2\mu_i}+\frac{\mu_i}{2}\right]
-\int^1_0d\beta\left(\frac{\sigma_1^2r^2}{2\nu}+\frac{\nu}{2}+\sigma_2 r\right)
+\frac12\mu(\dot{\ver}\ven)^2+\frac12\tilde{\mu}[\dot{\ver}\times\ven ]^2,
\label{L13}
\ee
where $\sigma_1=\sigma_H+\eta^2(\sigma_H-\sigma_E)$, $\sigma_2=2\eta(\sigma_E-\sigma_H)$ and
we have defined the reduced masses for the angular and radial motion separately:
\be
\mu=\frac{\mu_1\mu_2}{\mu_1+\mu_2},
\label{murad}
\ee
\be
\tilde{\mu}=\mu_1(1-\zeta_1)^2+\mu_2\zeta_1^2+\int^1_0(\beta-\zeta_1)^2\nu d\beta.
\label{muang}
\ee
The original form of the Lagrangian is
readily restored once extrema in all four einbeins, $\{\mu_1,\mu_2,\nu(\beta),\eta(\beta)\}$, are taken.
Generally speaking, the einbein fields appear in the Lagrangian and, as was mentioned before, even in absence of the corresponding
velocities, they can be considered as extra degrees of freedom introduced to the
system. The einbeins can be touched upon when proceeding from the Lagrangian of the system to its Hamiltonian and thus they mix with the
ordinary particles coordinates and momenta. Besides, in order to
preserve the number of physical degrees of freedom, constrains are to be imposed on the system and then the formalism of constrained systems quantisation
\cite{Dirac} is operative (see, for example, \cite{ein3} for the open straight--line QCD string quantisation using this formalism).
A nontrivial algebra of constraints and the process of disentangling the physical degrees of freedom and non-physical ones make the problem very
complicated. In the meantime, a simpler approach to einbeins exists which amounts to considering all (or some) of them as variational parameters and thus
to taking extrema in the einbeins either in the Hamiltonian or in its spectrum \cite{ein2,regge}. Being an approximate approach this technique
appears accurate enough (see, for example, \cite{KNS}) providing a simple but powerful and intuitive method of investigation.
The physical meaning of the variables $\mu_i$ ($i=1,2$) is the average kinetic energy of the $i$-th
particle in the given state, namely, $\mu_i=\lan\sqrt{\vep^2+m^2_i}\ran$ (see the discussion in
\cite{DKS,KNS}). The continuous einbein variable $\nu(\beta)$ has the meaning of the QCD string energy density
\cite{DKS}.

Following the standard procedure, we build now the canonical momentum as
\be
\vep=\frac{\partial L}{\partial \dot{\ver}}=\mu(\ven\dot\ver)\ven+\tilde{\mu}(\dot\ver-\ven(\ven\dot{\ver})),
\label{v16}
\ee
with its radial component and transverse component being
\be
(\ven\vep)=\mu(\ven\dot{\ver}),\quad [\ven\times\vep]=\tilde{\mu}[\ven\times\dot{\ver}],
\label{160}
\ee
respectively. Thus we arrive at the spin-independent part of the Hamiltonian \cite{DKS}:
\be
H=\sum_{i=1}^2\left[\frac{m_i^2}{2\mu_i}+\frac{\mu_i}{2}\right]+\int^1_0d\beta\left(\frac{\sigma_1^2r^2}{2\nu}+\frac{\nu}{2}+\sigma_2 r\right)
+\frac{p_r^2}{2\mu}+\frac{\veL^2}{2\tilde{\mu}r^2}.
\label{Hm}
\ee
Notice that the kinetic part of the Hamiltonian (\ref{Hm}) has a very clear structure: the radial motion of the quarks
happens with the effective mass $\mu$, whereas for the orbital motion the mass $\tilde{\mu}$ is somewhat different,
containing the contribution of the inertia of the string.
For $\sigma_E=\sigma_H=\sigma$, the field $\eta$ drops from the Hamiltonian and
the standard expression for the string with quarks at the ends \cite{DKS} readily comes out from (\ref{Hm}).

\subsection{Spin-dependent interactions in hadrons}

In heavy quarkonia (up to the order $1/m^2$) the standard Eichten--Feinberg decomposition is valid \cite{EF,12,EF2}:
\begin{eqnarray}
V_{SD}^{(0)}(r)&=&\left(\frac{\vesig_1\veL}{4m_1^2}+\frac{\vesig_2\veL}{4m_2^2}\right)\left[\frac1r\frac{dV_0}{dr}+\frac2r\frac{dV_1}{dr}\right]
+\frac{(\vesig_1+\vesig_2)\veL}{2m_1m_2}\frac1r\frac{dV_2}{dr}\nonumber\\
\label{SO0}\\
&+&\frac{(3(\vesig_1\ven)(\vesig_2\ven)-\vesig_1\vesig_2)}{12m_1m_2}V_3(r)+
\frac{\vesig_1\vesig_2 }{12m_1m_2}V_4(r)\nonumber,
\end{eqnarray}
where each potential $V_n(r)$ ($n=$0-4) contains both perturbative (P) and
nonperturbative (NP)
contributions: $V_n(r)=V_n^P(r)+V_n^{NP}(r)$.
The static interquark potential $V_0(r)$, together with the potentials $V_1(r)$ and $V_2(r)$, satisfies the Gromes relation \cite{Gr},
\be
V_0'(r)+V_1'(r)-V_2'(r)=0.
\label{Ge}
\ee
Notice that this relation refers both to the perturbative and nonperturbative parts of the potentials $V_n(r)$ $(n=0,1,2)$.

We now return to the Green's function of the quark--antiquark system (\ref{1}) and restore spin--dependent terms in order to derive the
generalisation of (\ref{SO0}).

Let us quote here without derivation the full set of spin-dependent potentials, both perturbative and nonperturbative,
obtained in the framework of FCM and with the string rotation
taken into account (an interested reader can find the details of the derivation in \cite{BNS3}):
\be
\begin{array}{l}
\ds\left(\frac{{\bm \sigma}_1\veL}{4{\bar m}_1^2}+\frac{{\bm \sigma}_2\veL}{4{\bar m}_2^2}\right)\frac1r\frac{dV_0}{dr}\to
\ds\frac{1}{2r}\int_0^\infty d\nu\int_0^r d\lambda \left[\vphantom{\frac12}D+D_1+(\lambda^2+\nu^2)\frac{\partial
D_1}{\partial\nu^2}\right]\\[3mm]
\hspace*{3cm}\ds\times\left[(1-\zeta_1)\frac{{\bm \sigma}_1\veL}{\mu_1\tilde{\mu}}+(1-\zeta_2)\frac{{\bm
\sigma}_2\veL}{\mu_2\tilde{\mu}}\right],\\[3mm]
\ds\left(\frac{{\bm \sigma}_1\veL}{4{\bar m}_1^2}+\frac{{\bm \sigma}_2\veL}{4{\bar m}_2^2}\right)\frac2r\frac{dV_1}{dr}\to-
\frac1r\int^\infty_0d\nu\int^r_0d\lambda D
\ds\left(1-\frac{\lambda}{r}\right)\\[3mm]
\hspace*{3cm}\ds\times\left[(1-\zeta_1)\frac{{\bm
\sigma}_1\veL}{\mu_1\tilde{\mu}}+(1-\zeta_2)\frac{{\bm
\sigma}_2\veL}{\mu_2\tilde{\mu}}\right],\\[3mm]
\ds\frac{({\bm \sigma}_1+{\bm \sigma}_2)\veL}{2{\bar m}_1{\bar m}_2}\frac1r\frac{dV_2}{dr}\to
\ds\frac{1}{r^2}\int^\infty_0d\nu\int^r_0\lambda d\lambda \left[\vphantom{\frac12}D+D_1
+\lambda^2\frac{\partial D_1}{\partial\lambda^2}\right]\frac{(\vesig_1+\vesig_2)\veL}{\tilde{\mu}}\left(\frac{\zeta_1}{\mu_1}\right),\\[3mm]
\ds\frac{(3(\vesig_1\ven)(\vesig_2\ven)-\vesig_1\vesig_2)}{12\bar{m}_1\bar{m}_2}V_3(r)\to
-2r^2\frac{\partial}{\partial r^2}\int_0^\infty d\nu D_1(r,\nu)\frac{(3(\vesig_1\ven)(\vesig_2\ven)-\vesig_1\vesig_2)}{12\mu_1\mu_2},\\[3mm]
\ds\frac{\vesig_1\vesig_2}{12\bar{m}_1\bar{m}_2}V_4(r)\to6\int_0^\infty d\nu\left[D(r,\nu)+\left[1+\frac23r^2\frac{\partial}{\partial\nu^2}\right]D_1(r,\nu)\right]
\ds\frac{\vesig_1\vesig_2}{12\mu_1\mu_2},
\end{array}
\label{SDints}
\ee
where the masses $m_i$ are replaced by $\mu_i$ and $\tilde\mu$, which makes this result applicable also to light quarks.
Notice that this result \cite{242,5*}, is not due to the $1/m$ expansion, but is obtained with the only
approximation made being the Gaussian approximation for field correlators (corrections may come from triple and quartic correlators).
Accuracy of this approximation was checked both at $T=0$ \cite{11}
and at $T>T_c$ \cite{24} to be of the order of one percent.

With this explicit form of the potentials and taking the limit of heavy quarks
one can check the Gromes relation (\ref{Ge}), which now reads:
\begin{eqnarray}
V_0'(r)+V_1'(r)-V_2'(r)=2\int_0^\infty d\nu\int_0^r d\lambda [D^E(\lambda,\nu)-D^H(\lambda,\nu)]\nonumber\\[-3mm]
\label{Ge2}\\[-3mm]
+r\int_0^\infty d\nu[D_1^E(r,\nu)-D_1^H(r,\nu)]\nonumber.
\label{20s}
\end{eqnarray}
At $T=0$, $D^E=D^H$ and $D_1^E=D_1^H$, so that the Gromes relation (\ref{Ge}) is satisfied. 
In the next Section we shall discuss the FCM at nonzero temperatures, in particular at $T>T_c$, where the
latter equalities between colour-electric and colour-magnetic correlators do not hold, and the Gromes
relation is therefore violated.

\section{FCM at $T>0$}\label{T0}

\subsection{The winding measure of integration}

Now  we turn to the case $T>0$ and use Matsubara technique for the path integrals in the FFSR, first introduced in \cite{5},
\begin{eqnarray}
(Dz)_{xy}^w&&=\lim_{N\to \infty}\prod^N_{m=1}\frac{d^4\xi(m)}{(4\pi\varepsilon)^2}\nonumber
\\
&&\times\sum_{n=0,\pm 1,\ldots}\frac{d^4p}{(2\pi)^4}\exp
\left[ip\left(\sum^N_{m=1}\zeta(m)-(x-y)-n\beta\delta_{\mu
4}\right)\right]. \label{33a}
\end{eqnarray}
where $\zeta(k)=z(k)-z(k-1)$ and $\varepsilon=t/N$, and $\beta=1/T$.
It is easy to check that, with such a measure, for example, the free massless propagator takes its standard form:
\begin{eqnarray}
(-\partial^2)^{-1}_{xy}&=&\int^\infty_0 dt\exp\left[-\sum_{m=1}^N\frac{\zeta^2(m)}{4\varepsilon}\right ]\prod_m
\frac{d\zeta(m)}{(4\pi\varepsilon)^2}\sum_n\frac{d^4p}{(2\pi)^4} \nonumber\\
&\times&\exp\left[ip\left(\sum\zeta(m)-(x-y)-n\beta\delta_{\mu4}\right)\right]\label{34a}\\
&=&\sum_n\int_0^\infty\exp\left[-p^2t-ip(x-z)-ip_4n\beta\right]dt\frac{d^4p}{(2\pi)^4}\nonumber\\
&=&\sum_{k=0,\pm1,\ldots}\int\frac{Td^3p}{(2\pi)^3}\frac{\exp[-ip_i(x-y)_i-i2\pi kT(x_4-y_4)]}{ p^2_i+(2\pi kT)^2},\nonumber
\end{eqnarray}
where the Poisson summation formula was used to obtain:
\be
\sum_{n=0,\pm1,\ldots}\exp(ip_4 n\beta)=\sum_{k=0,\pm1,\ldots}2\pi\delta(p_4\beta-2\pi k).
\label{35a}
\ee

For quarks one is to take into account the antisymmetric nature of the fermionic fields, which yields instead of (\ref{33a})
\begin{eqnarray}
\overline{(Dz)}^w_{xy}&=&\lim_{N\to \infty}\prod^N_{m=1}\frac{d^4\zeta(m)}{(4\pi\varepsilon)^2}\nonumber\\
&\times&\sum_{n=0,\pm1,\dots}(-1)^n\frac{d^4p}{(2\pi)^4}\exp\left[ip\left(\sum^N_{m=1}\zeta(m)-(x-y)-n\beta\delta_{\mu 4}\right)\right].
\label{45a}
\end{eqnarray}

Therefore, in order to proceed from $T=0$ to $T>0$ we simply replace $(Dz)_{xy}$ by $(Dz)^w_{xy}$ or $\overline{(Dz)}^w_{xy}$
in all expressions.

We now come to the point where we need to separate the quantum gluon field $a_\mu$ from the vacuum background field $B_\mu$, so that
$A_\mu=B_\mu+a_\mu$, both satisfying periodic boundary conditions,
\be
B_\mu(z_4,\vez)=B_\mu(z_4+n\beta,\vez),\quad a_\mu(z_4,\vez)=a_\mu(z_4+n\beta,\vez),
\label{24a}
\ee
where $n$ is an integer. The partition function can be written as
\be
Z(V,T)=\lan Z(B)\ran_B,\quad Z(B)=N\int D\{\phi\} \exp \left(-\int^\beta_0 d\tau \int d^3x L_{\rm tot} (x,\tau)\right)
\label{25a}
\ee
where $N$ is a normalization constant and $\phi$ denotes the entire set of fields $a_\mu$, $\Psi$, $\Psi^+$.
The explicit form of the Lagrangian $L_{\rm tot}$ is given in \cite{5}.
Furthermore, in (\ref{25a}) $\lan\ldots\ran_B$ stands for the averaging over the (nonperturbative) background fields $B_\mu$.
For our purposes one needs not know the details of this averaging.

Integration over the ghost and gluon degrees of freedom in (\ref{25a}) yields the same answer as for the case of $T=0$,
but now all the fields are subject to the periodic boundary conditions (\ref{24a}). Disregarding, for the sake of simplicity,
quark and source terms in $L_{\rm tot}$, one obtains
\begin{eqnarray}
Z(B)&=&N'(\det W(B))_{reg}^{-1/2}[\det(-D_\mu(B)D_\mu(B+a))]_{a=\delta/\delta J}\nonumber\\
&\times& \left \{1+\sum^\infty_{l=1}\frac{1}{l!}S_{int}\left(a=
\frac{\delta}{\delta J}\right)^l\right\}\exp\left(-\frac12JGJ\right)_{J_\mu=D_\mu(B)F_{\mu\nu}(B)},
\label{26a}
\end{eqnarray}
where the valence-gluon Green's function $G\equiv W^{-1}$ is $G_{\mu\nu}=(\tilde{D}^2_\lambda\cdot\hat{1}+2ig\tilde{F}_{\mu\nu})^{-1}$.
We can consider strong background fields, so that $gB_\mu$ is large (as compared
to $\Lambda^2_{QCD}$), while $\alpha_s=g^2/(4\pi)$ in this strong background is small at all distances. Moreover, it was shown that
$\alpha_s$ is frozen at large distances \cite{Fr}. In this case (\ref{26a}) is a perturbative sum in powers of $g^n$, arising from
the expansion in $(ga_\mu)^n$.

In what follows we shall discuss the Feynman graphs for the thermodynamic potential $F(T,\mu)$, related to $Z(B)$ via
\be
F(T,\mu)=-T\ln \lan Z(B)\ran_B.
\label{27a}
\ee

\subsection{Nonperturbative dynamics of quarks and gluons in the single-line approximation}

In this chapter we come to explicit calculations of various properties of the quark-gluon plasma.
It is clear that (\ref{27a}) contains all possible interactions, perturbative and nonperturbative, between quarks and gluons and, in
particular, creation and dissociation of bound states. It is impossible to take into account all possible subsystems and
interaction between them, so it is imperative to choose the strategy of approximations for the quark-gluon plasma as a
deconfined state of quarks and gluons.

We assume that the whole system of $N_g$ gluons, $N_q$ quarks, and
$N_{\bar q}$ antiquarks remains gauge-invariant at $T>T_c$, as it was at $T<T_c$. However, in case of
deconfinement and neglecting (in the first approximation) all
perturbative and nonperturbative interactions, any white system
possesses the same energy, depending only on the number and type
of constituents. In this case the partition function factorises
into a product of one-gluon and one-quark (antiquark)
contributions, and one can calculate the corresponding
thermodynamic potential. This first step is called the Single Line
Approximation (SLA) and below we shall calculate the results for
the gas of quarks and gluons in the strong vacuum fields using our
path-integral formalism. We follow \cite{5,an1}$^-$\cite{23}.

In this chapter we consider gluons and quarks in SLA. Notice that, although the interaction between quarks and gluons is switched
off in this approximation, there exists a strong interaction of quarks and gluons with the nonperturbative vacuum fields.
We therefore keep only the lowest order in $(ga_\mu)$ but take into account interaction with the nonperturbative background vacuum
fields to all orders, using the field correlator technique. As a result one can write
\be
Z_0=e^{-F_0(T)/T}=N'\lan \exp (-F_0(B)/T)\ran_B.
\label{28a}
\ee
Then single-gluon and single-quark contributions to the free energy can be written in the form:
\begin{eqnarray}
\frac{1}{T}F_0^{gl}(B)&=&\frac12\ln\det G^{-1}-\ln\det(-D^2(B))=\nonumber\\
&=&\Tr\left\{-\frac12\int^\infty_0\xi(s)\frac{ds}{s}e^{-sG^{-1}}+\int^\infty_0\xi(s)\frac{ds}{s}e^{sD^2(B)}\right\}
\label{29a}
\end{eqnarray}
and
\be
\frac{1}{T}F^q_0(B')=\frac12\ln\det (m_q^2-\hat{D}^2(B'))=-\frac12\Tr\int^\infty_0\xi(s)\frac{ds}{s} e^{-sm_q^2+s\hat D^2(B')},
\label{42a}
\ee
respectively, where $B'_\nu =B_\nu-i(\mu/g)\delta_{\nu 4}$, $\mu$ is the quark chemical potential, and $m_q$ is
the current quark mass (the pole mass when next terms in $\alpha_s$ are considered). The operator $\hat{D}^2(B')$ can be presented as:
\be
\hat{D}^2(B')=(D_\mu\gamma_\mu)^2=D^2_\mu(B')-gF_{\mu\nu}\sigma_{\mu\nu}\equiv D^2-g\sigma F.
\ee

Then, using the FFSR, one can write
\be
\lan F_0^{gl}(B)\ran_B=-T\int\frac{ds}{s}\xi(s)d^4x(Dz)^w_{xx}e^{-K}\left[\frac12\Tr\lan\tilde{\Phi}_F(x,x,s)\ran_B-
\Tr\lan\tilde{\Phi}(x,x)\ran_B\right],
\label{39a}
\ee
where taking trace implies summation over the Lorentz and colour indices and $\tilde{\Phi}_F$ is defined in (\ref{13}).

Similarly, one has for the quarks:
\be
\lan F^q_0(B')\ran_B=-\frac{T}{2}\int_0^\infty\frac{ds}{s}\xi(s)d^4x\overline{(Dz)}^w_{xx}e^{-K-sm^2}\Tr\lan\Phi_\sigma(x,x,s)\ran_B,
\label{44a}
\ee
where $\Tr$ implies summation over the spin and colour indices and the operator $\Phi_\sigma$ is given in (\ref{Ps}).

For future convenience let us define the proper-time gluon and quark kernels,
\be
G(s)\equiv\int(Dz)^w e^{-K}\Tr\lan W_\Sigma(C,s)\ran,\quad
S(s)\equiv\int\overline{(Dz)}^w e^{-K}\Tr\lan W_\sigma(C,s)\ran
\label{P3}
\ee
where
\be
W_\Sigma=\frac{1}{(N^2_c-1)}\left[\frac12\tilde{\Phi}_F-\tilde{\Phi}\right],\quad W_\sigma=\frac{1}{N_c}\Phi_\sigma.
\label{P5}
\ee

Then, using the relation between the pressure and the free energy, $PV_3=-\lan F_0(B)\ran_B$,
one can find (the $T$-independent term with $n=0$ is subtracted, $V_3$ is set equal to unity)
\be
P_{gl}={(N^2_c-1)}\int^\infty_0\frac{ds}{s}\sum_{n\neq 0}G^{(n)}(s)
\label{P1}
\ee
for gluons and
\be
P_q={2N_c}\int^\infty_0 \frac{ds}{s} e^{-m_q^2s}\sum^\infty_{n=1}(-1)^{n+1} [S^{(n)} (s)+ S^{(-n)}(s)],
\label{P2}
\ee
for quarks, respectively.

Therefore, in order to calculate $G^{(n)}(s)$ and $S^{(n)}(s)$, one needs to evaluate the
Wilson loop operators $\lan W_{\Sigma,\sigma}(C_n)\ran$.
First, we neglect the spin-dependent terms in these Wilson loops
since spin-dependent interactions at $T=0$ were considered before,
while their generalisation for $T>0$ is straightforward. Thus we deal with the operator $\lan W(C)\ran$.
Let us look more carefully into the topology of this Wilson loop.
It is important to bear in mind that the basis states $|k\ran$ with the quark-gluon numbers
$N_g^{(k)}$, $N^{(k)}_q$, $N_{\bar q}^{(k)}$ entering the partition function,
\be
Z=\sum_n \lan n|e^{-H/T}|n\ran,
\label{26}
\ee
are gauge-invariant, and one can use
a generic decomposition of the type: $|k\ran=|(gg)(gg)(ggg)(q\bar q)\ldots\ran$,
where particles in parentheses form white combinations.
Then, neglecting interactions between white subsystems, one finds that the contribution of
such white subsystems, for example, $(gg)$ and $(q\bar q)$, to be
$$
Z_{(12)}=\int d\Gamma_1 d\Gamma_2 \lan \Tr W(C^{(1)}_n, C^{(2)}_n)\ran,\quad
d\Gamma_i=ds_i Dz_i e^{-K_i},
$$
where $W(C^{(1)}_n C_n^{(2)})$ is a closed Wilson loop made of paths
of the two gluons, of the $q\bar q$ pair, and, finally, of the parallel transporters (Schwinger
lines) in the initial and final states --- the latter are necessary to make the states gauge-invariant.

To proceed we apply the non-Abelian Stokes theorem and the Gaussian
approximation (see \cite{10,11} for the discussion and relevant
references) to arrive at (\ref{8}), and it is important now to
specify the surface $S_n$ with the surface elements
$ds_{\mu\nu}(u)$ in the integral on the r.h.s. of (\ref{8}). For
this analysis let us to consider colour-electric and
colour-magnetic contributions separately. For the field correlator
we use formulae (\ref{Hs0}).

Consider first the colour-electric piece with $\nu=\sigma=4$ (note
that, by definition, $\mu<\nu$ and $\lambda<\rho$):
\be
J_n^E\equiv\frac12\int D_{i4k4}(u-v)d\sigma_{i4}(u)d\sigma_{k4}(v).
\label{P9}
\ee

The surface $S_n$ is inside the winding Wilson loop for the gauge-invariant $q\bar q$ system.
Since the correlator $D^E(u)$ vanishes at $T>T_c$, while the contribution of the correlator $D^E_1$
leads, at large interquark distances, to a sum of terms for the quarks and antiquarks separately (we shall study the interaction of quarks in
chapter~\ref{BSLA} below). Then for a single quark one obtains in this way
$$
J_n^E=\frac12\int^{n\beta}_0 du_4 \int^{n\beta}_0 dv_4
\int^\infty_0 \xi d\xi D_1^E (\sqrt{(u_4-v_4)^2+\xi^2})=$$
\be
=\frac12 n\beta \int^{n\beta}_0 d\nu
\left(1-\frac{\nu}{n\beta}\right) \int^\infty_0
\xi d\xi D_1^E (\sqrt{\nu^2+\xi^2}).
\label{P10}
\ee

Note that, for $n=1$, one recovers the expression for the Polyakov
loop obtained in \cite{an1,23}, namely \be L_{\rm
fund}=\exp\left(-\frac{1}{2T}V_1(\infty)\right),\quad\frac{1}{2T}V_1(\infty)=J^E_1.
\label{P11} \ee

At this point an important simplification occurs. Originally the path $C_n$ is given by a complicated
function $z_\mu(\tau)$ in four dimensions. However the integral over $d\xi$ in (\ref{P10}) is always
from a point $z_\mu(\tau)$ on $C_n$ to infinity and it does not depend on the particular form of $C_n$ ---
it is the same as for the straight-line Polyakov loop. This is true, however, only for the $D_{i4k4}$
and not for $D_{iklm}$.

Another important observation is that the correlator $D_{\mu\nu\lambda\rho}(u-v)$ is not a periodic function of
$(u_4-v_4)$, in contrast to the fields $F_{\mu\nu}(u)$ and $F_{\lambda\sigma}(v)$.
This will be true also for the path integrals from any given point $x$ to an arbitrary point $y$,
and is a consequence of the vacuum average of the parallel transporter $\Phi(u,v)$ present in
$D_{\mu\nu\lambda\rho}(u-v)$.

Notice that the mixed contributions in (\ref{P9}),
\be
J^{EH}_n\equiv\frac12\int D_{i4kl}(u-v)d\sigma_{i4}(u)d\sigma_{kl}(v),
\label{P12}
\ee
is $T$-independent, so we shall neglect it in what follows.

We finally turn to the spatial term, which will play a special role in what follows.
Here the main contribution comes from the colour-magnetic correlator $D^H(u-v)$,
which provides the area law for the (closed) spatial projection $A_3$ of the surface $S_n$.
Correspondingly we shall write this term as
(for $ A_3\gg T_g^2$, where $T_g$ is the gluon correlation length,
$D^H(x)\sim\exp(-x/T_g)$):
\be
\lan W_3(C_n)\ran=\exp(-\sigma_sA_3),
\label{P13}
\ee
where the spatial string tension is defined in (\ref{sigmas})
and $A_3$ is the minimal area of the spacial projection of the
surface $S_n$. Similarly to the case of the $J^E_n$, one should start with white states of
$q\bar q$, $gg$ or $3q$. As will be seen, the colour-magnetic vacuum $(D^H)$ acts in the centre-of-mass frame
only in the states of the system with $L\neq 0$ and the resulting contribution of the $D^H$ does not separate into single-line terms,
but it rather acts pairwise (triple-wise for $3q$). Therefore one can account for colour-magnetic
interaction in the higher (2-line or 3-line) terms. We shall neglect colour-magnetic
interaction in the first approximation, as it is weak as compared to the colour-electric one. As a result, the 4D dynamics in (\ref{P3})
separates into 3D and 1D ones, and it is possible to write
\be
G^{(n)}(s)\equiv G_4^{(n)}(s)G_3(s)=\int (Dz_4)^w e^{-K_4-J_n^E}G_3(s)
\label{P15}
\ee
and, similarly, $S^{(n)}(s)=S_4^{(n)}(s)S_3(s)$, with the free quark (gluon) factors
\be
S^{(0)}_3 (s)=G_3^{(0)}(s)=\int (D^3z)e^{-K_3}=\frac{1}{(4\pi s)^{3/2}}.
\label{P16}
\ee

One should notice at this point that $G^{(n)}(s)$ and $S^{(n)}(s)$ differ in the spin factors
and in the colour group representation. Neglecting, in the lowest approximation, the spin factors,
one obtains $S^{(n)}$ from $G^{(n)}$ by simply replacing adjoint quantities by fundamental
ones. Note that one can use Casimir scaling \cite{11,Cas} to write
$\sigma_s(\mbox{\rm adj})=\frac94\sigma_s(\mbox{\rm fund})$.

To compute $G_4^{(n)}(s)$ one can take into account that $J_n^E$ does not depend on $z_4$ and can be
pulled out of the integral, while the remaining integral over $z_4$ can be taken exactly. In particular,
splitting the proper-time interval $N\varepsilon=s$, $N\to\infty$ one can write:
$$
\int(Dz_4)^w e^{-K_4}=\prod^N_{k=1}\left(\frac{d\Delta z_4(k)}{\sqrt{4\pi\varepsilon}}e^{-\frac{(\Delta z_4
(k))^2}{4\varepsilon} } e^{ip_4 \Delta z_4 (k)}\right)e^{-ip_4n\beta} \frac{dp_4}{2\pi}=
$$
\be
=\int\frac{dp_4}{2\pi}e^{-ip_4 n\beta-p^2_4 s}=\frac{1}{2\sqrt{\pi s}}e^{-\frac{n^2 \beta^2}{4s}}.
\label{P17}
\ee
As a result, one arrives at
\be
G^{(n)}(s)=\frac{1}{2\sqrt{\pi s}}e^{-\frac{n^2\beta^2}{4s}-\tilde{J}^E_n}G_3^{(0)}(s)
\label{P18}
\ee
$$
S^{(n)}(s)=\frac{1}{2\sqrt{\pi s}} e^{-\frac{n^2\beta^2}{4s}- J^E_n}S_3^{(0)}(s),\quad
\tilde{J}^E_n=\frac94 J^E_n.
$$

Substituting (\ref{P15}) -- (\ref{P18}) into (\ref{P1}) and (\ref{P2}), one arrives at the expressions for
the pressure with the only unknown quantity being the fundamental Polyakov line $J^E_n$. In the next chapter
we discuss in detail the behaviour of the Polyakov lines at finite temperatures.

\subsection{Polyakov lines at finite temperatures}

An important result of the previous chapter is that colour-electric correlator $D^E_1$ yields
factorised contribution, that is $Z_{(12)}=Z_1Z_2$, where each individual factor $Z_i$ contains
only the part of the common loop in the form of the Polyakov loop factor (with the singlet free
energy in the exponent). The contribution of colour-magnetic fields does not factorize, however,
this contribution can be considered as a correction, so it will be neglected.
In addition, the average of the Polyakov line operator was related to the colour-electric correlator
$J_1^E$ and to the static potential $V_1(\infty)$. In this chapter we discuss the behaviour of the Polyakov
lines and static potentials in the Gaussian approximation to the QCD vacuum.

At $T_c>T>0$ one has four Gaussian correlators $D^E(x)$,
$D_1^E(x)$, $D^H(x)$, and $D_1^H(x)$, with the string tensions
$\sigma^{E,H}$ given by (\ref{sigmas}). At $T>T_c$ the correlator
$D^E$ (and, consequently, $\sigma^E$) vanishes, as was suggested
in \cite{6,7} and then was proved on the lattice \cite{7*}. The
three remaining correlators are nonzero, moreover the spatial
string tension $\sigma_s\equiv\sigma^H$ grows with the temperature
in the dimensionally reduced limit \cite{Red}. This fact explains
also the growth with temperature of the Debye mass, $m_D\cong
2\sqrt{\sigma_s}$ \cite{25,13}, which is known from the lattice
data \cite{24}. Apart from this quantity, we shall not use below,
in this section, the colour-magnetic correlators, since they do
not produce static potentials between particles in the relative
$S$-wave. Therefore we are interested only in the colour-electric
correlators $D^E(x)$ (inside the confined phase bounded by the
curve $T_c(\mu)$) and $D_1^E(x)$ (in the whole $\mu, T$ plane).

Let us stress once more that, in our approach, only
gauge-invariant states $|n\ran$ are to be considered in the
partition function (\ref{26}) at $T>0$. This is evident in the
confined phase, since any coloured part of the given
gauge-invariant system is connected to other parts by strings. In
absence of colour-electric strings in the deconfined phase the
necessity of using gauge-invariant amplitudes is less evident,
except for world-lines in the spatial directions, where
colour-magnetic confinement with a nonzero $\sigma_s$ is
operating. Our use of gauge-invariant amplitudes, which factorise
at large interparticle distances in the deconfined phase, leads to
an explicit prediction for the EoS with modulus of phase factors,
which approximately equal to modulus of Polyakov lines.

Below we use, as in \cite{an1,an2}, gauge-invariant states, $|n\ran$ at all temperatures and chemical potentials
and describe the interparticle dynamics in terms of gauge-invariant quantities, like pair-wise or triple static potentials.
The large-distance limit of these potentials yields one-particle characteristics --- the self-energy parts of quarks, antiquarks,
gluons, and so on. One can use those to study thermodynamics of QGP in the one-particle, or SLA \cite{an1,an2}. It is rewarding,
that the field correlator method is a natural instrument in describing this deconfined dynamics since, in absence of the $D^E$,
the correlator $D^E_1$ has the form of the full derivative and produces gauge-invariant one-particle pieces --- self-energy
parts --- automatically (in addition to the interparticle interaction decreasing at large distances).

Gauge-invariant states $|n\ran$, formed with the help of parallel
transporters (Schwinger lines) (\ref{PA}), create, as shown in
\cite{an1}, Wilson loops $W(C)$ for $q\bar q$, $qqq$, or other
quark systems. Lattice data allow one to extract static interquark
potentials in such multiquark systems \cite{9,26}. When treating
coloured systems, like $(qq)$, taken as a part of a
gauge-invariant system ($qq\bar q\bar q$ in this case), the pairs
$(qq)$ and $(\bar{q}\bar{q})$ are separated at a large distance
and the potential $V(qq,\bar q\bar q$) is neglected.

We start from the colour-singlet $q\bar q$ system and write
contributions of the $D^E$, and $D^E_1$ to the static potentials
at a nonzero $T=1/\beta$ \cite{9}: 
\be 
V_0(r,T)=2\int^\beta_0 d\nu (1-\nu T) \int^r_0 (r-\xi) d\xi D^E(\sqrt{\xi^2+\nu^2}),
\label{28} 
\ee 
\be 
V_1(r,T)=\int^\beta_0 d\nu (1-\nu T) \int^r_0\xi d\xi D^E_1(\sqrt{\xi^2+\nu^2}), 
\label{27} 
\ee 
which contribute to the modulus of the Polyakov loops as  
\be 
L_{\rm fund}^{(V)}=\exp\left(-\frac{V_1(T)+2V_0}{2T}\right),\quad L_{\rm adj}^{(V)}=\left(L_{\rm fund}^{(V)}\right)^{9/4} 
\label{29} 
\ee
where $V_1(T)\equiv V_1(\infty, T)$, $V_0\equiv V_0(r^*, T)$, and
$r^*$ is an average distance between the heavy-quark line and
light antiquark (for $n_f>0$) or between the ``heavy-gluon line''
and a gluon for $L_{\rm adj}$. The Casimir scaling relation
(\ref{29}) predicted in \cite{11} is in good agreement with
lattice data \cite{24}, as well as vanishing of the $L_{\rm fund}$
for $T\leq T_c$, $n_f=0$ and the strong decrease of $L_{\rm adj}$
for $T\leq T_c$. Indeed, for $T \leq T_c$ and $n_f=0$ one has
$r^*\rightarrow \infty$ and $V_0\rightarrow \infty$, explaining
vanishing of the $L_{\rm fund}^{(V)}$. For the $L_{\rm adj}^{(V)}$
in this region one can take into account the kinetic energy of the
gluon in the system consisting of an adjoint source plus a gluon
(gluelump). This yields an estimate $L_{\rm adj}(T\leq
T_c)=\exp(-m_{glp}/T)$, where $m_{glp}$ was computed in \cite{13}
to be of order 1~GeV.

Notice that the Polyakov lines measured on the lattice are expressed through the (singlet) free energy of a $Q\bar Q$ system at large
distances $F^1_{Q\bar Q}(\infty, T)$, in the same way as in (\ref{29}),
\be
L_{\rm fund}^{(F)}=\exp\left(-\frac{F^1_{Q\bar Q}(\infty, T)}{2T}\right).
\label{30}
\ee
In order to relate $L_j^{(F)}$ and $L_j^{(V)}$ one can use the standard representation for the free energy $F^1_{Q\bar Q}(r, T)$,
\be
\exp \left(-\frac{F^1_{Q\bar Q} (r,T)}{T}\right)=\sum_{n(Q\bar Q)} c_n\exp \left(-\frac{V_n^{Q\bar Q}(r,T)}{T}\right),
\label{31}
\ee
where the summation covers all excited and bound states involving $Q\bar Q$, and $V_n^{Q\bar Q}(r,T)$ is the energy term
of such a state $n$ with the distance between the static charges $Q$ and $\bar Q$  being equal to $r$.
It is clear that $L_j^{(V)}$ coincides with $L_J^{(F)}$ when all states $n$, except for the ground state $n=0$, are neglected.
In this case $V_0^{Q\bar Q}(r,T)$ coincides with $V_1(r, T)$ and, hence, with the $F^1_{Q\bar Q}(r, T)$.
Note that $V_1(r, T)$ in (\ref{27}) does not depend on $T$ in the string limit of QCD, when the vacuum correlation length $T_g$
tends to zero.

In the general case all states $n(Q\bar Q)$ contribute and therefore $(c_n>0)$ one has the following inequality:
\be
V_1(r, T) \geq F^1_{Q\bar Q} (r, T)
\label{32}
\ee

In order to define $V_1$ and $L_{\rm fund}$ properly one needs to
separate their perturbative and nonperturbative parts and to
renormalise $V_1$ to get rid of perimeter divergences. The
separation in $D^E_1(x)$ can be seen at small $x$ \cite{anal} 
\be
D_1^E(x)=\frac{4C_F\alpha_s}{\pi}\left[\frac{1+O(\alpha_s \ln^k x)}{x^4}+\frac{\pi^2G_2}{24N_c}+\ldots\right]
=\left(D_1^E\right)^{\rm pert}+\left(D_1^E\right)^{\rm np},
\label{33} 
\ee 
and at large $x$, where $\left(D_1^E\right)^{\rm np}(x)$ behaves as \cite{anal}: 
\be 
\left(D_1^E\right)^{\rm np}(x)\cong A_1\frac{e^{-M_0|x|}}{|x|},\quad A_1=2C_F\alpha_s M_0\sigma_{\rm adj}, 
\label{34} 
\ee 
where $M_0$ is the lowest gluelump mass \cite{13}, $M_0 \approx 1$ GeV,
$C_F=(N_C^2-1)/(2N_C)=4/3$ is the fundamental Casimir operator,
and $G_2$ is the gluon condensate, $G_2=(\alpha_s/\pi)\langle
F^a_{\mu\nu}F^a_{\mu\nu}\rangle$.

The corresponding separation of the $V_1(r,T)$ is done in \cite{9} and reads:
\be
V_1(r,T)=V_1^{\rm pert}(r,T)+V_1^{\rm np}(r,T)+V_1^{\rm div}(a),
\label{35}
\ee
where
\be
V_1^{\rm pert}(r,T)=-\frac{C_F\alpha_s}{r}e^{-m_Dr}(1+O(rT)),
\label{36}
\ee
$V_1^{\rm np}$ is given by (\ref{27}), with the substitution $D^E_1\to (D_1^E)^{\rm np}$, and
\be
V_1^{\rm div}(a)\cong\frac{2C_F\alpha_s}{\pi}\left(\frac{1}{a}+O(T\ln{a})\right).
\label{37}
\ee
Here $m_D=m_D(T)\approx  2\sqrt{\sigma_s}$ is the nonperturbative Debye mass \cite{25} and $a$ is the lattice spacing (cut-off).

\begin{figure}[t]
\begin{center}
\includegraphics[width=8cm]{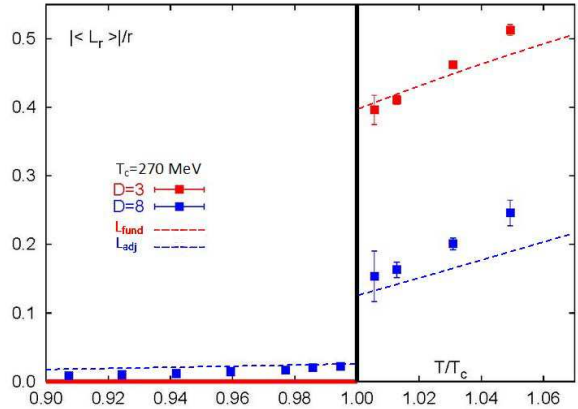}
\end{center}
\caption{Shown on the figure are curves of $L_{\rm adj}$ (blue
dashed) and $L_{\rm fund}$ (red dashed) compared to the ones taken
from \protect\cite{24}. In the $T<T_c$ region the
$M(\bar{\alpha}_s=0.195)=0.982$ GeV gluelump mass was used. In the
deconfinement region the fit (\ref{39}) was used with $T_c=270$
MeV for $L_{\rm fund}$ and the Casimir scaled value for $L_{\rm
adj}$.}\label{f1}
\end{figure}

The renormalisation procedure suggested in \cite{9} amounts to
discarding $V_1^{\rm div}(a)$, and this is in agreement with the
lattice renormalisation used in \cite{26}, where $F^1_{Q\bar
Q}(r,T)$ was adjusted to the form $V_1^{\rm pert}(r, T)$ at small
$r$ and $T$. Note, that $V_1^{\rm np}(r,T)\sim O(r^2)$ in this
region and the procedure indeed allows one to eliminate the
constant term $V_1^{\rm div}(a)$.

Let us discuss now the contribution of the $V_1$ to the interaction in $Q\bar Q$, $QQ$, and $3Q$ systems.
We start with the one-particle limit of the $V_1(r,T)$ and the corresponding contribution to the $L_{\rm fund}^{(V)}$.
According to the discussion above, one defines the renormalised Polyakov loop as in (\ref{29}), (\ref{35}), with
$V_1(T)\equiv V_1^{\rm np}(\infty,T)$. We shall neglect the difference between $L_j^{(V)}$ and $L_j^{(F)}$
(this difference becomes important at large $T$'s, when $L_{\rm fund}^{(F)}>1$, while $L_{\rm fund}^{(V)}<1$).

From (\ref{34}) one has (at $T\leq T_c$): 
\be 
V_1^{\rm np}(\infty,T)=\frac{A_1}{M_0^2}\left[1-\frac{T}{M_0}\left(1-e^{-\frac{M_0}{T}}\right)\right], 
\label{38}
\ee 
so that, for $M_0 \approx 1$ GeV, 
\be 
V_1^{\rm np}(\infty,T_c)\approx \frac{6\alpha_s(M_0)\sigma_{\rm f}}{M_0}\approx 0.5~\mbox{GeV}. 
\ee 
A similar estimate one obtains from the lattice data \cite{27} which, at $T\sim T_c$, can be
parameterize as 
\be 
F^1_{Q\bar Q}(\infty,T)\approx\frac{0.175}{1.35\left(\frac{T}{T_c}\right)-1},\quad
F^1_{Q\bar Q}(\infty,T_c)\approx 0.5~\mbox{GeV}. 
\label{39} 
\ee

Thus quarks (and antiquarks) have self-energy parts which, at
$T\approx T_c$, are given by $\kappa_q(T)=\kappa_{\bar q}(T)=V_1(T)/2\approx F^1_{Q\bar Q}(\infty, T)/2\approx 0.25$ GeV.

To illustrate the discussion of the $V_0$, $V_1$, and $L_{\rm fund}$, $L_{\rm adj}$ we compare in Fig.~\ref{f1} our theoretical
curves for the $L_{\rm fund}$ and $L_{\rm adj}$ (dashed curves)
with the lattice data taken from \cite{24} (dots). The theoretical
curves follow from (\ref{29}), with $V_1(\infty,T)=F_{Q\bar
Q}^1(\infty,T)$ taken from (\ref{39}) for $T>T_c$ and $L_{\rm
adj}=\exp{(-\frac{M_0}{T})}$ for $T \leq T_c$. From Fig.~\ref{f1}
one can see a good agreement between our theoretical predictions
and the lattice data.

Similarly, for gluons, one has at $T\sim T_c$ that $\kappa_g(T)=(9/4)\kappa_q(T)\approx 0.56$ GeV.

Let us turn now to the $r$-dependence of the interaction. The perturbative part has a standard screened Coulomb behaviour
(\ref{36}), while the nonperturbative part vanishes at small $r$'s:
\be
V_1^{\rm np}(r, T)\mathop{\sim}\limits_{r\to 0}\mbox{const}\cdot r^2.
\label{40}
\ee

From (\ref{27}) and (\ref{34}) one has \cite{9}
\be
V_1^{\rm np}(r,T)=V_1^{\rm np}(\infty,T)-\frac{A_1}{M^2_0}K_1(M_0 r)M_0r+O\left(\frac{T}{M_0}\right)\equiv V_1^{\rm np}(\infty,T)+v(r,T).
\label{41}
\ee
Therefore the nonperturbative interaction in the white system $Q\bar Q$ changes from $V_1^{\rm np}(\infty, T)\approx 0.5$ GeV at large
$r$'s to zero at small $r$'s. The same (multiplied by the factor 9/4) is true for the white $gg$ system.

We conclude this chapter by a discussion of the role of excited
states in the definition of the $F^1_{Q\bar Q}$ and of a possible
violation of the Casimir scaling for the $L_{\rm fund}$ and
$L_{\rm adj}$. It is clear that in the $F^1_{Q\bar Q}$ for $n_f=0$
the only possible excited states consist of gluons: $(Qg)(\bar
Qg)$, $(Qgg)(\bar Qgg)$, and so on. As it was shown in \cite{8},
the weakly bound states $(Qg)$ are indeed supported by $V_1(r,T)$
and, in neglect of the small binding energy, the total energy of
these states is roughly given by the sum of the self-energy parts
$\kappa_Q$ and $\kappa_g$, 
\be
E_{Qg}\approx\frac{1}{2}V_1(\infty,T)+\frac{9}{8}V_1(\infty,T)\approx 0.8~\mbox{GeV}(T\approx T_c). 
\label{42} 
\ee 
This should be compared with the possible bound state of an adjoint static source
$G$ plus gluon which, in the weakly binding limit, can be written as 
\be 
E_{Gg}\approx 2\times\frac{9}{8}V_1(\infty,T)\approx1.1~\mbox{GeV}(T\approx T_c). 
\label{43} 
\ee 
Notice that multiplicities of the states (\ref{42}) and (\ref{43}) are different, which leads to different
predictions for corrections to the $F^1_{Q\bar Q}$ and $F^1_{GG}$.
These corrections are not connected by the Casimir scaling, in
contrast to the main (ground-state) term, for which $V^1_{Q\bar
Q}=V_1(\infty,T)$ and $V^1_{GG}=\frac{9}{4}V_1(\infty,T)$.
Therefore one expects some violation of the Casimir scaling by
gluon-induced bound states in the $L_{\rm fund}$ and $L_{\rm adj}$, though high precision lattice data \cite{24} indicate a
small role of such bound states.

\begin{figure}[t]
\begin{center}
\includegraphics[width=8cm]{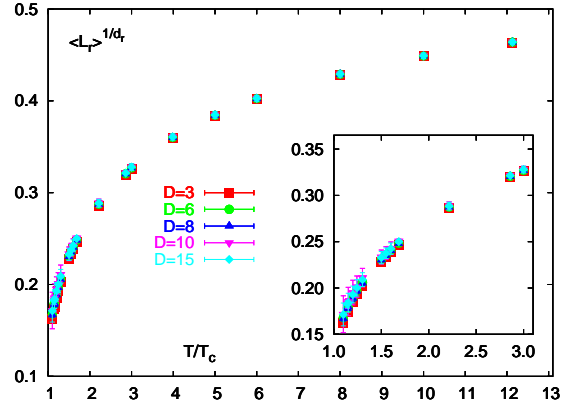}
\end{center}
\caption{The Casimir-scaled bare Polyakov loops for different
representations $D$ measured on $32^3\times 4$ lattices
\protect\cite{24}. } \label{fig:pol_bare_T}
\end{figure}

\subsection{Equation of state}

Here we follow the approach to study the QGP dynamics
\cite{an1,an2}, where the main emphasis was done on the vacuum
fields, and the resulting modification of quark and gluon
propagators was considered as the first and the basic step in the
nonperturbative treatment of QGP
--- the SLA discussed before. As a result one obtains the
nonperturbative EoS of QGP in the form of free quark and gluon
terms multiplied by the vacuum induced factors. The latter are
expressed via the only (nonconfining) colour-electric correlator
$D_1^E(x)$ and happened to be approximately equal to the absolute
values of Polyakov loops $L_{\rm fund}$, $L_{\rm adj}$ for quarks
and gluons, respectively. Furthermore, due to the Casimir scaling
property \cite{24,11}, these Polyakov loops are related to one
another, as $L_{\rm adj}=(L_{\rm fund})^{9/4}$.

Moreover, the phase diagram was calculated in the SLA assuming
that the phase transition is vacuum dominated, that is, a
transition from the confining vacuum with vacuum energy density
$\varepsilon_{\rm conf}\cong-\frac{\beta_0}{32}G_2({\rm conf})$ to
the nonconfining vacuum with $\varepsilon_{\rm dec}\cong
-\frac{\beta_0}{32} G_2({\rm dec})$. The resulting phase curve
$T_c(\mu)$ in \cite{ST} depends on $\Delta G_2=G_2({\rm
conf})-G_2({\rm dec})$ and it was found to be in good agreement
with lattice data \cite{0701210} for standard values of the
$G_2({\rm conf})$ and $\Delta G_2\approx 0.5 G_2$ --- this will be
discussed in detail in the next chapter.

Thus the SLA is a reasonable starting point with no fitting or
model parameters, since $L_{\rm fund}$ can be computed
analytically \cite{anal} or on the lattice \cite{24}, and $\Delta
G_2 \approx 0.5 G_2$ is a fundamental parameter of QCD. This
picture of the QCD phase transition was called in \cite{ST} the
Vacuum Dominance Model (VDM) originally proposed in \cite{6,8*} in
a simplified form (sometimes called the Evaporation Model).

\begin{figure}[t]
\begin{center}
\includegraphics[width=57mm]{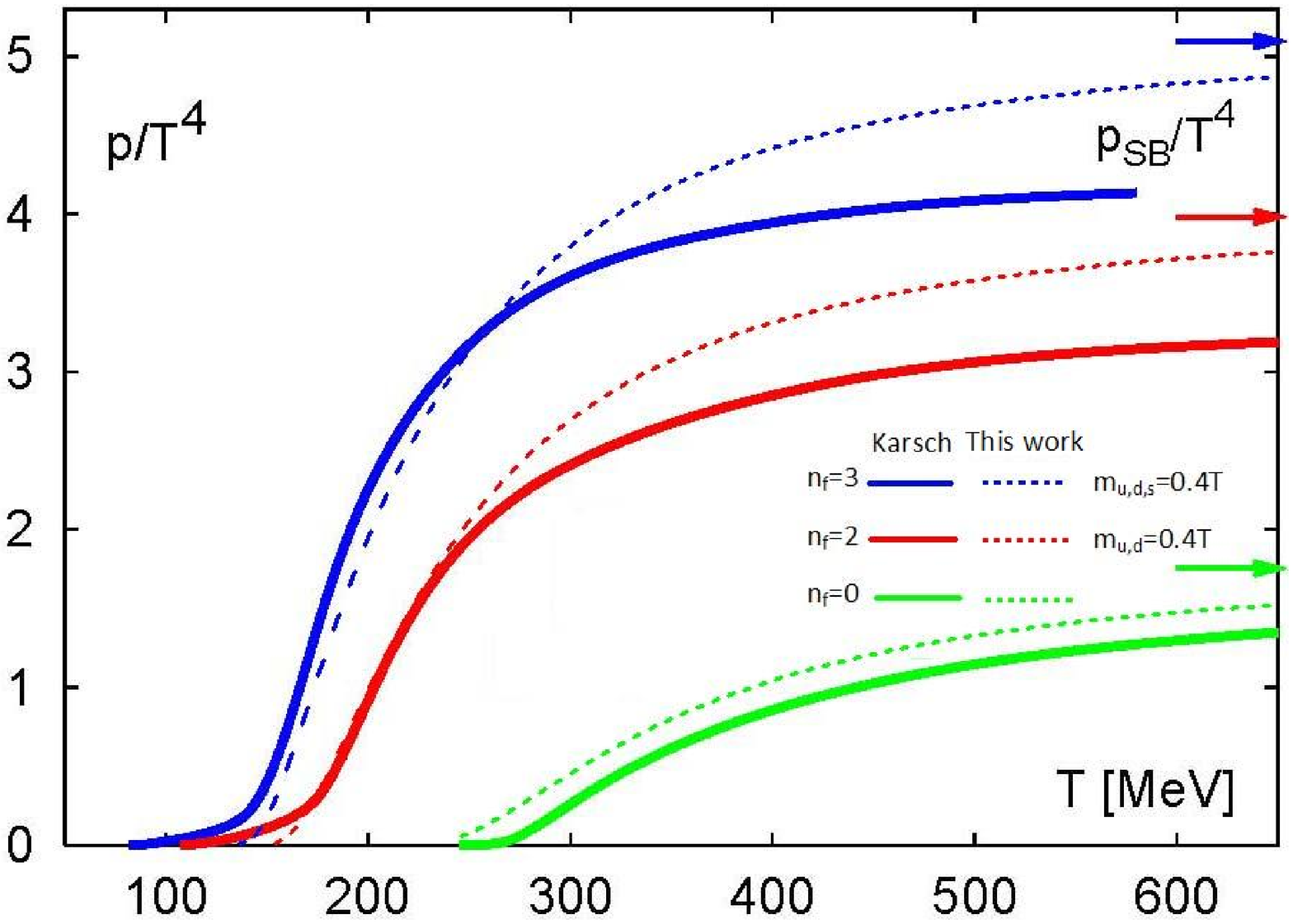}\hspace*{5mm}
\includegraphics[width=63mm]{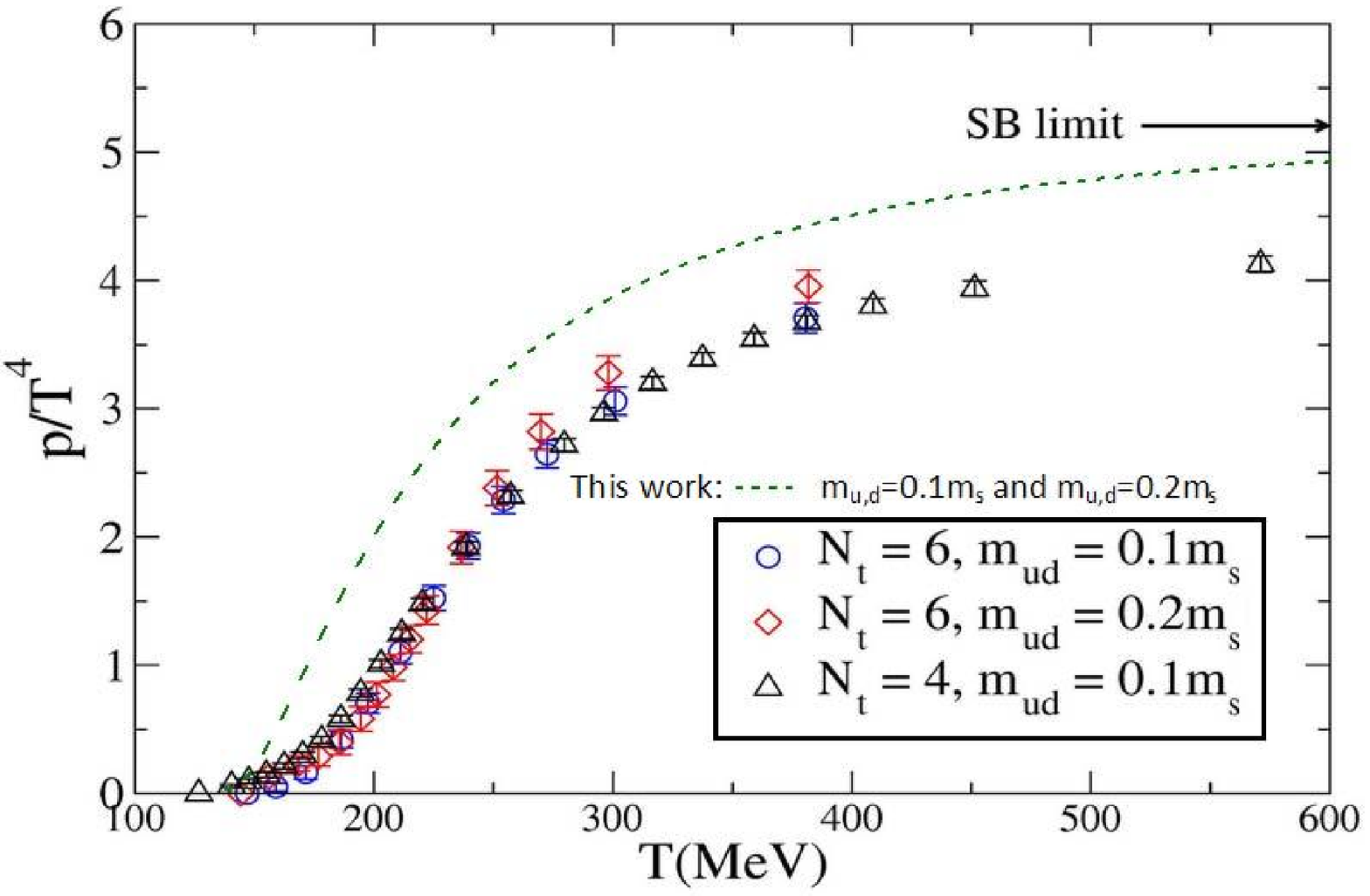}
\end{center}
\caption{Pressure $\frac{P}{T^4}$ as function of temperature $T$.
Shown on the left figure is a comparison of the analytical
calculus (\ref{51})(dashed lines) with the lattice results (bold
lines) \protect\cite{0608003} for the case $n_f=0,2,3$. Shown on
the right figure is the case of $n_f=2+1$. Green dashed line is
the analytical calculation (\ref{48}) compared to the lattice one
from \protect\cite{0610017}.}\label{f10}
\end{figure}

\begin{figure}[t]
\begin{center}
\includegraphics[width=6cm]{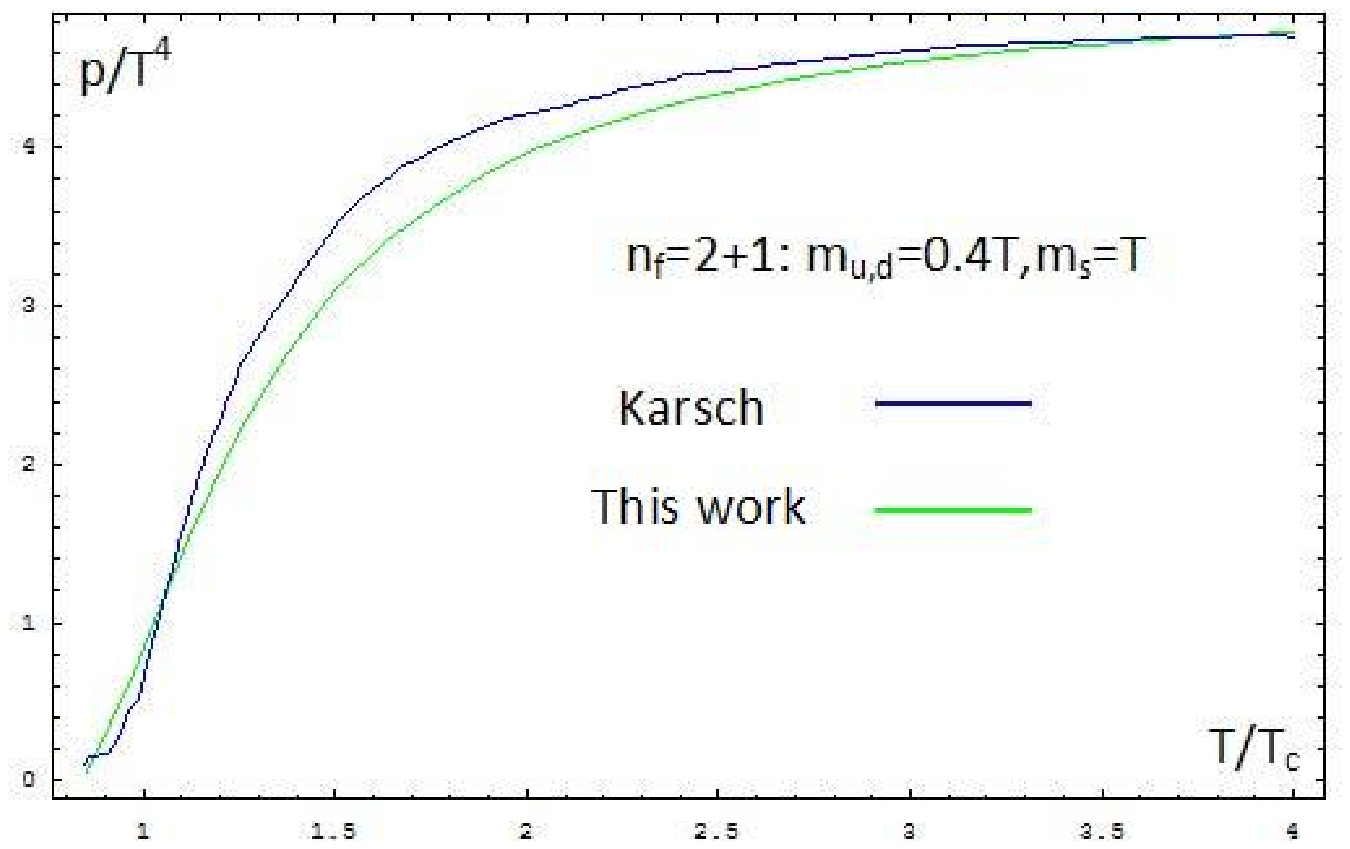}\hspace*{5mm}
\includegraphics[width=6cm]{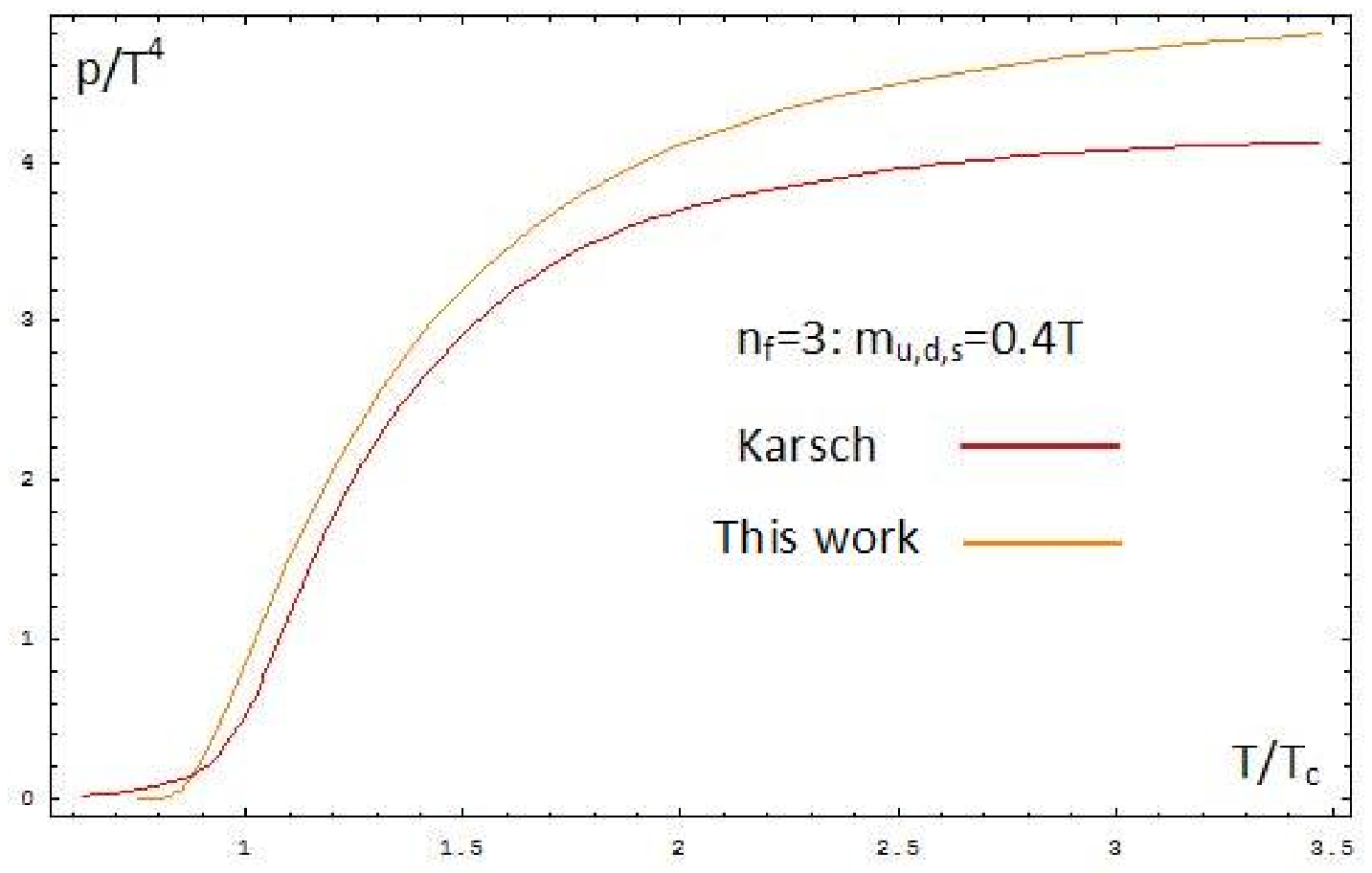}
\end{center}
\caption{Pressure $\frac{P}{T^4}$ as function of temperature  $T$.
The case of $n_f=2+1$ (left figure) and $n_f=3$ (right figure)
(\ref{51}). Lattice results were taken from
\protect\cite{0701210}.}\label{f11}
\end{figure}

In this chapter we exploit the reduced pressure $p=\frac{P}{T^4}$,
and combine (\ref{P1}),(\ref{P2}) and (\ref{P17}),(\ref{P18}) to
obtain a simple form, which can be written as:
\begin{equation}
p_q=\frac{P_q^{SLA}}{T^4}=\frac{4N_c n_f}{\pi^2}\sum_{n=1}^{\infty} \frac{(-1)^{n+1}}{n^4} L_{\rm fund}^n\varphi_q^{(n)}
\label{44}
\end{equation}
and
\begin{equation}
p_{gl}=\frac{P^{SLA}_{gl}}{T^4}=\frac{2(N_c^2-1)}{\pi^2}\sum_{n=1}^{\infty}\frac{1}{n^4}L_{\rm adj}^n
\label{45},
\end{equation}
for quarks and gluons, respectively. Here $\varphi_q^{(n)}$ is
\begin{equation}
\varphi_q^{(n)}=\frac{n^2 m_q^2}{2 T^2} K_2\left(\frac{nm_q}{T}\right)=
\frac{n^4}{6}\int_{0}^{\infty} \frac{z^4}{\sqrt{z^2+\nu^2}}e^{-n \sqrt{z^2+\nu^2}}dz,\quad \nu=\frac{m_q}{T},
\label{47}
\end{equation}
where an integral representation of the Bessel function $K_2$ was used in the second equality.
Then both sums (\ref{44}) and (\ref{45}) can be evaluated explicitly to yield:
\begin{equation}
p_q=\frac{2N_c}{3}\frac{n_f}{\pi^2}\int_0^{\infty} \frac{z^4}{\sqrt{z^2+\nu^2}}\frac{dz}{e^{\sqrt{z^2+\nu^2}+a}+1},\quad
p_{gl}=\frac{N_c^2-1}{3\pi^2}\int_{0}^{\infty}\frac{z^3dz}{e^{z+a_{gl}}-1},
\label{48}
\end{equation}
with $a_q=V_1(T)/2T$, $a_{gl}=\frac{9}{4} a_q$.

Here we consider the case of $\mu=0$ (generalisation to the case of nonzero chemical potentials will be discussed in chapter~\ref{Disc})
and characteristic temperature region of $T\approx T_c$ ($T_c=170\div270$ MeV) where quark masses do not affect the thermodynamical
functions appreciably. This is due to the fast convergence of the sum over $n$ at large $n$'s ensured by the factors $1/n^4$ and
$L^n$ ($L<1$), while $\varphi^{(n)}_q\approx 1$ for $n\simeq 1$. Indeed, for $m_q=0$ $\varphi_q^{(n)}=1$, whereas for $m_q=0.4T$
one has $\varphi_q^{(1)}=0.96$, while $\varphi^{(15)}_q=0.03$. Therefore one can neglect, with a good accuracy, the masses in
(\ref{48}), arriving at:
\begin{equation}
p_q=\frac{2n_f}{\pi^2}\int_0^{\infty} \frac{z^3 dz}{e^{z+a_q}+1},\quad
p_{gl}=\frac{8}{3\pi^2}\int_0^{\infty} \frac{z^3 dz}{e^{z+a_{gl}}-1}.
\label{51}
\end{equation}

The results presented in (\ref{51}) are compared with the lattice pressure data, in Figs.~\ref{f10},
for $n_f=2+1$ (left) and $n_f=3$ (right figure). In Fig.~\ref{f11} we show our calculated curves for the cases $n_f=2+1$ (left
part) and $n_f=3$ (right part), which are compared with lattice data from \cite{0701210}.

As a further simplification, one can use, instead of (\ref{51}), the first terms of the expansion in (\ref{44}), (\ref{45}),
namely:
\be
p_q=\frac{12 n_f}{\pi^2}L_{\rm fund},\quad p_g=\frac{16}{\pi^2} L_{\rm adj}.
\label{53}
\ee

Other useful quantities to compare with the lattice data are the internal energy density and the ``nonideality'' of the QGP:
\begin{equation}
\varepsilon=T^2\frac{\partial}{\partial T}\left(\frac{P}{T}\right)_V=\varepsilon_q+\varepsilon_{gl}
\label{55}
\end{equation}
and
\begin{equation}
I(T)=\frac{\varepsilon-3P}{T^4}=T \frac{\partial p}{\partial T},
\label{58}
\end{equation}
respectively.

\begin{figure}[t]
\begin{center}
\includegraphics[width=6cm]{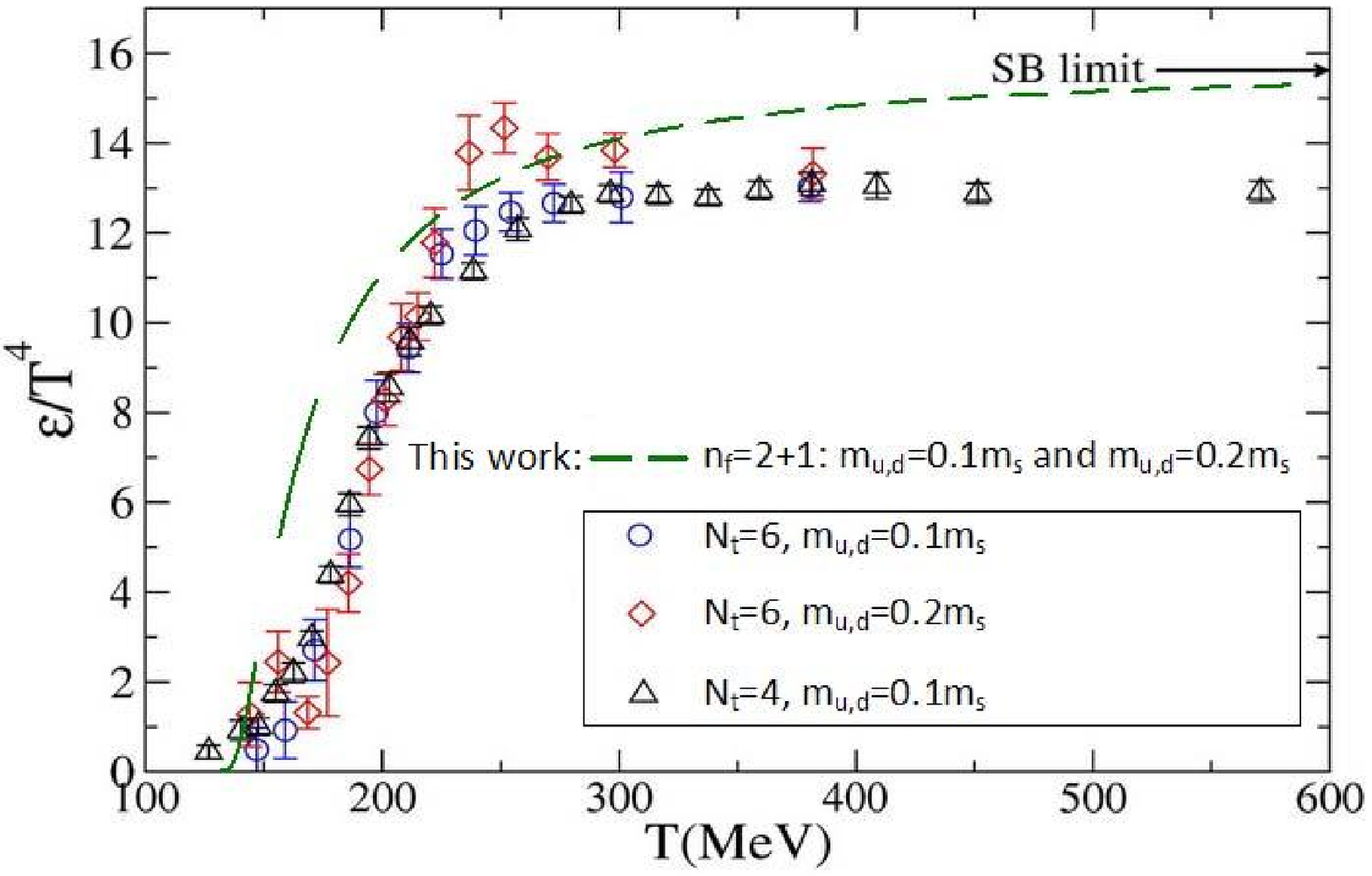}
\includegraphics[width=6cm]{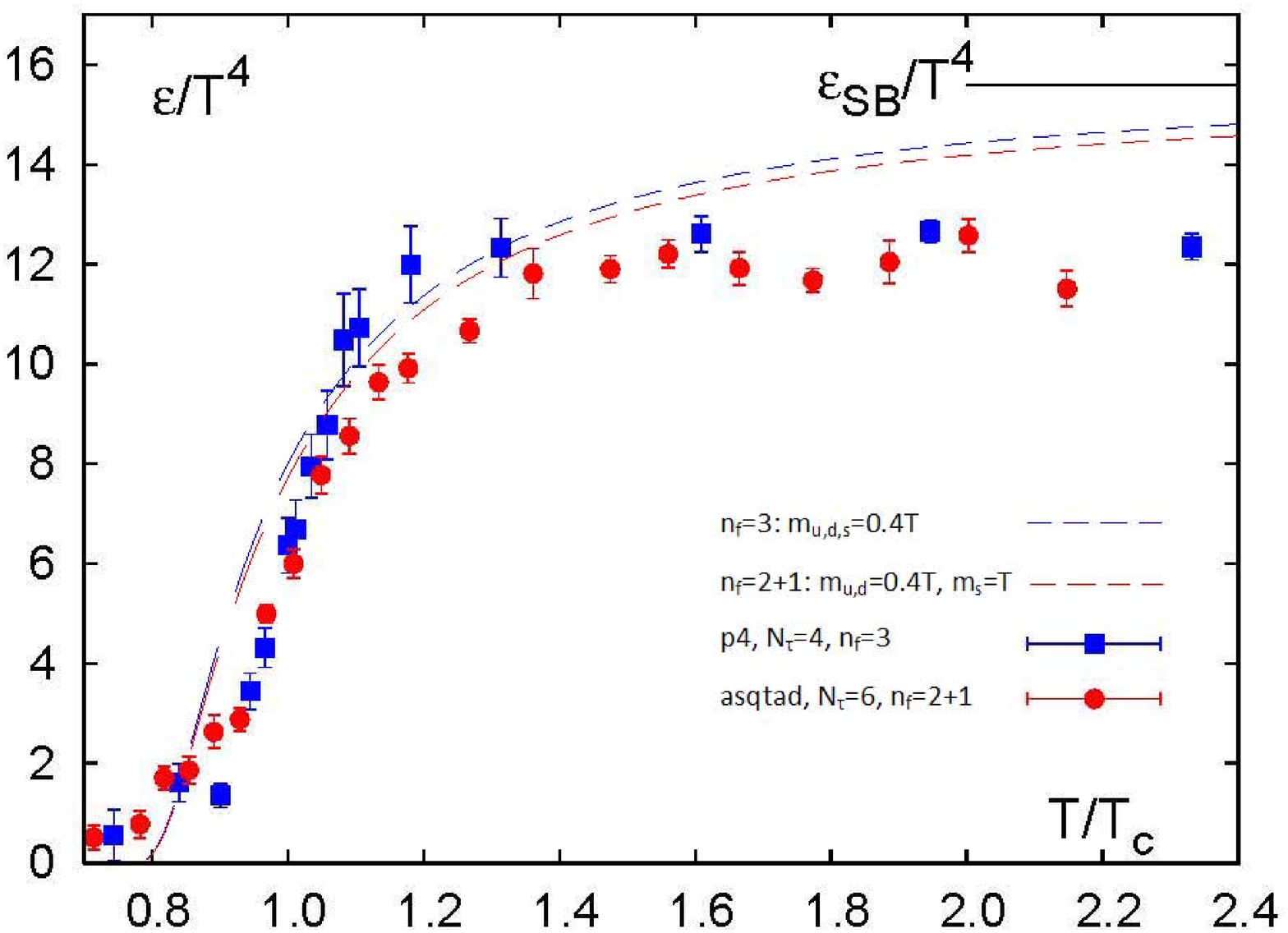}
\end{center}
\caption{Energy density $\frac{\varepsilon}{T^4}$ as function of
temperature  $T$. The case  $n_f=2+1$ with $m_{u,d}=0.1m_s$ and
$m_{u,d}=0.2m_s$ (green dashed curve) (\ref{55}) is compared to
lattice data from \protect\cite{0610017}(left fig.). The case
$n_f=2+1$ with $m_{u,d}=0.4T$, $m_s=T$ (red dashed curve) and
$n_f=3$ with $m_q=0.4T$ (blue dashed curve) (\ref{55}) are
compared to lattice data from \protect\cite{0608003}(right
fig.).}\label{f12}
\end{figure}

\begin{figure}[t]
\begin{center}
\includegraphics[width=8cm]{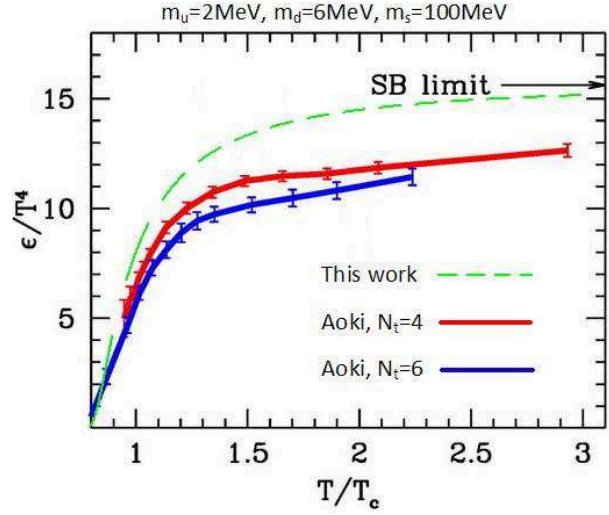}
\end{center}
\caption{Energy density $\frac{\varepsilon}{T^4}$ as function of
temperature  $T$. The curve for $n_f=3$ with $m_{u}=2$~MeV,
$m_{d}=6$~MeV, $m_{s}=100$~MeV (green dashed) (\ref{55}) is
compared to lattice data from \protect\cite{0510084}.}\label{f13}
\end{figure}

Using (\ref{48}) and (\ref{53}) one has
\begin{equation}
\varepsilon^{(0)}_{q}=\sum_{n_f}\frac{2}{\pi^2}T^2\frac{d}{dT}\left(T^3 \int_0^{\infty}
\frac{z^4}{\sqrt{z^2+\nu^2}}\frac{dz}{e^{\sqrt{z^2+\nu^2}+a_q}+1}\right),
\label{56}
\end{equation}
\begin{equation}
\varepsilon^{(0)}_{gl}=\frac{3}{3\pi^2}T^2\frac{d}{dT}\left(T^3 \int_0^{\infty} \frac{z^3dz}{e^{z+a_{gl}}+1}\right),
\label{57}
\end{equation}
and
\be
I(T)=\frac{12n_f}{\pi^2}T\frac{dL_{\rm fund}}{dT}+\frac{16}{\pi^2}T\frac{dL_{\rm adj}}{dT}
\label{59}
\ee

\begin{figure}[t]
\begin{center}
\includegraphics[width=6cm]{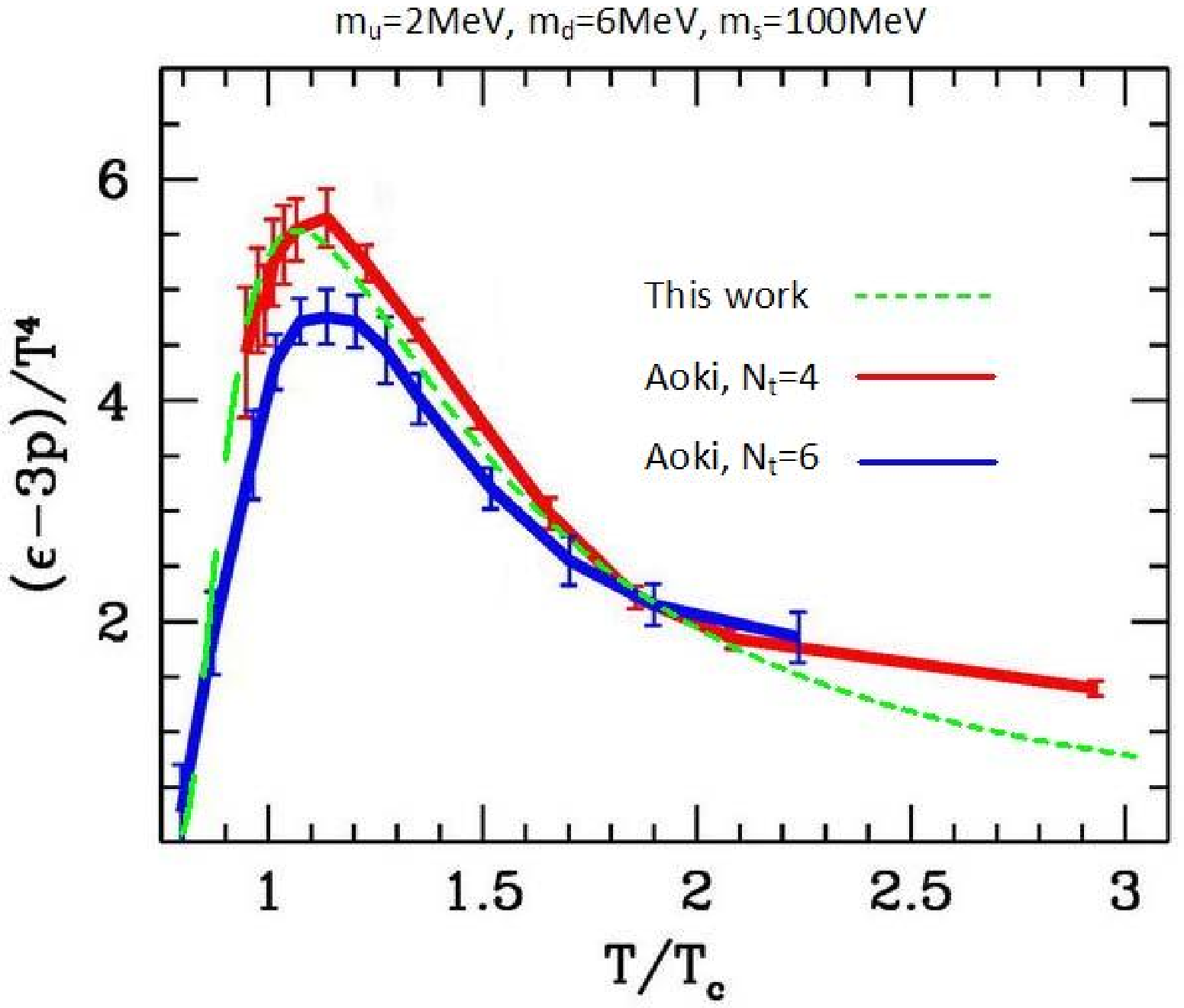}\hspace*{5mm}
\includegraphics[width=6cm]{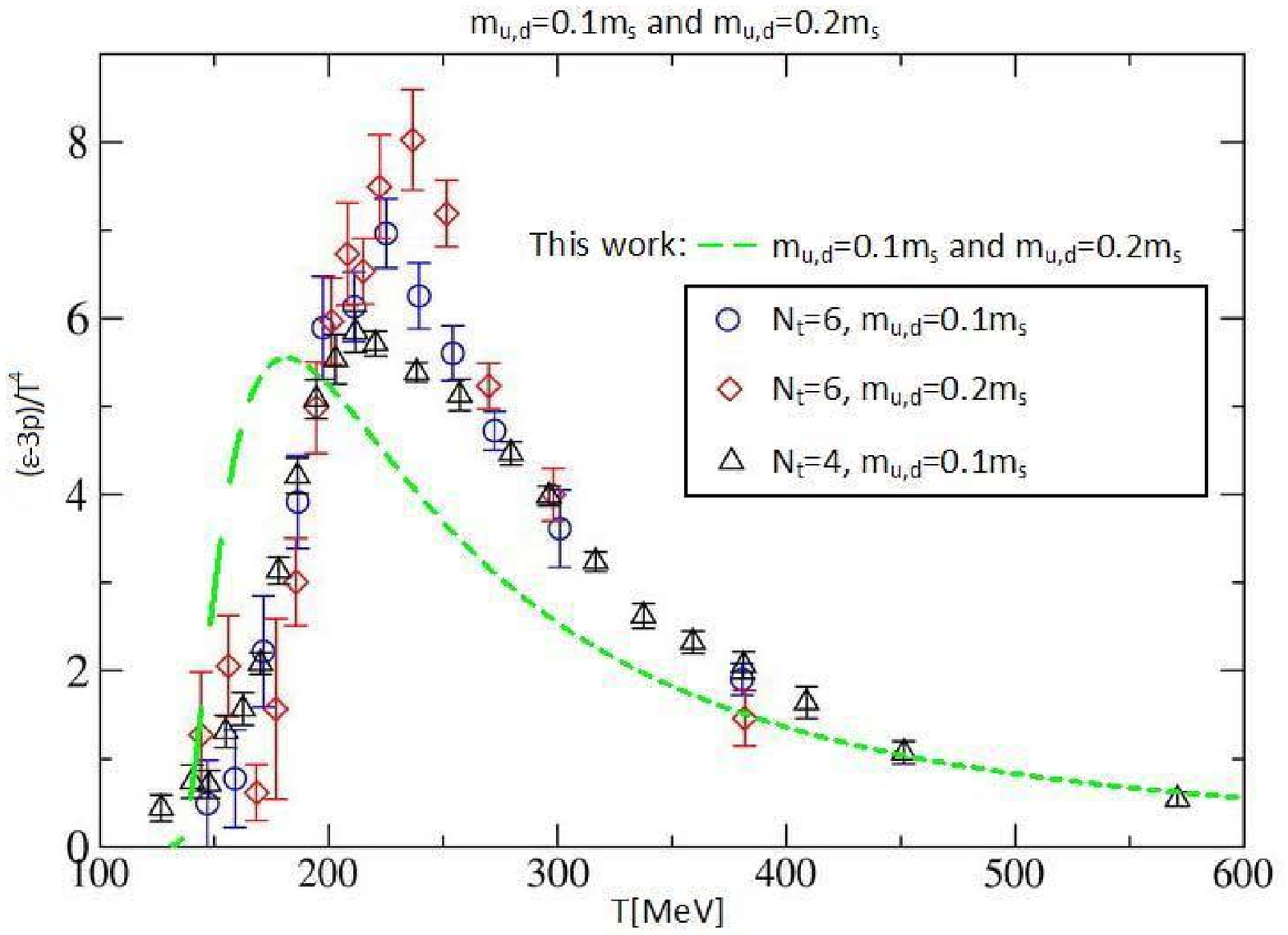}
\end{center}
\caption{"Nonideality" of QGP $(\varepsilon-3p)/T^4$. Shown are
the curves for (left fig.) $n_f=3$ with $m_u=2$ MeV, $m_d=6$ MeV,
$m_s=100$ MeV (green dashed line) compared to
\protect\cite{0510084} and (right fig.) for $n_f=2+1$ with
$m_{u,d}=0.1m_s$ and $m_{u,d}=0.2m_s$ compared to
\protect\cite{0610017}. Analytical calculations are done using
(\ref{48}) and (\ref{55}).}\label{f14}
\end{figure}

The results of the calculations for $\frac{\varepsilon}{T^4}$ are compared, in Fig.~\ref{f12}, \ref{f13},
with three different sets of the lattice data: \cite{0608003},\cite{0610017}, and \cite{0510084}. In Fig.~\ref{f14} we
demonstrate the $I(T)$, computed with the help of (\ref{59}) and (\ref{51}), with lattice data for 2+1 flavours
from \cite{0510084} (left curve) and from \cite{0610017} (right curve).

It is instructive to estimate the contribution of the $q\bar{q}$ and $gg$ interactions to the pressure. Writing the virial
coefficient in the form $P_j=P_j^{(0)}(1+(P_j^{(0)}/T)B_j(T)+\ldots)$, where
$P_j^{(0)}=P_q$ or $P_{gl}$ in the SLA, with
\be
B_j(T)=\frac{1}{2}\int\left(1-e^{U_j(r,T)/T}\right)dV,\quad j=\mbox{fund, adj},
\label{60}
\ee
and taking the interaction terms $U_{\rm fund}$ and $U_{\rm adj}$, at large $T$, as $U_j(r,T)=Tu_j(rT)$,
one obtains a corrected pressure in the form:
\be
P=P_q^{(0)}(1-c_q)+P_{gl}^{(0)}(1-c_{gl})
\label{61}
\ee
where
\be
c_{gl}\cong \frac{16}{\pi^2}\int_0^\infty \rho^2 d\rho(e^{|u_{\rm adj}(\rho)|}-1),\quad
c_{q}\cong \frac{12n_f}{\pi^2}\int_0^\infty \rho^2 d\rho (e^{|u_{\rm fund}(\rho)|}-1).
\ee

Note that the $q\bar q$ and $gg$ interactions in the singlet colour state is attractive, so that $|U_j|=-U_j$.
The dependence of the $u_j$ on $rT$ occurs at large $T$ (in the dimensionally reduced regime),
when the dynamical dimensional quantities are the spatial string tension $\sigma_H=\mbox{const}\cdot T^2$ and the Debye mass
$m_D(T)\cong 2\sqrt{\sigma_H}=\mbox{const}\cdot T$.

Thus one expects that: i) the corrected pressure (\ref{61}) is smaller than the SLA prediction and
ii) the large-$T$ behaviour of the $P(T)$ is below the Stefan-Boltzmann values (modulo logarithmic factors). Both
features are clearly seen in Figs.~\ref{f11}, \ref{f12}.

\section{Beyond Single Line Approximation: Interacting quarks and gluons}\label{BSLA}

In the previous Sections one-particle contributions to the
partition function were considered in the so-called SLA. It was
stressed however that, at $T>T_c$, some interactions, like the
residual colour-electric and colour-magnetic one, acts in white
ensembles of quarks and gluons. Such interactions were neglected
before, except for the contributions of the correlator $D_1^E$
which produced effectively single-line term $V_1(\infty)$. In
addition, $D^E_1$ as well as $D^H$, and $D^H_1$ produce bound
systems of quarks and gluons \cite{Nef,9,8} which may affect
strongly the dynamics of the QGP at $T\geq T_c$. Such bound states
will be discussed below. To this end we start from a Green's
function of the given quark(gluon) system and extract, in the same
way as it was done before for the quark-antiquark meson at $T=0$,
the corresponding Hamiltonian $H$, \be
G(\mbox{out}|\mbox{in})=\lan \mbox{out}|e^{-H/T}|\mbox{in}\ran.
\label{6.2} \ee

Our purpose in this Section is to discuss the properties of this
resulting Hamiltonian and possible consequences for the
quark-gluon thermodynamics. Bound states in the $q\bar q$, $gg$,
and $gq$ systems were studied in the framework of the FCM
\cite{9,8} and on the lattice --- see \cite{latbound} and
references in \cite{9,8}. One can consider two distinct dynamics:
the colour-electric one, due to $D_1^E$, and the colour-magnetic
one, due to $D^H$ and $D_1^H$.

\subsection{Bound states at $T>T_c$ due to colour-electric forces}

In the colour-electric case the interaction can be written as (see \cite{9,8} for the discussion): 
\be 
V_{q\bar q}^E(r,T)\equiv v_1(r,T)\equiv V_1(r,T)-V_1(\infty,T) 
\label{6.3} 
\ee 
where 
\be
V_1(r,T)=\int^{1/T}_0(1-\nu T)d\nu\int^r_0\xi d\xi D_1^E(\sqrt{\nu^2+\xi^2}). 
\label{6.4} 
\ee 
Notice that, in (\ref{6.3}), we subtracted $V_1(\infty, T)$ which was already
accounted in the SLA, in the form of the Polyakov lines --- see
(\ref{P11}).

Characteristic feature of $v_1 (r,T)$ is that it is short-ranged,
$r_{eff}\sim 0.3$ fm, and it can support bound $S$-wave states in
$(c\bar c)_1$, $(gg)_1$, $(cg)_3$, and $(gg)_8$ (the subscript
stands for the representation of the colour group). The binding is
weak, $|\varepsilon|\lesssim 0.14$ GeV for $T\lesssim 1.5 T_c$. On
the lattice, in addition to $S$-wave states $(c\bar c)_1$, $(b\bar
b)_1$, and light $(q\bar q)$, the lowest $P$-wave state is claimed
to exist (see \cite{latbound} for a review).

It is clear that weakly bound states should dissociate fast in the dense QGP and therefore they hardly produce any significant
effect on the production rate. They are important, however, for the kinetic coefficients.

First, we generalise the static interaction (\ref{6.4}) to the case, when two colour objects, $A$ and $B$,
combine into a common colour state $D$ \footnote{The construction is similar to that found in \cite{36} for the coefficient
$\bar b$ of the $V_1^{(Q\bar Q)}(r,T)$ but differs in the presence of the first term proportional to $\bar a$,
since in \cite{36} the constant term was not taken into account.}
$$
V_1^{(AB,D)}(r,T)=\frac{1}{2C_F}\{C_DV_1^{(Q\bar Q)}(\infty,T)+(C_A+C_B-C_D)V_1^{(Q\bar Q)}(r,T)\}
$$
\be
\equiv \bar{a}V_1^{(Q\bar Q)}(\infty,T)+\bar{b}V_1^{(Q\bar Q)}(r,T).
\label{039}
\ee
Here $C_F=\frac43$ is the Casimir operator for the fundamental charges, while $C_A$, $C_B$, and $C_D$ are that for the representations $A$,
$B$, and $D$, respectively. In Table~\ref{t1} we illustrate (\ref{039}).

\begin{table}[t]
\begin{center}
\begin{tabular}{|c|c|c|c|c|c|c|c|c|}
\hline
& $(Q\bar Q)_1$&$(Q\bar Q)_8$&$(Q Q)_3$&$(Q Q)_6$&$(Qg)_3$&$(Q g)_6$&$(gg)_1$&$(gg)_8$\\
\hline
$a$&0&9/8&1/2&5/4&1/2&5/4&0&9/8\\
\hline
$b$&1&-1/8&1/2&-1/4&9/8&3/8&9/4&1\\
\hline
\end{tabular}
\end{center}
\caption{Parameters of the static potential of binary systems $(Q\bar Q)_D$, $(QQ)_D$, $(Qg)_D$, and $(gg)_D$
in different colour representations $D$.}\label{t1}
\end{table}

For multicomponent systems one can similarly find,
\begin{eqnarray}
V_1^{(ABC,D)}(\ver_1,\ver_2,\ver_3,T)&=&\frac{1}{2C_F}\{C_DV_1^{(Q\bar Q)}(\infty,T)\nonumber\\
\label{040}\\
&+&\frac13(C_A+C_B+C_C-C_D)\sum_{i>j} V_1^{(Q\bar Q)}(\ver_i-\ver_j, T)\}.\nonumber
\end{eqnarray}

In particular,
\begin{eqnarray}
V_1^{(QQQ)}&=&\frac12\sum_{i>j}V_1^{(Q\bar Q)}(\ver_i-\ver_j,T)\label{041}\\
V_{10}^{(QQQ)}&=&\frac94 V_1^{(Q\bar Q)}(\infty,T)-\frac14\sum_{i>j}V_1^{(Q\bar Q)}(\ver_i-\ver_j,T)\label{042}\\
V_{8}^{(QQQ)}&=&\frac98 V_1^{(Q\bar Q)} (\infty,T)+\frac18\sum_{i>j}V_1^{(Q\bar Q)}(\ver_i-\ver_j,T).\label{043}
\end{eqnarray}

It is easy to check that, at large distances, all systems consisting of $n_Q$ quarks or antiquarks and $n_g$ gluons tend to the constant
limit, independent of $D$,
\be
V_1^{(n_Q Q,n_g g)}(|\ver_i-\ver_j|\to\infty)=E_Q n_Q+E_g n_g,
\label{044}
\ee
where
\be
E_Q=\frac12 V_1^{(q\bar Q)}(\infty, T),\quad E_g=\frac98 V_1^{(Q\bar Q)}(\infty, T).
\ee

Since the nonperturbative part of the $V_1^{(Q\bar Q)}(r,T)$ behaves as $O(r^2)$ at $r\to 0$, one obtains the lower bound
on the nonperturbative part of $V^{(n_Q Q, n_gg)} (r_{ij}, T)$: $V^{(n_Q Q, n_gg)} (r_{ij}, T)\gtrsim
(1/2)(C_D/C_F)V_1^{(Q\bar Q)}(\infty,T)$. As a consequence, one can predict the absence of bound states in some
particular channels, for example, in $(Q\bar Q)_8$, $(QQ)_6$, $(QQQ)_{10}$, and so on.

\begin{figure}
\begin{center}
\includegraphics[width=8cm]{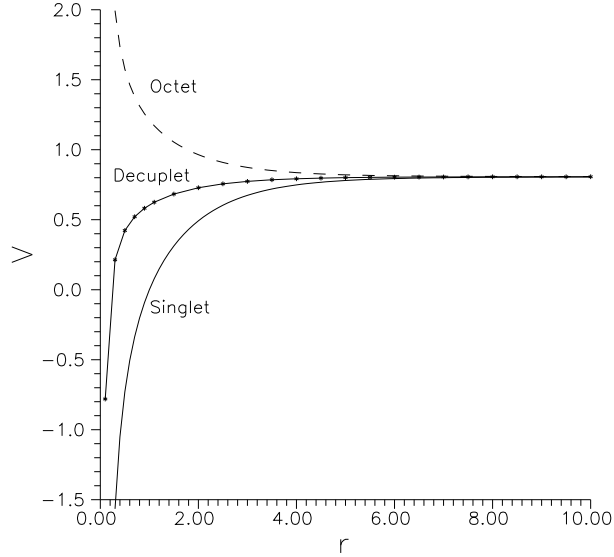}
\end{center}
\caption{The three-quark potential $V_1^{(QQQ)}(r,T)$ (in GeV) in different colour states for quarks placed at the vertices of
an equilateral triangle with sides $r$ versus $r$ (in GeV$^{-1}$).}\label{f2}
\end{figure}

As an application of the general relation (\ref{40}) we show, in Fig.~\ref{f2}, the static potentials
$V^{(QQQ)}(\ver_1,\ver_2,\ver_3,T)=V^{(D)}(r,T)$ of three static fundamental quarks in three different representations $D$:
singlet $(C_D=0)$, octet $(C_D=3)$ and decuplet, $(C_D=6)$. The quarks form a symmetric configuration of a equilateral triangle
with the sides $r$. From Fig.~\ref{f2} one can see that, indeed, all three potentials tend to the same limit
$(3/2)V_1^{(Q\bar Q)}(r,T)$ at large $r$'s, in agreement with (\ref{044}), while deviations from this asymptotic
at all distances are proportional to $\left(1/2,-1/4,1/8\right)$ for the singlet, decuplet, and octet, respectively, as
prescribed by (\ref{041})-(\ref{043}). This is in agreement with lattice calculations of free energies $F_{qqq}^{(D)}(r,T)$
presented in \cite{32}.

Having constructed static potentials, we can now exploit the
relativistic Hamiltonian technique, developed in \cite{ein2,DKS} and
successfully used for mesons, baryons, glueballs, and hybrids in
the confinement phase (see \cite{5*,6*} for a review). This
technique does not take into account chiral degrees of freedom.
Therefore, below we stick to the Hamiltonian technique of \cite{DKS}
and consider heavy quarkonia and baryons.

\begin{figure}[t]
\begin{center}
\includegraphics[width=6cm]{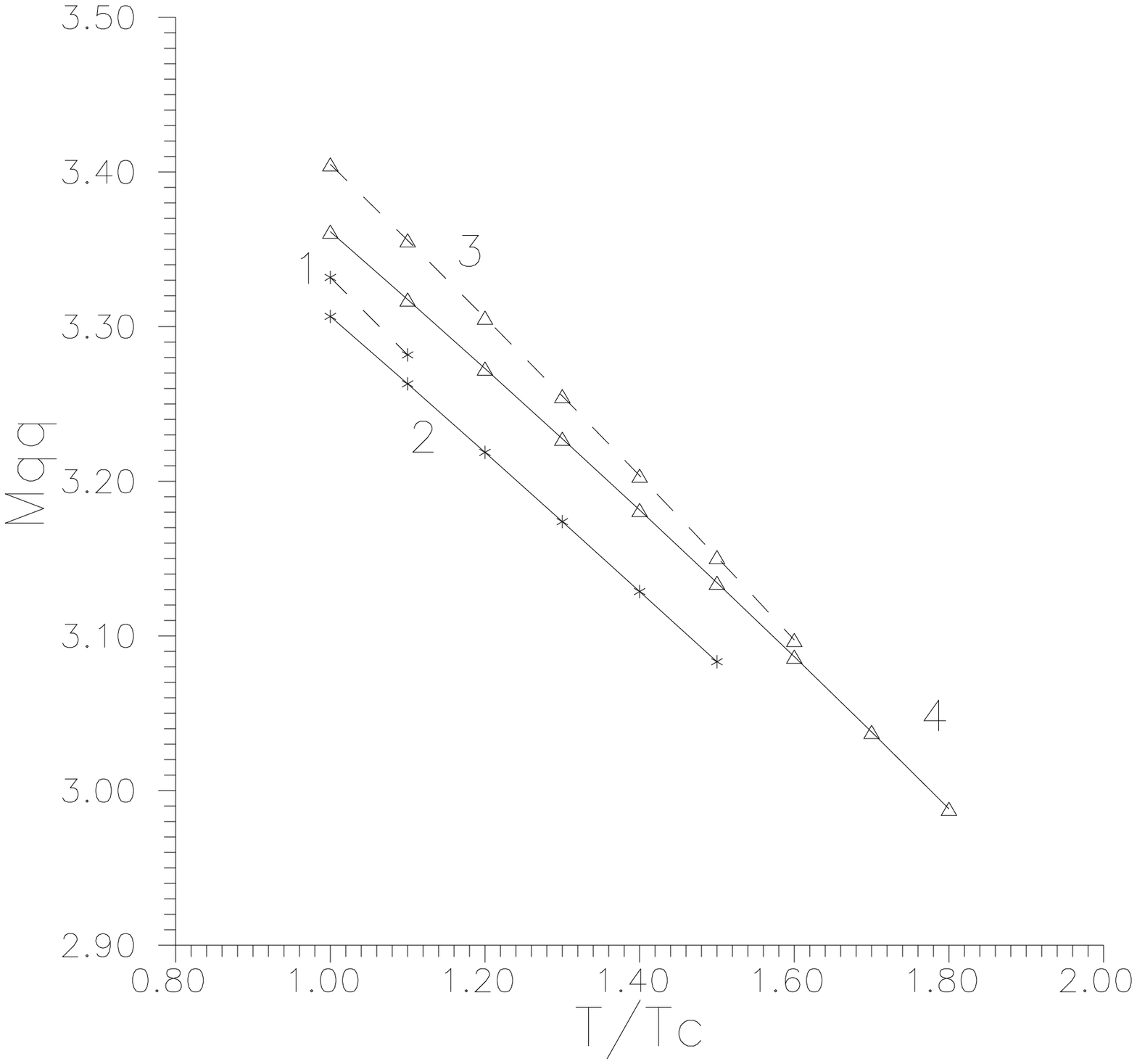}
\includegraphics[width=6cm]{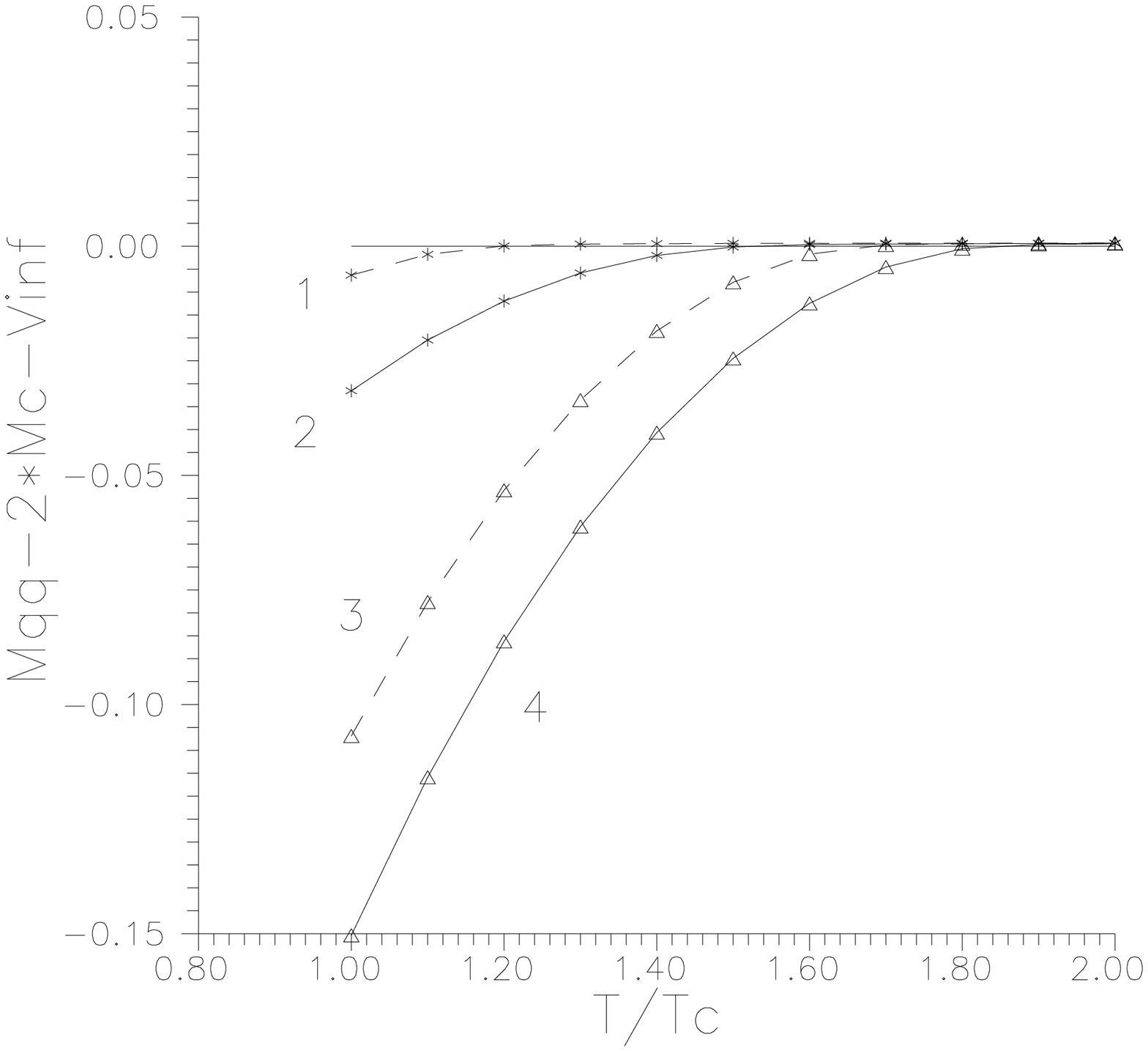}
\end{center}
\caption{Masses (first plot) and binding energies (second plot) of the $c\bar c$ colour-singlet bound states (in GeV) as
functions of $T/T_c$. The curves are numbered in accordance with the data sets.}\label{f4}
\end{figure}

\begin{figure}[t]
\begin{center}
\includegraphics[width=6cm]{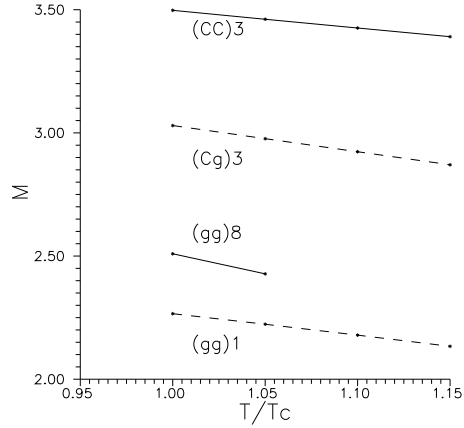}
\end{center}
\caption{Masses of bound states (in GeV) for the systems $(c\bar c)_3$, $(cg)_3$, $(gg)_1$, and $(gg)_8$ as
functions of $T/T_c$.}\label{f5}
\end{figure}

Similarly to the case of the quark--antiquark meson discussed in
detail before the bound-state problem for a generic
multiquark(gluon) system can be formulated in the form (we
introduce the einbein variables $\mu_i$ for quarks and gluons):
\be 
H\psi_n=\varepsilon_n\psi_n,\quad H=H_0+H_S+H_{SE},
\label{046} 
\ee 
where 
\be
H_0=\sum^{n_Q+n_g}_{i=1}\frac{\vep^2_i}{2\mu_i}+V_1^{(n_Qq,n_gg)}(\ver_{ij})
\label{047} 
\ee 
and $H_S, H_{SE}$ are the spin-dependent and self-energy parts of Hamiltonian defined in terms of field
correlators. Then the mass of the system is given by 
\be
M_n=\min_\mu\left\{\sum^{n_Q+n_g}_{i=1}\left(\frac{m^2_i}{2\mu_i}+\frac{\mu_i}{2}\right)+
\varepsilon_n(\mu_1,...\mu_{n_Q+n_g})+\bar{a}V_1^{(Q\bar Q)}(\infty,T)\right\}, 
\label{045} 
\ee 
after the minimisation performed with respect to all einbeins $\{\mu_i\}$. The stationary
values $\{\mu_i^{(0)}\}$ play the role of the constituent masses.
In \cite{8} a search for bound quark--antiquark states was
performed in the potential given by the sum of the nonperturbative
colour-electric interaction and the screened Coulomb interaction
(see (\ref{27}), (\ref{34}), (\ref{36}) above and the discussion
in \cite{8}), 
\be 
V_1^{Q\bar Q}(r,T)=V_1^{\rm np}(r,T)+V_1^C(r,T).
\label{048} 
\ee 
For the gluelump mass $M_0$ and the Debye mass ---
see (\ref{34}) and (\ref{36}) --- several data sets were used: in
set~I $M_0=1.0$ GeV, $m_D$=0.69 GeV, in set~II $M_0=1.0$ GeV,
$m_D$=0.2 GeV, in set~III $M_0=0.69$ GeV, $m_D$=0.69 GeV, in
set~IV $M_0=0.69$ GeV, $m_D$=0.2 GeV, in set~V $M_0=1.044$ GeV,
$m_D$=0.2 GeV, and in set~VI $M_0=1.044$ GeV, $m_D$=1.044 GeV. The
results are presented at Figs.~\ref{f4}, \ref{f5}. From these
figures one can see that, indeed, colour-electric interactions
above the deconfinement temperature are strong enough to maintain
bound states of quarks and gluons. Further details of the
calculations can be found in \cite{8}.

\subsection{Bound states at $T>T_c$ due to colour-magnetic forces}

In this chapter we investigate bound states of quarks (generalisation to gluons is trivial) above the $T_c$
which appear due to the residual nonperturbative colour-magnetic interactions. We follow \cite{Nef}.

We start from the Hamiltonian of a quark--antiquark meson (\ref{Hm}). Two particular cases are of most
interest. The first such case corresponds to equal masses ($m_1=m_2=m$ and therefore $\mu_1=\mu_2=\mu$),
whereas in the other case one mass is assumed infinitely large ($m_1\to\infty$, $m_2=m$ and $\mu_1\to\infty$, $\mu_2=\mu$).
Then the Hamiltonian reads:
\be
H=\frac{1}{\xi}\left(\frac{p_r^2+m^2}{\mu}+\mu\right)+\int^1_0d\beta\left(\frac{\sigma_1^2r^2}{2\nu}+
\frac{\nu}{2}+\sigma_2 r\right)+\frac{{\bm L}^2}{r^2[\xi\mu+2\int^1_0d\beta\nu(\beta-\xi/2)^2]},
\label{Hgen}
\ee
where $\xi=1$ for the case of equal masses and $\xi=2$ for the case of the heavy--light system.

We take the extrema in $\nu(\beta)$ and $\eta(\beta)$ now, approximating $\eta(\beta)$ by a uniform in $\beta$ distribution.
Then the Hamiltonian (\ref{Hgen}) takes the form
\be
H=\frac{1}{\xi}\left(\frac{p_r^2+m^2}{\mu}+\mu\right)+V_{\rm SI}(r),
\label{Hllhl}
\ee
and the spin--independent potential reads:
\be
V_{\rm SI}(r)=\eta_0\sigma_E r+\left(\frac{1}{\eta_0}-\eta_0\right)\sigma_H r+\frac{1}{\xi} \mu y^2,\quad \eta_0=\frac{y}{\arcsin y},
\label{VSI}
\ee
with $y$ being the solution of the transcendental equation
\be
\frac{\sqrt{L(L+1)}}{\sigma_H r^2}=\frac{\xi}{4y}\left(1+\eta_0^2\left(1-\frac{\sigma_E}{\sigma_H}\right)\right)\left(\frac{1}{\eta_0}-
\sqrt{1-y^2}\right)+\frac{\mu y}{\sigma_H r}.
\label{yeqgen}
\ee
The interested reader can find the details of a similar evaluation performed for the Hamiltonian (\ref{Hgen})
with $\sigma_E=\sigma_H=\sigma$ in \cite{regge,BS0}.

The remaining einbein $\mu$ is to be considered as the variational parameter to minimise the spectrum of the Hamiltonian (\ref{Hllhl}).
Obviously, the extremal value of $\mu$ depends on quantum numbers and acquires two contributions: one coming from the current quark mass $m$
and the other, purely dynamical, contribution coming from the mean value of the radial component of the momentum $p_r$.

It is instructive to pinpoint the difference in the potential (\ref{VSI}) below and above the $T_c$.
At small $r$'s, the potential (\ref{VSI}) turns to the centrifugal barrier
$L(L+1)/(\xi\mu r^2)$, whereas its large--$r$ behaviour differs dramatically for the temperatures
below and above the $T_c$. Indeed, the leading large-$r$ contribution to the inter-quark potential corresponds to $y\ll 1$ and,
for $T<T_c$, reads:
\be
V_{\rm conf}(r)=\sigma_E r.
\label{lin}
\ee
This is the linear confinement which is of a purely colour--electric nature and which admits angular--momentum--dependent corrections (see
\cite{DKS,regge}).

In the deconfinement phase, at $T>T_c$, the colour--electric part of the potential (\ref{VSI})
vanishes, the leading long--range term coming from the angular--momentum--dependent part of the interaction:
\be
V_{\rm SI}(r)=\frac{3L(L+1)}{\xi^2\sigma_H r^3}+\ldots.
\label{VSI20}
\ee
Interestingly, in the deconfinement phase in absence of the confining potential,
the spin--independent interaction becomes short--ranged decreasing as $1/r^3$
at large inter-quark separations. This feature means the full compensation of the centrifugal barrier which
would naively behave as $1/r^2$ instead. The reason is obvious: at large inter-quark separations, the effective quark mass $\mu$ is to be compared to
the ``mass" of the string $\sigma r$. The bound--state problem solved in the potential (\ref{lin})
gives a large value $\langle p_r\rangle\propto \sigma_E\langle r\rangle$, so that, even for light (massless) quarks, their effective mass $\mu$
appears quite large ($\mu\gg m$).
On the contrary, for light quarks and in absence of the strong confining interaction (\ref{lin}), the values of $\mu$ are
small ($\mu\approx m$) and can be neglected as compared to the string contribution $\sigma_H r$.
This makes the spin--dependent terms in the effective inter-quark interaction important in
this regime, as opposed to the confinement phase, where they give only small corrections to the bound states formed in the confining
potential (\ref{lin}).

Now we turn to the derivation of spin--dependent contributions to the inter-quark potential. The full set of
spin-dependent interactions is given in (\ref{SDints}). The leading
spin--dependent term is the spin--orbit interaction (we omit contributions of $D_1^{E,H}$ which bring about
short--range terms ${\cal O}(r^{-3})$),
\be
V_{SO}(r)=\left(\frac{{\bm S}_1\veL_1}{2\mu_1^2}-\frac{{\bm S}_2\veL_2}{2\mu_2^2}\right)
\left(\frac{1}{r}\frac{d V_0}{dr}+\frac{2}{r}\frac{dV_1}{dr}\right),
\label{Vls1}
\ee
where $V_i(r)$ can be expressed through the colour--electric and colour--magnetic field correlators,
\be
\frac{1}{r}\frac{d V_0}{dr}=\frac{2}{r}\int_0^\infty d\tau\int_0^r d\lambda D^E(\tau,\lambda),\quad
\frac{2}{r}\frac{dV_1}{dr}=-\frac{4}{r}\int_0^\infty d\tau\int_0^r d\lambda\left(1-\frac{\lambda}{r}\right)D^H(\tau,\lambda).
\label{SIall}
\ee

Only the $V_1$ potential survives above the $T_c$, so that the resulting spin--dependent potential reads:
\be
V_{SO}(r)=-\frac{2\xi{\bm S}{\bm L}}{\mu r(\xi\mu+2\langle\nu(\beta-\xi/2)^2\rangle)}\int_0^\infty d\tau\int_0^r
d\lambda D^H(\tau,\lambda)\left(1-\frac{\lambda}{r}\right),
\label{SOgen}
\ee
where
\be
\langle\nu(\beta-\xi/2)^2\rangle\equiv\int_0^1d\beta\nu(\beta-\xi/2)^2=
\frac{\sigma_H r}{2\xi^2y^2}\left(1+\eta_0^2\right)\left(\frac{1}{\eta_0}-\sqrt{1-y^2}\right),
\label{yeqgen2}
\ee
and ${\bm S}={\bm S}_1+{\bm S_2}$, for the light--light system, and ${\bm S}$ is the light--quark spin, for the heavy--light quarkonium.

\begin{figure}[t]
\begin{center}
\includegraphics[width=8cm]{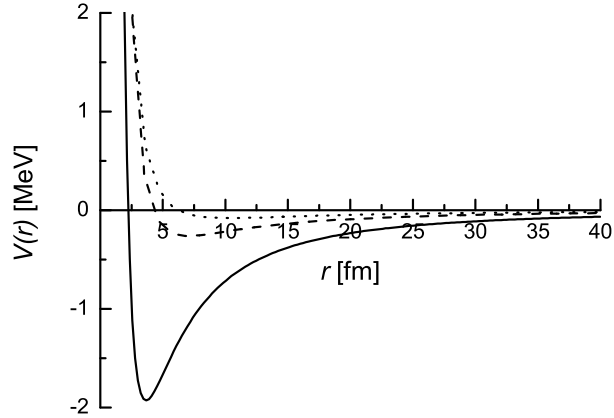}
\end{center}
\caption{The profile of the effective potential (\ref{Vf}) for
$m=1$~GeV (solid line), $m=2$~GeV (dashed line), and $m=3$~GeV
(dotted line).}\label{pots}
\end{figure}

It follows from (\ref{SOgen}) that above the deconfinement temperature,
for the states with the total momentum $J=L+S$, ${\bm S}{\bm L}>0$ and the potential $V_{SO}(r)$
becomes attractive, with a possibility to maintain bound states. Furthermore, its
slow decrease as $r\to\infty$ suggests that an infinite number of bound states exists, with the binding energies asymptotically
approaching zero. Let us study these bound states in more detail. Hereafter in this chapter
$\sigma_E=0$ and we use the notation $\sigma$ for the magnetic tension $\sigma_H$.

In view of an obvious similarity of the light--light and heavy--light cases
(the difference manifesting itself only in numerical coefficients), we investigate numerically
only the light--light system, as a paradigmatic example. Furthermore, for $r\gg T_g$, the potential (\ref{SOgen}) does not depend on
the form of the correlator $D^H$ since
\be
2\int_0^\infty d\tau\int_0^r d\lambda D^H(\tau,\lambda)\left(1-\frac{\lambda}{r}\right)
\mathop{\approx}\limits_{r\gg T_g}\sigma.
\ee
Finally, we neglect the perturbative part of the inter-quark interaction for it is screened to a large extent contributing to short--ranged
forces only whereas the effect discussed in this work is essentially a long--ranged effect.

Therefore we study the spectrum of bound states in the potential
\be
V(r)=\left(\frac{\arcsin y}{y}-\frac{y}{\arcsin y}\right)\sigma r+\mu y^2
-\frac{\sigma l}{\mu r(\mu+2\langle\nu(\beta-1/2)^2\rangle)},
\label{Vf}
\ee
which is the sum of the spin--independent term (\ref{VSI}) and the spin--orbital term (\ref{SOgen});
$y$ is the solution of (\ref{yeqgen}) with $\sigma_E=0$.
In Fig.~\ref{pots} we plot the effective potential (\ref{Vf}) for three values of the quark mass:
$m=1$GeV, $2$GeV, and $3$GeV.

The resulting eigenenergy $\varepsilon_{n_rl}(\mu)$ is added then to the free part of the
Hamiltonian (\ref{Hgen}),
\be
M_{n_rl}(\mu)=\frac{m^2}{\mu}+\mu+\varepsilon_{n_rl}(\mu),
\label{Mmin}
\ee
and this sum is minimised with respect to the einbein $\mu$,
\be
\left.\frac{\partial M_{n_rl}(\mu)}{\partial\mu}\right|_{\mu=\mu_0}=0,\quad M_{n_rl}=M_{n_rl}(\mu_0).
\ee

In Table~\ref{t6} we present the set of parameters used in our numerical calculations,
whereas in Table~\ref{t7} we give the results for the binding energy
for the $b\bar{b}$, $c\bar{c}$, and $s\bar{s}$ quarkonia above the $T_c$ for $l=1$ and $n_r=0,1$.
We ensure therefore that for $L\neq 0$ the potential (\ref{Vf}) does support bound states. The binding
energy is small ($|E_{n_rL}|\ll T$ for the $b$ and $c$ quarks and $|E_{n_rL}|\lesssim T$ for $s$ quarks) so
these bound states can dissociate easily.

\begin{table}[t]
\begin{center}
\begin{tabular}{|c|c|c|c|c|c|}
\hline
Parameter&$m_b$, GeV& $m_c$, GeV& $m_s$, GeV&$\sigma$, GeV$^2$& $T_g$, fm\\
\hline
Value&4.8&1.44&0.22&0.2&0.2\\
\hline
\end{tabular}
\end{center}
\caption{The set of parameters used for the numerical evaluation.}\label{t6}
\end{table}

\begin{table}[t]
\begin{center}
\begin{tabular}{|c|c|c|c|}
\hline
&$b\bar{b}$&$c\bar{c}$&$s\bar{s}$\\
\hline
$n_r=0$&-0.007&-0.19&-45\\
\hline
$n_r=1$&-5$\times 10^{-4}$&-0.015&-2.7\\
\hline
\end{tabular}
\end{center}
\caption{The binding energy $E_{n_rL}\equiv M_{n_rL}-2m$ (in MeV) for the ground state and for the first radial excitation
in the potential (\ref{Vf}) with $L=1$ for the $b\bar{b}$, $c\bar{c}$, and $s\bar{s}$ quarkonia.}\label{t7}
\end{table}

\begin{figure}[t]
\includegraphics[width=55mm]{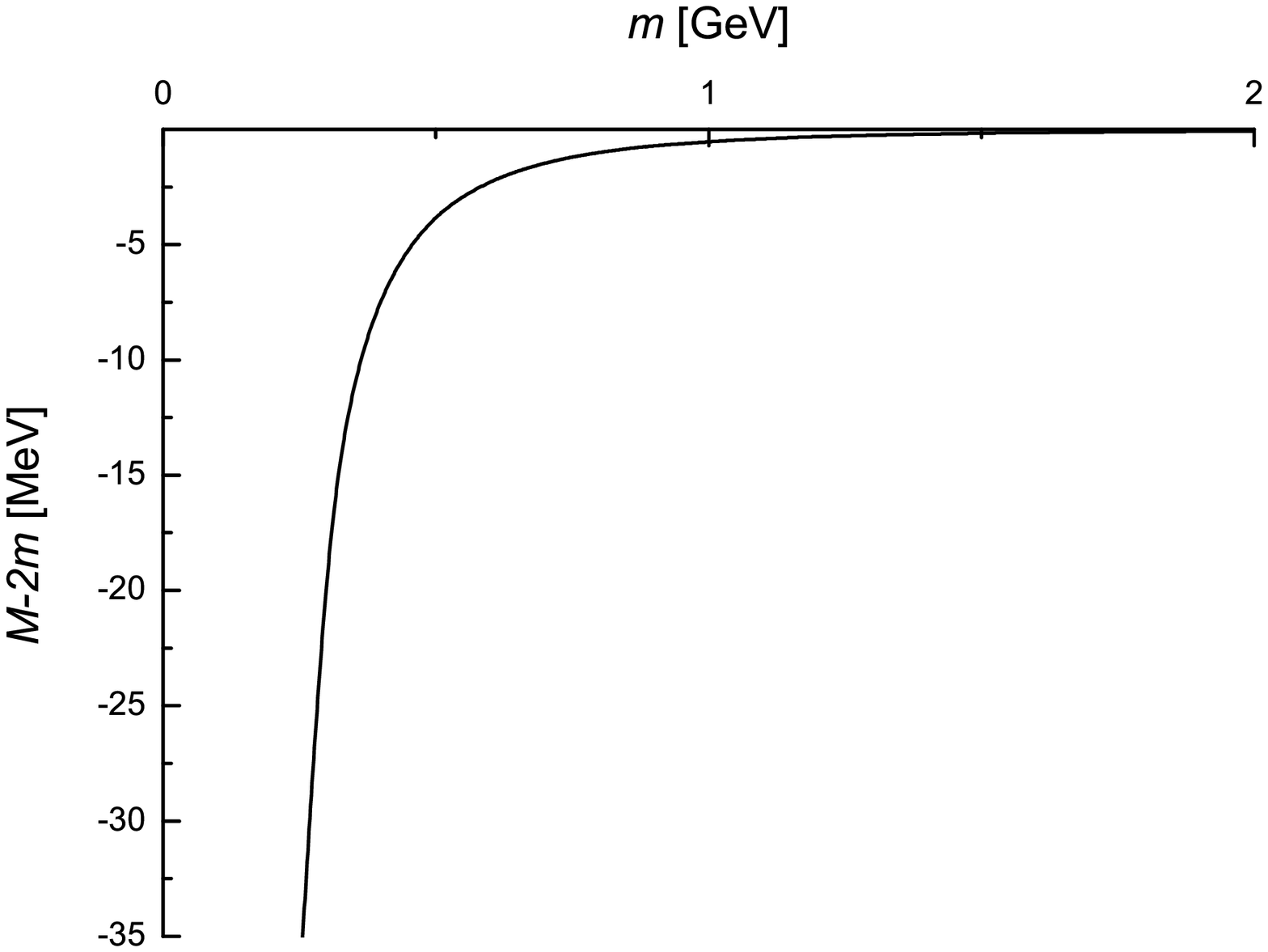}\hphantom{aa}
\includegraphics[width=55mm]{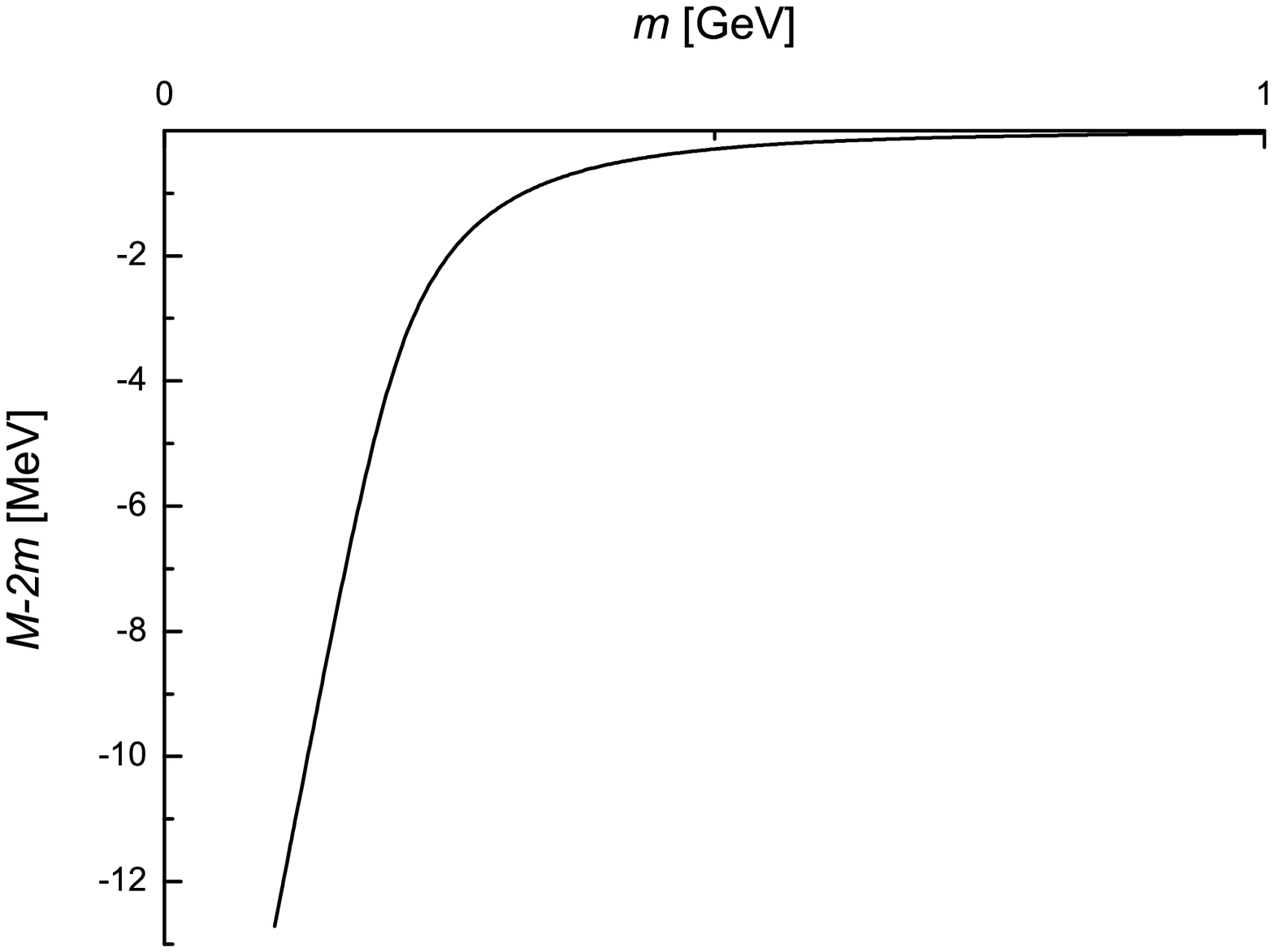}
\caption{The binding energy of the quark--antiquark system versus
the mass of the quark for $L=1$ and $n_r=0$ (first plot) and
$n_r=1$ (second plot).}\label{masses}
\end{figure}

Let us discuss the problem of binding of light quarks.
The effective potential (\ref{Vf}) admits different forms at different inter-quark separations,
depending on which contribution, of the quark mass term  $\mu$ or of the ``string mass"
$2\langle\nu(\beta-1/2)^2\rangle)$,
gives the dominating contribution, that is for $\mu\gg\sigma r$ and $\mu\ll\sigma r$.
If large distances contribute most to the bound state
formation (the latter case), then $V(r)={\cal O}(L(L+1)/(\sigma r^3))+{\cal O}(L/(\mu r^2))$, where the first
term comes from the spin--independent interaction (see (\ref{SIall})) and the other stems from the
spin--orbit potential. The dependence of the binding energy on $\mu$ is expected then to be rather
moderate, approximately as $1/\mu$.

On the contrary, in the former case with the string dynamics giving a correction to the quark mass term,
the potential (\ref{Vf}) can be approximated as
\be
V(r)\approx\frac{L(L+1)}{\mu r^2}-\frac{\sigma l}{\mu^2 r},
\ee
that is by the sum of the centrifugal barrier and the attractive Coulomb-like potential with the effective coupling
\be
\alpha_{\rm eff}=\frac{\sigma L}{\mu^2}.
\label{aeff}
\ee
The corresponding eigenenergy can be found in any textbook in Quantum Mechanics and gives a stronger dependence on $\mu$,
\be
\varepsilon(\mu)\propto -\mu\alpha_{\rm eff}^2\propto -\frac{\sigma^2 L^2}{\mu^3}.
\label{mucoul}
\ee

Let us consider the states with $L=1$ and $n_r=0$.
We follow now the procedure described in detail before, that is we solve the full problem numerically,
and find the dependence of the eigenenergy $\varepsilon$ on the einbein $\mu$ to be
\be
\varepsilon(\mu)\propto -\frac{1}{\mu^{2.79}}.
\label{vemu}
\ee

Comparing this to (\ref{mucoul}) we find a good agreement, with the small deviation in the power resulting
from the proper string dynamics. We conclude therefore that the dynamics of the system develops at the
inter-quark separations $T_g\ll r\lesssim m/\sigma$ (since, for the set of parameters given in Table~\ref{t6}, $m\gtrsim\sigma T_g$ then there is room for such separations).

Let us discuss now the procedure of minimisation of the spectrum (\ref{Mmin}) in $\mu$. First of all,
let us notice that, in the einbein field formalism, the calculation of the
spectrum naively looks like a nonrelativistic calculation due to the ``nonrelativistic" form of the kinetic energy in
the Hamiltonian with the einbein field $\mu$ introduced. In the meantime, the full relativistic form of the
quark kinetic energy is readily restored as $\mu$ takes its extremal value and hence this is the procedure of
taking extremum in $\mu$ in the masses (\ref{Mmin}) to sum up an infinite series of relativistic corrections and thus
to restore the relativistic spectrum. For example, the relativistic
ground state eigenenergy $E_0=m\sqrt{1-(Z\alpha)^2}$ of the one--body Dirac equation with the Coulomb potential $-Z\alpha/r$ can be reproduced {\em exactly} with the help of the einbein technique.
Finally, one can visualise the form of $\mu$ considering the effective
Dirac equation for the light quark in the field of the static antiquark source. When written in the form of a second--order
differential equation, it
contains the spin--orbit term of the same form as given in (\ref{Vls1}) but with $\mu$ replaced by the
combination $\epsilon+m+U-V$, where $U$ and $V$ are scalar and vector potentials,
respectively. For light (massless) quarks this combination takes drastically different values below and above the deconfinement temperature. Indeed, in the confining phase of QCD, when spontaneous chiral symmetry breaking leads to a strong effective, dynamically generated scalar potential $U$, this effective ``$\mu$" is large. On the contrary, above the $T_c$, when $U$ is small ``$\mu$" is also small (it can be even negative since the eigenenergy $\epsilon$ may have any sign).

It has been found numerically that the extremum in $\mu$ for $M_{n_rL}(\mu)$, (\ref{Mmin}),
exists for $m$ exceeding the value of approximately $0.22$GeV (for the given $\sigma=0.2$GeV), and no extremum
exists for smaller values of the quark mass (see Fig.~\ref{masses} for the dependence of the binding energy
$E_{n_rL}$ on the quark mass). This property of the bound state spectrum can be easily
understood using the analogy with the bound state problem for the Dirac equation with the potential in the form of a
deep square well or Coulomb potential discussed above.
For example, for the Coulomb potential, a problem appears as the coupling
exceeds unity --- the well--known problem of $Z>137$. From (\ref{aeff}) we easily find this critical
phenomenon to happen at $m\approx\mu\lesssim \sqrt{\sigma}\approx 0.4$GeV.
This estimate is in good agreement with the result of
our direct numerical calculations quoted above.

Physically this situation means that many quark--antiquark and/or gluon pairs are formed and finally stabilise
the vacuum. Formally the problem is not anymore a two--body problem, but rather many--body, so that
many--body techniques are to be applied. For example, in electrodynamics with $Z>137$,
one can derive the resulting self-consistent field of the Thomas--Fermi type \cite{z137}.
A similar situation can be expected in the deconfinement phase of QCD. In absence of the linear potential,
the einbein $\mu$ (playing the role of the effective quark mass) is not anymore bounded from below by the
values of order $\sqrt{\sigma_E}\simeq 0.4$GeV coming from the binding energy in the linearly rising potential.
To see the onset of this phenomenon in the framework of our two--body (one--body for the heavy--light case)
Hamiltonian, one should take into account the negative--energy part of the spectrum, when the full matrix
form of the Hamiltonian is considered \cite{negmu}. Indeed, the matrix structure of the Hamiltonian occurs
in the path--integral formalism from the two--fold time--forward/backward motion described by the
positive/negative values of $\mu$. Off--diagonal terms in the matrix Hamiltonian produce the turning
points in the particle trajectory and result in Z--graphs.

Notice that the same is true for the glueballs and gluelumps since in this case
equations are the same as for light--light and heavy--light quarkonia, respectively, but with the quark spin
replaced by the gluon spin and $\sigma_H$ by $\frac94\sigma_H$.

It is important to notice that a separated quark--antiquark pair
was considered in this chapter. In reality such quark--antiquark
pairs are to be considered in the medium formed by other quarks
and gluons, that is as a part of the SQGP. As a measure of the
interaction in SQGP one can consider the ratio of the mean
potential energy to the mean kinetic energy of the particles in
the plasma, $\Gamma=\langle V\rangle/\langle K\rangle$. It is easy
to estimate that $\langle K\rangle\simeq T$ and $\langle
V\rangle\simeq \sigma_H/T$. This gives $\Gamma=\sigma_H/T^2$ and
so this parameter is large for quarks and it is several times
larger for gluons. Therefore, SQGP is a strongly interacting
medium which looks like a liquid, rather then as a gas. With the
growth of the temperature the medium becomes more dense, and the
mean distance between particles decreases. As this distance
becomes comparable to the radius of the bound states discussed in
this chapter, the latter will dissociate because of the screening
effects. In other words, the hot medium plays the role of a
natural cut--off for the effect of bound pair creation discussed
above. Notice however that, for the quark masses around 0.2GeV,
the radius of the bound state is of the order of  one fm and it is
expected to decrease further with the decrease of the quark mass,
even if the pair creation process is properly taken into account.
This means that indeed there is room for such bound states for the
temperatures above the $T_c$. Breakup, with the growth of the
temperature, of such high--$l$ states for quarks, and especially
for gluons which possess  more degrees of freedom than quarks, may
affect such characteristics of the plasma as its free energy and
it entropy (for a recent paper of explaining the near--$T_c$
behaviour of these characteristics see \cite{ant}).

Concluding this section one can say that colour--magnetic (spin--dependent) interaction acting on light quark
or gluonic systems enforces nonperturbative creation of light $q\bar{q}$ and $gg$ pairs.

\section{Generalisation to nonzero baryon densities}\label{Disc}

In this section we will extend the method to the case of nonzero
$\mu$ and nonzero interaction with vacuum $(V_1\neq 0)$ and to
find $T_c(\mu)$,  for $\mu>0$. The simple picture of VD dynamics
with the only input $L_{\rm fund}(T)$, taken from lattice or analytic
calculation suggested in \cite{an1}, which was discussed below as
the Vacuum Dominance Model (VDM), is adopted below and is shown to
produce surprisingly reasonable results, being in good agreement
with available lattice data for nonzero $\mu$, where these data
are reliable.

\subsection{Nonperturbative EoS for $\mu>0$}

The main idea of the  VDM  is that the most important part of
quark and gluon dynamics in the strongly interacting plasma is the
interaction of each individual quark or gluon with vacuum fields.
This interaction is derived from field correlators and is
rigorously proved to be embodied in   factors, which happen to
coincide with the modulus of Polyakov loop\footnote{We neglect in
this approximation the difference between $L_{\rm fund}$ expressed via
$V_1(\infty, T)$ and $L_{\rm fund}^{lat}$ where the role of $V_1(T)$
is played \cite{an1,9} by singlet $Q\bar Q$ free energy
$F^1_{Q\bar Q} (\infty, T)$. The latter quantity contains all
excited states, so that $V_1(T)\geq F^1_{Q\bar Q} (\infty, T)$.},
\be
L_{\rm fund} =\exp\left(-\frac{V_1 (T)+2V_D}{2T}\right),\quad L_{\rm adj}=\exp\left(-\frac94 \frac{V_1(T)+2V_D}{2T}\right),
\label{trusov1}
\ee
where $V_1(T)\equiv V_1 (\infty, T),$  $V_D\equiv V_D(r^*,T)$ and $V_1(r,T),
V_D$ found in \cite{9} to be  ($\beta\equiv 1/T)$ the same as for
$\mu=0$ case (see Eqs.~(\ref{28}),(\ref{27})) but with $D^E$,
$D_1^E$ in principle depending on $\mu$.

In $V_D, $  Eq.(\ref{trusov1}) $r^*$ is the average size of the
heavy-light $Q\bar q$ or its adjoint equivalent system; for
$T>T_c$ one has $D^E=V_D=0$. For $T<T_c$ one has
$r^*_{\rm fund}=\infty$ for $n_f=0$, yielding $L_{\rm fund}=0,$ however
$r^*_{\rm adj}\approx 0.4$ fm for any $n_f$ and gives nonzero $L_{\rm adj}
(T<T_c)$. The form (\ref{trusov1}) is in good agreement with
lattice data \cite{24} and also explains why $L_{\rm fund}$ is a good
order parameter (however approximate for $n_f\neq 0$). Note, that
only NP parts $D_1^E, D^E$ enter in (\ref{28}),(\ref{27}), see
\cite{9} for discussion of separation of these parts;
renormalisation procedure is discussed also in \cite{8}.

In the lowest NP approximation one neglects pair, triple etc.
interactions between quarks and gluons (which are important for
$T_c\leq T\leq 1.2 T_c$ where density is low and screening by
medium is  not yet operating, see \cite{Nef,an1})) and derives the
following EoS (this approximation is called in \cite{an1,an2} the
Single Line Approximation (SLA)) 
\be
p_q\equiv{\frac{P^{SLA}_{q}}{T^4}} = \frac{4N_cn_f}{{\pi}^2}
\sum^\infty_{n=1}\frac{ (-1)^{n+1}}{n^4} L^{n}_{\rm fund}
\varphi^{(n)}_q \cosh\frac{\mu n}{T}, 
\label{trusov4} 
\ee 
\be
p_{gl} \equiv \frac{P_{gl}^{SLA}}{T^4} =\frac{2(N^2_c-1)}{\pi^2}
\sum^\infty_{n=1} \frac{1}{n^4} L_{\rm adj}^n 
\label{trusov5}
\ee
with 
\be 
\varphi^{(n)}_q (T) =\frac{n^2m^2_q}{2T^2} K_2
\left(\frac{m_q n}{T}\right)\approx 1-\frac{1}{4} \left(\frac{n
m_q}{T}\right)^2+\ldots 
\label{trusov6} 
\ee

In (\ref{trusov4}), (\ref{trusov5}) it was assumed that $T\lesssim
\frac{1}{\lambda}\cong 1$ GeV, where $\lambda$ is the vacuum
correlation length, e.g. $D_1^{(E)} (x) \sim e^{-|x|/\lambda}$,
hence powers of $L^n_i$,  see \cite{an1} for details.

With few percent accuracy one can replace the sum in
(\ref{trusov5}) by the first term, $n=1$, and this form will be
used below for $p_{gl}$, while for $p_q$ this replacement is not
valid for large $\frac{\mu}{T}$, and one can use instead the form
equivalent to (\ref{trusov4}), 
\be 
p_q=\frac{n_f}{\pi^2}\left[\Phi_\nu \left( \frac{\mu-\frac{V_1}{2}}{T}\right)+\Phi_\nu
\left(-\frac{\mu+\frac{V_1}{2}}{T}\right)\right]
\label{trusov7}
\ee 
where $\nu=m_q/T$ and 
\be 
\Phi_\nu (a)=\int^\infty_0  \frac{z^4 dz}{\sqrt{z^2+\nu^2}}\frac{1}{(e^{\sqrt{z^2+\nu^2}-a}+1)}.
\label{trusov8}
\ee

Eqs.~(\ref{trusov7}), (\ref{trusov5}) define $p_q, p_{gl}$ for all
$T, \mu$ and $m_q$, which is the current (pole) quark mass at the
scale of the order of $T$.

To draw $p_q, p_{gl}$ and $p\equiv  p_q+p_{gl}$ as functions of
$T, \mu$ one needs explicit form of $V_1(T)$. This was obtained
analytically and discussed in \cite{9}; another form was found
from $D_1^E(x)$ measured on the lattice in \cite{trusov8} and is
given in \cite{8}.

Also from lattice   correlator  studies \cite{7*,trusov8}
$V_1(T=T_c)$ is (with $\sim 10\% $ accuracy) 0.5 GeV and is
decreasing with the growth of $T$ (cf. Fig. 2 of \cite{9} and Fig.
1 of \cite{8}). This behaviour is similar to that found repeatedly
on the lattice  direct measurement of $F^{(1)}_\infty$, see e.g.
Fig. 2 of \cite{27} where $F_\infty^{(1)} =V_1(T)$ is given for
$n_f =0,2,3$. In what follows we shall exploit the latter curves
parameterizing them for $T\geq T_c$ and all $n_f$ as 
\be 
V_1(T)=\frac{0.175~{\rm GeV}}{1.35\left(\frac{T}{T_c}\right) -1},\quad
V_1(T_c)\approx 0.5 {~\rm GeV}.
\label{trusov9}
\ee 
For $\mu>0$ one can expect a $\mu$-dependence of $V_1$, however it should be weak
for values of $\mu$ much smaller than the scale of change of
vacuum fields. The latter scale can  be identified with the
dilaton mass $m_d$, which is of the  order of the lowest glueball
mass, i.e. $\sim  1.5$ GeV $\equiv  m_d$. Hence one can expect,
that $V_1$ in the lowest approximation does not depend on $\mu$.
This is supported by the lattice measurements in \cite{trusov20},
where for $\frac{T}{T_c}=1.5$ and $\frac{\mu}{T} =0.8$ the values
of $F^{(1)}_\infty$ are almost indistinguishable from the case of
$\frac{\mu}{T}=0$.

\begin{figure}
\includegraphics[width=12cm]{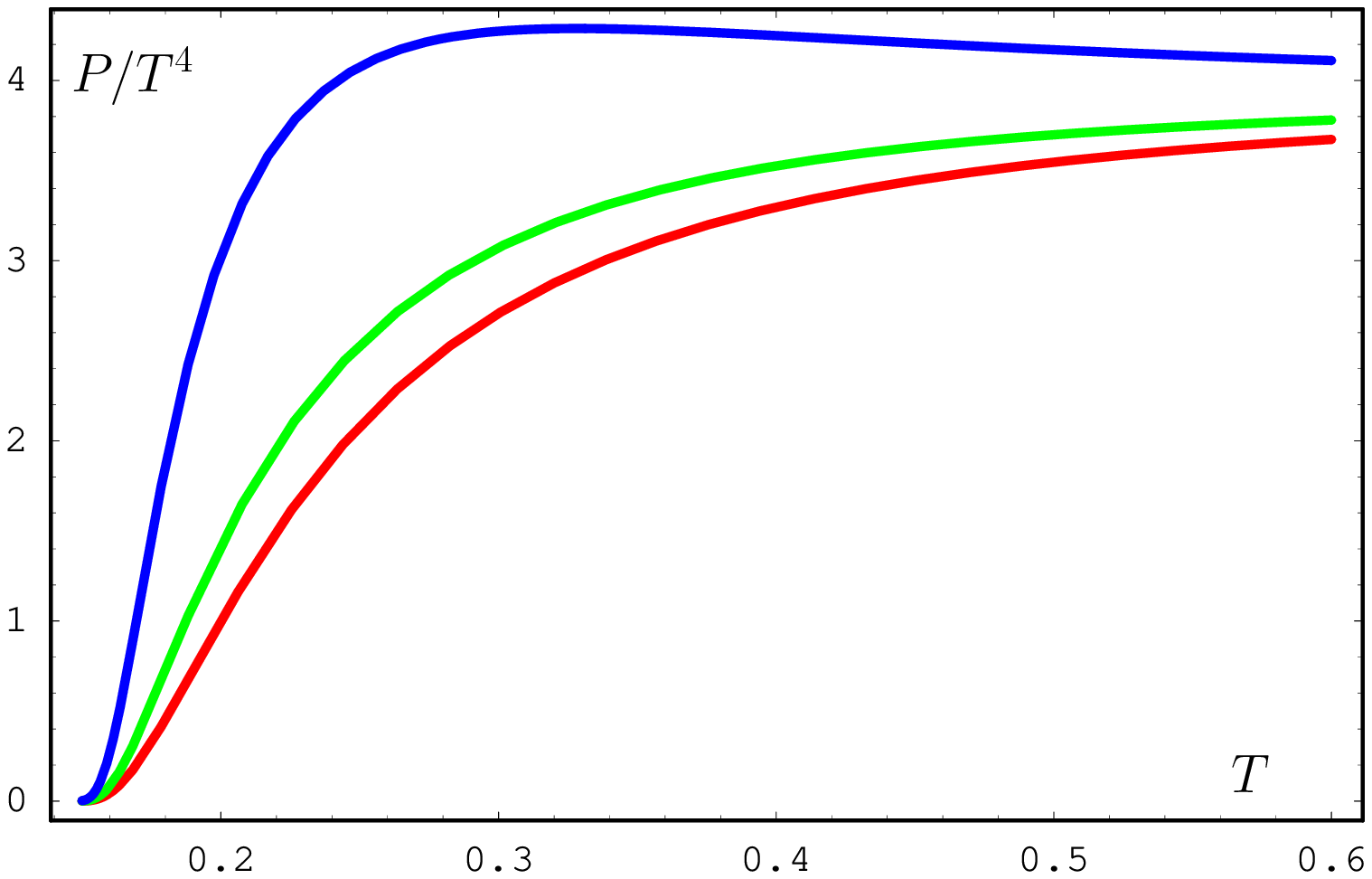}
\caption{Pressure $\frac{P}{T^4}$ from
Eq.(\ref{trusov4},\ref{trusov5}) as function of temperature  $T$
(in GeV) for $ n_f=2$, $\mu=0$, $0.2$, $0.4$~GeV (bottom to top),
and $\Delta G_2=0.0034$ GeV$^4$.} \label{f1trusov}
\end{figure}

To give an illustration of the resulting EoS we draw in
Fig.~\ref{f1trusov} the pressure $p$ for the cases $\mu=0$, $0.2$,
$0.4$~GeV and $n_f=2$. One can see a reasonable behaviour  similar
to the  lattice data,  see \cite{trusov25} for a review and
discussion.

\subsection{Phase transition for nonzero $\mu$}

Here we extend the  VD mechanism suggested  in \cite{6} to the
case of nonzero $\mu$ and $V_1$. We assume as was said in
Introduction that the phase transition occurs from the full
confining vacuum with all correlators $D^E, D_1^E, D^H, D^H_1$
present to the deconfined vacuum where $D^E$ vanishes. The basics
of our physical picture is  that all fields (and correlators) do
not change for $\mu, T$ changing in a wide interval, unless $\mu,
T$ become comparable to the dilaton mass $m_d \approx M$(glueball
$0^+)\approx 1.5$ GeV\footnote{Note that $G_2$ does  not depend on
$\mu, n_f$ in the leading order of the $1/N_c$ expansion and one
expects a growth of the magnetic part of $G_2$ at $T>T_{dim.red}
\approx 2 T_c$, $G_2^{mag} \approx O(T^4)$, in the regime of
dimensional reduction. }. Therefore correlators and $\sigma^E,
\sigma^H$ are almost constant till $T=T_c$ and  at $T\geq T_c$ a
new vacuum phase with $D^E=\sigma^E =0$ is realised, which yields
lower thermodynamic potential (higher pressure). Lattice
measurements \cite{7*,trusov8} support this picture. The crucial
step is that one should take into account in the free energy $F$
of the system also the free energy of the vacuum, i.e. vacuum
energy density $\varepsilon_{vac} =\frac{1}{4} \theta_{\mu\mu} =
\frac{\beta(\alpha_s)}{16\alpha_s} \lan (F^a_{\mu\nu})^2\ran=
-\frac{(11-\frac23 n_f)}{32} G_2$, which can be estimated via the
standard gluonic  condensate \cite{trusov16} $G_2\equiv
\frac{\alpha_s}{\pi} \lan (F^a_{\mu\nu})^2\ran\approx 0.012$
GeV$^4$.

Hence for the pressure $P=-F$ one can write in  the phase $I$
(confined) 
\be 
P_I =|\varepsilon_{vac}|+\chi_1(T)
\label{trusov10}
\ee 
where $\chi(T)$ is the hadronic gas pressure, starting with pions, $\chi_{pion} \cong \frac{\pi^2}{30}
T^4$. In the deconfined phase one can write 
\be 
P_{II}=|\varepsilon^{dec}_{vac}|+(p_{gl} + p_q)T^4
\label{trusov11}
\ee
where $|\varepsilon^{dec}_{vac}|$ is  the vacuum energy density in
the deconfined phase, which is mostly (apart from $D^E_1(0)\approx
0.2 D^E(0)$, see \cite{7*,trusov9}) colourmagnetic energy density
and by the same reasoning as before we take it as for $T=0$, i.e.
$|\varepsilon^{dec}_{vac}| \cong 0.5|\varepsilon_{vac}|$.

Equalizing $P_I$ and $P_{II}$ at  $T=T_c(\mu)$ one obtains the
equation for $T_c$ 
\be 
T_c(\mu)=\left(\frac{\Delta|\varepsilon_{vac}|+\chi(T)}{p_{gl}+p_q}\right)^{1/4}
\label{trusov12}
\ee 
where $\Delta|\varepsilon_{vac}| =|\varepsilon_{vac} | - | \varepsilon_{vac}^{dec}| \approx
\frac{(11-\frac23 n_f)}{32} \Delta G_2 $;~ $\Delta G_2\approx
\frac12 G_2;$ $p_{gl}$ and $p_q$ are given in (\ref{trusov5}),
(\ref{trusov7}) respectively and depend on both $T_c$ and $\mu$.

In this letter we shall consider the simplest case when the
contribution of hadronic gas $\chi_1(T)$ can be  neglected in the
first approximation. Indeed, pionic gas yields only $\sim 7$\%
correction to the numerator of (\ref{trusov12}) at $T\approx T_c$,
and from \cite{trusov22a} one concludes that $\chi(T_c)\lesssim
0.5 T^4_c$, which yields a $\lesssim 10\%$ increase of $T_c$ for
$G_2\sim 0.01$ GeV.

From  the expression for $T_c(\mu)$ (\ref{trusov12}) one can find limiting
behaviour of $T_c(\mu\to 0)$ and $\mu_c(T\to 0)$. For the first
one can use for $p_q$ and $p_g$ (\ref{trusov4}) and (\ref{trusov5}) and
expand r.h.s. of (\ref{trusov12}) in ratio $p_g/p_q$ with the result.
\be 
T_c=T^{(0)}\left(1+\frac{V_1(T_c)}{8T_c}+O\left(\left(\frac{V_1(T_c)}{8T_c}\right)^2\right)\right)
\label{trusov13}
\ee
where the last term yields a 3\% correction, and 
$T^{(0)}=\left(\frac{(11-\frac23 n_f) \pi^2\Delta G_2}{32\cdot 12n_f}\right)^{1/4}$. Solving (\ref{trusov13}) for $T_c$ one has 
\be
T_c=\frac12 T^{(0)} \left(1+\sqrt{1+\frac{\kappa}{T^{(0)}}}\right) \left(1+\frac{m^2_q}{16T^2_c}\right)
\label{trusov14} 
\ee 
with $\kappa\equiv \frac12 V_1(T_c)$. From (\ref{trusov12}), (\ref{trusov13}) one can compute
expansion $T_c(\mu)$ in powers of $\mu$, 
$$ 
T_c(\mu_B)=T_c(0)\left(1-C\frac{\mu^2_B}{T^2_c(0)}\right),\quad\mu_B =3\mu.
$$
$$
C=\frac{1+\sqrt{1+\frac{\kappa}{T^{(0)}}}}{144\sqrt{1+\frac{\kappa}{T^{(0)}}}}=0.0110(3)~~{\rm for}~n_f=2,3,4.
$$

One can see, that $C$ practically does not depend on $n_f$ and is
in the same ballpark as  the values  found by lattice
calculations, see \cite{trusov3,trusov27} for reviews and
references.

Another end point of the phase curve, $\mu_c(T\to 0)$, is found
from (\ref{trusov12}) when one takes into account asymptotic
$\Phi_0 (a \to \infty) =\frac{a^4}{4} +\frac{\pi^2}{2} a^2 +
\frac{7\pi^4}{60}+...$, which yields (for small $\frac{m_q}{\mu}$)
\be 
\mu_c(T\to 0)=\frac{V_1(T_c)}{2}+(48)^{1/4} T^{(0)}\left( 1+ \frac{3m^2_q}{4\mu^2_c}\right)\left( 1-\frac{\pi^2}{2}
\frac{T^2}{\left(\mu_c-\frac{V_1(T_c)}{2}\right)^2}\right)
\label{trusov15}
\ee

\begin{figure}[htb]
\includegraphics[width=60mm]{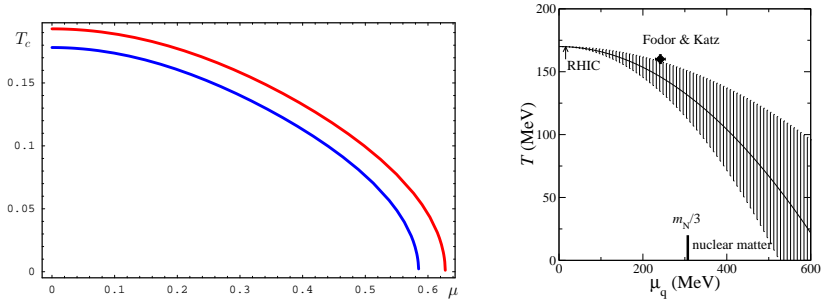}\hphantom{aaaa}
\includegraphics[width=40mm]{ptline_0318.eps}
\caption{The phase transition curve $T_c(\mu)$ from
Eq.(\ref{trusov12}) (in GeV) as function of quark chemical
potential $\mu$ (in GeV) for $n_f=2$ (upper curve ) and $ n_f=3$
(lower curve) and $\Delta G_2=0.0034$ GeV$^4$ (left side) in
comparison with lattice results \protect\cite{trusov22} (right
side)}. \label{f2trusov}
\end{figure}

The resulting curve $T_c(\mu)$ according to Eq.(\ref{trusov12})
with $\chi_1\equiv 0$ is given  on Fig.~\ref{f2trusov} (left side)
for $\Delta G_2 =0.00341 $GeV$^4$, $n_f=2,3$ and $m_q=0$. To
compare, we have shown on the same figure (right side) the lattice
data \cite{trusov22} obtained in the reweighting technique.

\subsection{Discussion of results}

The prediction  of $T_c(\mu)$ depends only on two numbers: 1) the value of
gluonic condensate $ \Delta G_2$ and 2) the value of
$V_1(T_c)=0.5$ GeV taken from lattice data \cite{27} (and quantitatively close to
the value from the analytic form \cite{9}).

We take $G_2$ in the limits 0.004 GeV$^4 \leq G_2\leq0.015$
GeV$^4$, the value  $G_2=0.008 $GeV$^4$ ($\Delta G_2=0.0034 $ GeV$^4$)  being
in agreement with lattice data of $T_c(0)$, for $n_f=0,2,3$, see Table~\ref{table_trusov}.

\begin{table}
\caption{The values of  $T_c(\mu=0)$ and $\mu_c(T=0)$ computed
using (\ref{trusov14}) and (\ref{trusov15}) for several values of
$\Delta G_2$ and $ n_f=0,2,3$.} \label{table_trusov}
\begin{center}
\begin{tabular}{|l|l|l|l|l|}
\hline
&&&&\\
$\Delta G_2/(0.01~{\rm  GeV}^4)$& 0.191&0.341&0.57&~~~1\\
&&&&\\
\hline
&&&&\\
$T_c({\rm GeV})$~~ $n_f=0$ &0.246&0.273&0.298&0.328\\ &&&&\\
\hline
&&&&\\
$T_c({\rm GeV})$~~ $n_f=2$ &0.168&0.19&0.21&0.236\\ &&&&\\
\hline
&&&&\\
$T_c({\rm GeV})$~~  $n_f=3$ &0.154&0.172&0.191&0.214\\&&&&\\
\hline
&&&&\\
$\mu_c({\rm GeV})$ ~~ $n_f=2$ & 0.576& 0.626&0.68&0.742\\&&&&\\
\hline
&&&&\\
$\mu_c({\rm GeV})$ ~~ $n_f=3$ & 0.539& 0.581&0.629&0.686\\&&&&\\
\hline
\end{tabular}
\end{center}
\end{table}

Note that $T_c$ at $n_f=0$ in Table~\ref{table_trusov}  is
obtained not from (\ref{trusov14}), but directly from
(\ref{trusov12}) with $n_f=0, \chi(T)=0$.

The curve in Fig.~\ref{f2trusov} has the expected form, which agrees with the
curve, obtained in \cite{trusov22} by the reweighting technique and
agrees  for $\mu<300$ MeV with  that, obtained by the density
of state method \cite{trusov26}, and by the imaginary $\mu$ method \cite{trusov25}.

An analysis of the integral (\ref{trusov8}) for $\nu=0$ reveals that it
has a mild singular point at $\mu_{sing} =\frac{V_1}{2} \pm i\pi T$,
which may show up in derivatives in $\frac{\mu}{T}$. At
$T=0, \mu_{sing} = \frac{V_1}{2} \cong 0.25$ GeV and is  close
to the point where one expects irregularities on the phase
diagram  \cite{trusov26,trusov28}.

The limit of small $\mu$ is given in (\ref{trusov14}). Taking $V_1
(T_c)\approx 0.5$ GeV as follows from lattice and analytic
estimates, one obtains with $\sim 3\%$ accuracy the values of
$T_c$ given  on Fig.~\ref{f2trusov} for $n_f=2,3$ and $\Delta
G_2=0.0034$ GeV$^4$.

The values of $\mu_c, $  from (\ref{trusov15})  are given in Table~\ref{table_trusov}  and are in
agreement with the curves shown in Fig.~\ref{f2trusov}. Note that
$\chi(T=0)=0$ and (\ref{trusov15}) holds also in the case, when
hadron (and baryon) gas is taken into account.

At this point one should stress that our calculation  in VDM of
$p(T)$ does not contain  model parameters and the only
approximation is the neglect of interparticle interaction as
compared to the interaction of each one with  the vacuum (apart
from neglect of $\chi(T)$). Fig.~\ref{f1trusov} demonstrates that
this approximation is reasonably good and one expects some
10$\div15\%$ accuracy in prediction of $T_c(\mu)$. Note that in
VDM the phase transition is of the first order, which is supported
for $n_f=0$ by  lattice data, see e.g. \cite{24} however for
$n_f=2,3$ lattice results disagree, see \cite{trusov29,trusov30}
for a possible preference of the first order transition. One
however should have in mind that the final conclusion for lattice
data depends on input  quark masses and continuum limit.

The ``weakening'' of  the phase transition for $n_f>0$ and nonzero
quark  masses is explained in our approach by the flattening of
the curve $P(T)$ at $T\approx T_c$ when hadronic gas $\chi(T)$ is
taken into account,  since $\chi(T)$ for  $n_f=0$ is much smaller
than for $n_f=2$.

Finally, as seen from our expressions for $p_q, p_g$
(\ref{trusov4}), (\ref{trusov7}), our EoS is independent of
$Z(N_c)$ factors and $Z(N_c)$ symmetry  is irrelevant for the NP
dynamics in our approach. This result is a consequence of more
general property -- the gauge invariance of the partition function
for all $T$, $\mu$, which requires that only closed Wilson loops
appear in the resulting expressions, yielding finally  only
absolute value of the Polyakov loop, or in other words, is
expressed via only singlet free energy of quark and  antiquark
$F_{Q\bar Q}^{(1)} \approx V_1$.

\section{Concluding discussions} \label{Concl}

\subsection{Comparison to other approaches}

Our resulting EoS, in particular $P(T,\mu)$ in the leading (single
line) approximation has the form of ideal gas  factors multiplied
by moduli of Polyakov Lines (PL). In the next orders also
interaction between quark and gluon lines is included, which leads
to the decreasing of pressure, so the final results are in better
agreement with lattice data, see Figs. 3--6.

The interaction between quarks and gluons has both colourelectric
and colourmagnetic components and was discussed in section 4.

Thus it is clear that PL play a very important dynamical role in
the FCM. This role is in creation of $T$-dependent self-energy
term for each individual quark or gluon, which suppresses their
contribution to the pressure. One can say therefore, that quarks
and gluons acquire vacuum induced effective masses.

From this point of view it is reasonable to discuss other
approaches in the literature, where PL  play important role.

One should start with the formalism of effective action for
Polyakov lines, existing for almost  30 years
\cite{myst1m}$^-$\cite{myst3m}, see also \cite{myst4m} for later
development.

Here PL and $Z(N)$ symmetry are mostly used conceptually as an
order  parameter for the deconfinement transition, and the
effective potential was calculated for purely gluon and
quark-gluon system to study potential extrema  in terms of PL
moduli and phases. From the discussion in sections 1-4 it is
clear however, that only moduli of PL enter EoS  in our approach and $Z(N)$
symmetry is irrelevant for dynamical properties of QGP.
Moreover, quarks violate $Z(N)$ symmetry and the latter has no
connection with the deconfinement in the unquenched case.
Instead, moduli of PL are crucially important as dynamical input in
establishing EoS and the QCD phase diagram at nonzero $\mu$.

Indeed, it is the modulus of $L(T)$, which defines in Eq.~(\ref{trusov12}) the
transition temperature $T_v(\mu)$ in the SLA.

There is another approximation \cite{myst5m,myst6m} in the theory of
QGP, which is closer to ours, since it also uses PL as a kind
of dynamical input. It is called PNJL and is based on the
Nambu-Jona-Lasinio Lagrangian with quark operators  augmented
by PL. Here one obtains EoS and effective  quark and gluon masses
in qualitative agreement with lattice data.  However, since NJL
Lagrangian is not  confining, the detailed picture of
deconfinement may be quite different.

The action density of the PNJL model can be written as (see
R\"{o}ssner et al in \cite{myst6m} for a review and additional
references)
\be 
S=\psi^+(\partial_\tau -i \veal \mathbf{\nabla}+\gamma_4 m_0 -A_4 )\psi + N(\psi,\psi^+),
\label{1m}
\ee
where $N$ is quartic in $\psi$ as in NJL model, and  $A_4$ is
constant field entering PL, which therefore become complex in
general. This fact and the associated $Z(N_c)$ symmetry differs
from our results and our EoS, where only modulus  of PL enters.
Nevertheless one can obtain in this approach a reasonable
agreement for the first moments of pressure with respect to
$\mu/T$ \cite{myst6m}.

At this point it is useful to compare our results with the
phenomenological models of QGP, where quark and gluon masses
have been introduced as input and fitting parameters ,see e.g.
\cite{myst7m,myst8m}.

In \cite{myst7m,myst8m} the quasiparticle  propagators are
introduced with real and imaginary parts of self-energy (effective
mass) terms, which can be compared to our PL self-energies,
$\varepsilon_i=\frac12 c_i V_1 (\infty, T)$, $L_i =\exp
\left(-\frac{V_1(\infty, T)}{2T} c_i\right),~~ i=q,g,~~ c_i =1,
\frac94$.

It is interesting that the resulting effective masses for quarks
and gluons approximately satisfy Casimir scaling as in our
$c_iV_1(\infty, T)$, and have the same order of magnitude as in
our calculations. A more detailed comparison will be subject of
future publications.

We now turn to the important question of the density induced phase
transition, which may be important for several planned ion
machines and possible existence of quark cores in neutron stars.
As was discussed above in section 5, FCM predicts deconfinement
phase transition at large $\mu_q= \mu_{cr} \approx 0.6 $ GeV,
provided vacuum properties and in particular gluon condensate and
PL are not affected by $\mu_q$. It is important to study  what
happens in the region $\mu_N \leq\mu_q \leq\mu_{cr}$, where
$\mu_N$ is the chemical potential at normal nuclear density,
$\mu_N \approx 0.3$ GeV.

It is clear that nonzero $\mu_q$ can influence baryon structure,
and some change in the nucleon mass and radius was studied before
\cite{myst9m}. However much more drastic change due to $\mu_q$
appears in the interquark forces, as was found in \cite{myst10m}.
Here it was realized that light quark interaction and confinement
is obtained self-consistently from nonlinear equations, yielding
linear scalar confinement for vanishing $\mu_q$, but a deeper well
for $r<\frac{\mu_q}{\sigma}$ appears for nonzero $\mu_q$, which
makes baryon lighter and more compact. Moreover, $3nq$ systems may
become more stable for $n=1,2,...$ due to larger number of pairing
interquark forces. More detailed quantitative analysis is needed
for the final positive answer, however it seems feasible , that at
growing $\mu_q$ the nuclear matter becomes a dense  medium of
smaller size $3q$ with an admixture of $6q, 9q$,... bags, the
latter is growing in percentage with $\mu_q$, unless quarks find
themselves finally in the deconfined phase. In this scenario the
role of multiquark bags and, hence, of cumulant  mechanism is
growing with density, which can be measured experimentally in
future ion-ion experiments.

Till now nothing was said about $qq$ pairing and possible quark
superconductivity, widely discussed in literature --- see
\cite{myst11m} for reviews. The $np$ vacuum changes this picture in several
respects. The first one was discussed in \cite{myst12m}, where it
was stressed, that the minimum of thermodynamic potential
$\Omega_0$ (with the gap in the scalar diquark sector $\Delta_0
=(0.1\div 0.15)$ GeV is of the order of $10^{-3}$   GeV$^4$ or
less, and this value is of the order of $np$ vacuum density
$|\varepsilon|$, so the realization of the perturbative diquark
condensation crucially depends on this $np$ value and is
questionable when $|\Omega_0| < |\varepsilon|$.

Another $np$ feature of the QCD vacuum, which may prevent standard
diquark scenario, is the $qq$ interaction in the diquark sector,
discussed in section 4. As can be seen in Eq.~(\ref{039})
interaction in the $qq$ sector can be written as

\be 
V_{qq}=\bar a_d V_1(\infty, T)+\bar b_d V_1 (r,T),
\label{2m}
\ee 
where $\bar a_{\bar 3} =\frac12,~~ \bar a_6
=\frac54,~~ \bar b_{\bar 3} =\frac12, ~~\bar b_6=-\frac14$ to be
compared with $q\bar q$ system, where $\bar a_1 =0,~~ \bar b_1 =1$
and $V_1 (\infty, T_c) \approx 0.5$ GeV.

Here $V_1 (r, T)$ is the sum of perturbative and $np$
interactions, $V_1 (r,T)= V_1 (\infty, T) + v(r, T)$, and $v_{np}
(r, T)$ is negative everywhere.

As a result one has additional (constant) repulsion in the $qq$
system as compared to $q \bar q$, of the order of $\frac12 V_1
(\infty, T_c)\approx 0.25$ GeV, which is larger than  the assumed
gap $\Delta_0$. Therefore one can expect that pairing is destroyed
and one has only white correlations in the $q\bar q$ and $qqq$
systems which can appear as bound states, e.g. in the $S$ -wave
$c\bar c$ system, which is supported by lattice data
\cite{myst13m}.

Thus we are coming again to the `` quark-bag scenario'' discussed
above, where nucleons in nuclear matter become tighter and lighter
with increasing density and are accompanied with growing number of
$6q$, $9q$, $\ldots$ quark bags finally occupying the whole volume.

This scenario possibly implies a more hard EoS of nuclear matter,
than usually exploited and can lead to higher masses of neutron
stars with quark  matter cores, see in this respect  the
discussion in \cite{myst14m,concl15}.

Finally, we discuss in interesting idea of Dremin and
collaborators, about the possibility of Cherenkov emission of
gluons in QGP due to the effective creation of nonzero
dielectric permittivity $\varepsilon$ in a dense medium
\cite{myst15m}. As was shown there, this Cherenkov emission can
possibly explain the double-humped structure of the  away-side jet
at RHIC. We can add  here from the point of view of FCM, that
$\varepsilon$ is sensitive not only to the medium effects, but can
also get contributions from the change of QCD vacuum properties,
which happens at $T>T_c$. This topic is now under investigation.

\subsection{Mysteries of deconfinement}

The theory discussed in this review was able to address most important 
questions related to the QGP dynamics, namely:
\begin{enumerate}
\item Why does deconfinement take place at some temperature and
density? The theory based on the FCM shows that thermodynamic
potential (pressure) grows faster with the temperature $T$ in the
deconfined phase, as compared to the confined phase, so that the
(almost) constant difference in vacuum energy density between the
phases is overcome at some $T=T_c$. 
\item What is the main
dynamical difference between the phases? It is shown in the
review that, in the confined phase, both colourelectric (CE) and
colourmagnetic (CM) confinement coexist while, in the deconfined
phase, only the latter is retained \cite{concl1}. In the meantime,
CE forces due to the nonconfining correlator $D^E_1$ also survive
in the deconfined phase and they can support bound states at not too large $T$'s,
$T\lesssim 1.5 T_c$. 
\item What are manifestations of the CM
confinement? As was discussed in this review, the CM confinement
ensures strong interaction in the quark-gluon systems and can
support weakly bound states for nonzero angular momentum $L$.
Moreover, most part of the spin-dependent forces is due to the CM
confining correlator $D^H$ and thus it survives the deconfinement
transition \cite{concl2}. The CM confinement solves the old Linde
paradox since the 3D perturbative diagrams acquire Wilson loops
of the confining background field. The latter converge at large
distances as $\langle W\rangle\sim\exp(-\sigma_H\times\mbox{area}$) and
make all integrals convergent \cite{8*}. However, this is probably
only a small part of the story. Indeed, the very notion of
confinement implies that, when moving with some velocity, any coloured subsystem 
is not free but is rather connected, by a CM
string, to its colour partner, which makes the entire system white. As a
consequence, a gluon or a quark acquires a nonperturbative CM mass,
which is velocity-dependent. This effect was studied at the
example of the Debye mass, which was calculated in \cite{25} and was shown
to reproduce exactly the lattice data \cite{concl5} (in contrast
to a perturbative treatment, which is not gauge invariant and is about
50\% off the data \cite{concl5}). It is clear, that this CM
dynamics is of a general character and may be related to many
features ascribed to QGP, for example to jet quenching, collective
phenomena, and so on. This important topic requires a further careful investigation. 
\item What is the connection between deconfinement and Chiral Symmetry (CS)
restoration? On the theoretical side, it was found in the FCM that
the same correlator $D^E$, which ensures the CE confinement, enters
the kernel of the nonlinear equation, the latter demonstrating Spontaneous
Breaking of Chiral Symmetry  (SBCS) \cite{trusov31}. Thus
confinement entails SBCS, and both should disappear at the same
temperature. This behaviour is observed in most of lattice simulations
(see, for example, \cite{lat1}). However, in some papers (for example, in \cite{concl8}), a
difference $\Delta T\sim 20$ MeV is found between the temperatures of the deconfinement and CS restoration, 
though it is not clear whether it is a physical effect or just a result of lattice
approximations. One might inquire at this point of the role played by
the CM confinement. This problem can be connected with the value
of the chiral condensate $\langle \bar q q\rangle$, as a
function of temperature. It used to be believed and found on the
lattice that $\langle \bar q q\rangle$ disappeared for $T\geq
T_c$. However, it was found recently, in new accurate calculations \cite{concl9}
for the quenched SU(2) case with overlap quarks (not violating
CS), that almost a half of the $|\langle\bar q q\rangle|$ survives at
$1.5T_c\geq T\geq T_c$. This observation calls for further work both
on the lattice and in the analytic domain. In addition to the studies with
quarks in the fundamental representation, discussed above, there
appeared several works \cite{concl10} on the phase transition for
adjoint quarks, where it was found that $T_c< T_{SBCS}$. This
fact stresses again that we are still far from the detailed
understanding of the connection between confinement and SBCS.
\item An important role in comparison of the QGP properties and results
of RHIC experiments is played by the transport coefficient of the QGP, 
and especially by the shear viscosity $\eta$ and bulk viscosity
$\zeta$ \cite{concl11}. The existing calculations are not very reliable since they
exploit nonperturbative dynamics in typically Minkowski (not Euclidean) configurations. The FCM, supplied by a careful
analytic continuation from the Euclidean region, is able to
produce both $\eta$ and $\zeta$ as saturated by propagators of
two-gluon glueballs, coupled by the CM confinement. This work is
in progress now. 
\item The FCM approach to the phase transition at nonzero
$T$ and $\mu$ was given above, in section 5.1, based on the
assumption that finite density does not affect strongly the vacuum
energy gap, the latter being known from conformal anomaly. This is supported by
lattice calculations \cite{trusov20} of the free energy $F(T,\mu)$ and looks plausible, since the internal scale of the vacuum energy
density is connected to excitations with the mass of order of the lowest glueball mass, $M\approx 1.5$ GeV, whereas the
corresponding $\mu$ is much lower. However, this question requires further detailed studies since it may change drastically
the entire picture of the phase transition at low $T$'s due to density.
Another point which might invalidate the standard picture of the $qq$ pairing phases is the fact that vacuum fields, especially
measured by the CM correlators, are very strong and can deform or totally destroy this pairing, preferring the $q \bar q$ pairing.
\item A very important but still unresolved question is the
dimensional transition from 4D to 3D with the temperature increase. Studying lattice data for the spacial string tension
$\sigma_H (T)$ (see, for example, \cite{concl14} and refs. therein), one
can see a sharp change of its behaviour at $T=1.5 T_c$, for SU(2), and at
$T=1.2 T_c$, for SU(3), which is not explained by theory. At larger
$T$ one has a typical 3D behaviour $\sigma_H(T) \sim T^2$.
The coefficient was considered, for example, in \cite{concl**}. This point
of almost ``dimensional phase transition" deserves more study.
\item There is an interesting development in the theory of the QGP,
namely, QGP in an external field, which is totally beyond the
scope of this review. We refer the reader to \cite{concl***} and papers cited there
for details. 
\item Finally, let us mention the problem of the phase transition from nuclear
matter to quark matter. The EoS for the nuclear matter at
growing density cannot be obtained as an extrapolation of the EoS
for the normal nuclear density. As was mentioned in \cite{myst10m,myst12m},
density may strongly distort the usual confinement forces
between quarks in a nucleon, the latter may loose in its mass and
radius, and multiquark bags may be formated. This very
interesting development may be of a practical use and can be
connected both to the phenomenon of the so-called nuclear scaling
\cite{concl17} and to the hot point of quark stars --- see, for example, \cite{concl15}. 
\end{enumerate}
We stop at this point leaving, however, many interesting problems beyond the scope of this review.

\section*{Acknowledgements}\label{Acknowl}

This work was supported by the Federal Agen\-cy for Atomic Energy
of Russian Fe\-de\-ration, by the Federal Programme of the Russian
Ministry of Industry, Science, and Technology No. 40.052.1.1.1112,
by the grant NSh-4961.2008.2 for the leading scientific schools,
and by the grants RFFI-06-02-17012 and RFFI-06-02-17120. Work of
A. V. N. was supported by DFG-436 RUS 113/991/0-1(R), PTDC/FIS/70843/2006-Fisica, and by the non-profit
``Dynasty'' foundation and ICFPM. M. A. T. would like to
acknowledge the partial support from the President Grant No.
MK-2130.2008.2.


\begin{thebibliography}{999}
\bibitem{1} B. M\"{u}ller, J. L. Nagle, {\em Ann. Rev. Nucl. Part. Sci.} {\bf 56}
(2006) 93; D. d'Enterria, {\em J. Phys. G} {\bf 34} (2007) S53;
BRAHMS: {\em Nucl. Phys. A} {\bf 757} (2005) 1; PHENIX: {\em Nucl.
Phys. A} {\bf 757} (2005) 184; PHOBOS: {\em Nucl. Phys. A} {\bf
757} (2005) 28; STAR: {\em Nucl. Phys. A} {\bf 757} (2005) 102.
\bibitem{2} J. Kapusta, {\em Finite Temperature Field theory} (Cambridge University Press, Cambridge, 1989);
K.Yagi, T.Hatsuda and Y.Miake, {\em Quark-gluon Plasma} (Cambridge Univ. Press, Cambridge 2005).
\bibitem{3} G. J. Gross, R. D. Pisarski, and L. G. Yaffe, {\em Rev. Mod. Phys.} {\bf 53} (1981) 43;
L. G. Yaffe, {\em Nucl. Phys. B (Proc. Suppl.)} {\bf 106} (2002) 117; M. Gyulassy and L. McLerran, {\em Nucl. Phys. A} {\bf 750} (2005) 30.
\bibitem{4} H. G. Dosch, {\em Phys. Lett. B} {\bf 190} (1987) 177; H. G. Dosch and Yu.A.Simonov, {\em Phys. Lett. B} {\bf 205} (1988), 339;
Yu.A.Simonov, {\em Nucl. Phys. B} {\bf 307} (1988) 512.
\bibitem{4*} A. Di Giacomo, H. G. Dosch, V. I. Shevchenko, and Yu. A. Simonov, {\em Phys. Rep.} {\bf 372} (2002) 319.
\bibitem{5*} Yu. A. Simonov, {\em Proc. QCD: Perturbative or Nonperturbative}
eds. L.Ferreira., P.Nogueira, J.I.Silva-Marcos (World Scientific,
Singapore, 2001), p. 60; arXiv: hep-ph/9911327.
\bibitem{6*} A. M. Badalian, V. I. Shevchenko, and Yu. A. Simonov, {\em Yad. Fiz.} {\bf 69} (2006) 1818.
\bibitem{7*} M. D'Elia, A. Di Giacomo, and E. Meggiolaro, {\em Phys. Lett. B} {\bf 408} (1997) 315;
{\em Phys. Rev. D} {\bf 67} (2003) 114504; A. Di Giacomo, E. Meggiolaro, and H. Panagopoulos, {\em Nucl. Phys. B} {\bf 483} (1997) 371.
\bibitem{6} Yu. A. Simonov, {\em JETP Lett.} {\bf 55} (1992) 605.
\bibitem{8*} Yu. A. Simonov, {\em Proc. Varenna 1995, Selected Topics in Nonperturbative QCD}, p.319.
\bibitem{7} Yu. A. Simonov, {\em JETP Lett.} {\bf 54} (1991) 249.
\bibitem{5} Yu. A. Simonov, {\em Phys. At. Nucl.} {\bf 58} (1995) 309.
\bibitem{Nef} A. V. Nefediev and Yu. A. Simonov, {\em Phys. At. Nucl.} {\bf 71} (2008) 171.
\bibitem{FFSR} Yu. A. Simonov and J. A. Tjon, {\em Ann. Phys.} {\bf 223} (1993) 1, {\em ibid} {\bf 300} (2002) 54.
\bibitem{10}  M. Halpern, {\em Phys. Rev. D} {\bf 19} (1979) 517; I. Aref'eva, {\em Theor Math. Phys.} {\bf 43}, 353 (1980);
N. Bralic, {Phys. Rev. D} {\bf 22} (1980) 3090; Yu. A. Simonov, {\em Sov. J. Nucl. Phys.} {\bf 50} (1989) 134;
M. Hirayama and S. Matsubara, {\em Progr. Theor. Phys.} {\bf 99} (1998) 691.
\bibitem{11} Yu. A. Simonov, {\em JETP Lett.} {\bf 71} (2000) 127; V. I. Shevchenko and Yu. A. Simonov, {\em Phys. Rev. Lett.}
{\bf 85} (2000) 1811.
\bibitem{D} Yu. A. Simonov, {\em Z. Phys. C} {\bf 53} (1992) 419; A. M. Badalian, B. L. G. Bakker, and Yu. A. Simonov,
{\em Phys. Rev. D} {\bf 75} (2007) 116001.
\bibitem{Komas} Y. Koma and M. Koma, {\em Nucl. Phys. B} {\bf 769} (2007) 79.
\bibitem{BNS2} A. M. Badalian, A. V. Nefediev, and Yu. A. Simonov, {\em JETP Lett.} {\bf 88} (2008) 648.
\bibitem{ein} L. Brink, P. Di Vecchia, and P. Howe, {\em Nucl. Phys. B} {\bf 118} (1977) 76.
\bibitem{DKS} A. Yu. Dubin, A. B. Kaidalov, and Yu. A. Simonov, {\em Phys. Lett. B} {\bf 323} (1994) 41.
\bibitem{ein3} Yu. S. Kalashnikova and A. V. Nefediev, {\em Phys. At. Nucl.} {\bf 60} (1997) 1389.
\bibitem{ein2} Yu. A. Simonov, {\em Phys. Lett. B} {\bf 226} (1989) 151.
\bibitem{Dirac} P. A. M. Dirac, {\em Letures on Quantum Mechanics} (Belter Graduate School of Science, Yeshiva University, New York (1964)).
\bibitem{regge} V. L. Morgunov, A. V. Nefediev, and Yu. A. Simonov, {\em Phys. Lett. B} {\bf 459} (1999) 653.
\bibitem{KNS} Yu. S. Kalashnikova,  A. V. Nefediev, and Yu. A. Simonov, {\em Phys. Rev. D} {\bf 64} (2001) 014037.
\bibitem{EF} E. Eichten and F. L. Feinberg, {\em Phys. Rev. D} {\bf 23} (1981) 2724.
\bibitem{12} Yu. A. Simonov, {\em Nucl. Phys. B} {\bf 324} (1989) 67; A. M. Badalian and Yu. A. Simonov, {\em Phys. Atom. Nucl.} {\bf 59},
(1996) 2164, [{\em Yad. Fiz.} {\bf 59} (1996) 2247]; M. Schiestl and H. G. Dosch, {\em Phys. Lett.  B} {\bf 209} (1988) 85.
\bibitem{EF2} A. Barchielli, N. Brambilla, and G. M. Prosperi, {\em Nuovo Cim. A} {\bf 103} (1990) 59;
A. Pineda and A. Vairo, {\em Phys. Rev. D} {\bf 63} (2001) 054007; N. Brambilla and A. Vairo, {\em Phys. Rev. D} {\bf 55} (1997) 3974.
\bibitem{Gr} D. Gromes, {\em Z. Phys. C} {\bf 26} (1984) 401.
\bibitem{BNS3} A. M. Badalian, A. V. Nefediev, and Yu. A. Simonov, {\em Phys. Rev. D} {\bf 78} (2008) 114020.
\bibitem{242} Yu. A. Simonov, in: {\em "Sense of Beauty in Physics"} (Pisa Univ. Press, 2006), p.29;
A. M. Badalian and Yu. A. Simonov, {\em Yad. Phys.} {\bf 59} (1996) 2247.
\bibitem{24} S. Gupta, K. H\"{u}bner, and O. Kaczmarek, {\em Nucl. Phys. A} {\bf 785} (2007) 278; K. H{\"u}bner and C. Pica,
arXiv:0809.3933[hep-lat].
\bibitem{Fr} Yu. A. Simonov, {\em Phys. At. Nucl.} {\bf 58} (1995) 107; {\em ibid} {\bf 65} (2002) 135.
\bibitem{an1} Yu. A. Simonov, {\em Ann. Phys.} {\bf 323} (2008) 783.
\bibitem{an2} E. V. Komarov and Yu. A. Simonov, {\em Ann. Phys.} {\bf 323} (2008) 1230.
\bibitem{23} H. G. Dosch, H.-J. Pirner, and Yu. A. Simonov, Phys.Lett. B {\bf 349} (1995) 335.
\bibitem{Cas} G. S. Bali, {\em Phys. Rev. D} {\bf 62} (2000) 114503; S. Deldar, {\em Phys. Rev. D} {\bf 62} (2000) 034509.
\bibitem{Red} Yu. A. Simonov, {\em Phys. At. Nucl.} {\bf 70} (2007) 44.
\bibitem{25} N. O. Agasian and Yu. A. Simonov, {\em Phys. Lett. B} {\bf 639} (2006) 82.
\bibitem{13} Yu. A. Simonov, {\em Nucl. Phys. B} {\bf 592} (2001) 350.
\bibitem{9} Yu. A. Simonov, {\em Phys. Lett. B} {\bf 619} (2005) 293.
\bibitem{26} O. Kaczmarek, F. Karsch, P. Petreczky, and F. Zantow, {\em Phys. Lett. B} {\bf 543} (2002) 41.
\bibitem{anal} Yu. A. Simonov, {\em Phys. At. Nucl.} {\bf 69} (2006) 528.
\bibitem{27} O. Kaczmarek and F. Zantow, arXiv:hep-lat/0506019.
\bibitem{8} A. Di Giacomo {\em et al.}, {\em Phys. At. Nucl.} {\bf 70} (2007) 908.
\bibitem{0608003} F. Karsch, {\em J. Phys. Conf. Ser.} {\bf 46} (2006) 122.
\bibitem{0610017} C. Bernard, T. Burch, C. DeTar, S. Gottlieb, L. Levkova, U. M. Heller, J. E. Hetrick, D. B. Renner, D. Toussaint,
and R. Sugar, {\em PoSLAT2006} {\bf 139} (2006).
\bibitem{ST} Yu. A. Simonov and M. A. Trusov, {\em Phys. Lett. B} {\bf 650} (2007) 36; {\em JETP Lett.} {\bf 85} (2007) 730.
\bibitem{0701210} F. Karsch, arXiv:hep-ph/0701210.
\bibitem{0510084} Y. Aoki, Z. Fodor ,S. D. Katz, K. K. Szabo, {\em JHEP} {\bf 0601} (2006) 089.
\bibitem{latbound} P.Petreczky, hep-lat/0409139; {\em Nucl. Phys. A} {\bf 785} (2007) 10.
\bibitem{36} E. V. Shuryak and I. Zahed, {\em Phys. Rev. D} {\bf 70} (2004) 054507.
\bibitem{32} K. H{\"u}bner, O. Kaczmarek, F. Karsch, and O. Vogt, arXiv:hep-lat/0408031.
\bibitem{BS0} F. Buisseret and C. Semay, {\em Phys. Rev. D} {\bf 70} (2004) 077501.
\bibitem{z137} A. B. Migdal, D. N. Voskresenksy, and V. S. Popov, {\em Pis'ma v ZhETF}, {\bf 24} (1976) 186;
{\em ZhETF}, {\bf 72} (1977) 834.
\bibitem{negmu} Yu. A. Simonov, {\em Phys. Atom. Nucl.} {\bf 68} (2005) 709.
\bibitem{ant} D. Antonov, S. Domdey, and H.-J. Pirner, {\em Nucl. Phys. A} {\bf 789} (2007) 357.
\bibitem{trusov8} M. D'Elia, A. Di Giacomo and E. Meggiolaro, {\em Phys. Rev. D} {\bf 67} (2003) 114504.
\bibitem{trusov20}  M. D\"{o}ring, S. Ejiri, O. Kaczmarek,  F. Karsch, E. Laermann,
{\em Eur. Phys. J. C} {\bf 46} (2006) 179.
\bibitem{trusov25} M. P. Lombardo, {\em PoSCPOD2006} (2006) 003; M. D. 'Elia and M. P. Lombardo, {\em Phys. Rev. D} {\bf 67} (2003) 014505,
ibid {\bf 70} (2004) 074509; Ph. de Forcrand and O.Philipsen, {\em Nucl. Phys. B} {\bf 673} (2003) 170.
\bibitem{trusov16} M. Shifman, A. Vainshtein, V. Zakharov, {\em Nucl. Phys. B} {\bf 147} (1979) 385, 448.
\bibitem{trusov9} Yu. A. Simonov, {\em Phys. At. Nucl.} {\bf 69} (2006) 528.
\bibitem{trusov22a} F. Karsch, Prog. Theor. Phys. Suppl. {\bf 153} (2004) 106.
\bibitem{trusov3} C. Schmidt, hep-lat/0701019; C. Bernard {\em et al.}, {\em Phys. Rev. D} {\bf 75} (2007) 094505;
R. Gavai, {\em Pramana67} (2006) 885; U. M. Heller, {\em PoSLAT2006} (2006) 011; F. Karsch, hep-lat/0601013, {\em J. Phys. Conf. Ser.}
{\bf 46} (2006) 122.
\bibitem{trusov27} S. Aoki, {\em Int. J. Mod. Phys. A} {\bf 21} (2006) 682.
\bibitem{trusov22} C. R. Allton, S. Ejiri, S. J. Hands, {\em et al.}, {\em Nucl. Phys. Proc. Suppl.} {\bf 141} (2005) 186;
Nucl. Phys. Proc. Suppl. {\bf 141} (2008) 186.
\bibitem{trusov26} Z. Fodor, S. D. Katz, C. Schmidt, JHEP {\bf 0703} (2007) 121.
\bibitem{trusov28} M. Stephanov, {\em PoSLAT2006} (2006) 024.
\bibitem{trusov29} A. Di Giacomo, {\em Braz. J. Phys.} {\bf 37} (2007) 208.
\bibitem{trusov30} V. G. Bornyakov {\em et al.}, {\em PoSLAT2005} (2005) 157.
\bibitem{myst1m} D. Gross, R. D. Pisarski, L. G. Yaffe, {\em Rev. Mod. Phys.} {\bf 53} (1981) 43.
\bibitem{myst2m} L. Mc Lerran and B.Svetitsky, {\em Phys. Lett. B} {\bf 98} (1981) 195.
\bibitem{myst3m} N. Weiss, {\em Phys. Rev. D} {\bf 24} (1981) 475, {\em ibid} {\bf 25} (1982) 2667.
\bibitem{myst4m} M. Ogilvie amd P. N. Meisinger, arXiv:0811.2025[hep-lat].
\bibitem{myst5m} K. Fukushima, {\em Phys. Lett. B} {\bf 553} (2003) 38;
{\em Phys. Rev.  D} {\bf 68} (2003) 045004; {\em Phys. Lett. B} {\bf 591} (2004) 277;
Y. Hatta and K. Fukushima, {\em Phys. Rev.  D} {\bf 69} (2004) 097502.
\bibitem{myst6m} C. Sasaki, B. Friman, K. Redlich, {\em Phys. Rev. D}  {\bf 75} (2007) 054026 (2007);
S. R\"{o}ssner, T. Hell, C. Ratti, and W. Weise, arXiv:0712.3152[hep-ph]; S. K. Ghosh, T. K. Mukherjee, M. G. Mustafa, and R. Ray,
arXiv:0805.4690.
\bibitem{myst7m} A. Peshier, {\em Phys. Rev. D} {\bf 70} (2004) 034016; {\em J. Phys. G} {\bf 31} (2005) 5371.
\bibitem{myst8m} W. Cassing, {\em Nucl. Phys. A} {\bf 791} (2007) 365.
\bibitem{myst9m} N. O. Agasian, {\em JETP Lett.} {\bf 57} (1993) 208; N. O. Agasian, D. Ebert, and E. Ilgenfritz,
{\em Nucl. Phys. A} {\bf 637} (1998) 135.
\bibitem{myst10m} Yu. A. Simonov, arXiv:hep-ph/0605022; Yu. A. Simonov and M. A. Trusov (in preparation).
\bibitem{myst11m} see {\em e.g.} F. Wilczek and  K. Rajagopal, in ``At the Frontier of Particla
Physics. Handbook of QCD'', v. 3, ed. M.Shifman (World Scientific, Singapore, 2001); B. O. Kerbikov, {\em Phys. Rev. D} {\bf 70} (2004)
057503; {\em Phys. At. Nucl.} {\bf 68} (2005) 916.
\bibitem{myst12m} N. O. Agasian, B. O. Kerbikov, and V. I. Shevchenko, {\em Phys. Rept.} {\bf 320} (1999) 131.
\bibitem{myst13m} M. D\"{o}ring {\em et al.}, {\em Phys. Rev. D} {\bf 75} (2007) 054504;
G. Aarts {\em et al.}, {\em Phys. Rev. D.} {\bf 76} (2007) 094513.
\bibitem{myst14m} N. K. Glendenning, {\em Phys. Rev. D} {\bf 46} (1992) 1274;
F. \"{O}zel, {\em Nature} {\bf 441} (2006) 1115;
G. H. Bordbar, M. Bigdeli, and T. Yazdizadeh, {\em Int. J. Mod. Phys. A} {\bf 21} (2006) 5991.
\bibitem{concl15} M. Alford {\em et al.}, {\em Nature} {\bf 445} (2007) E7.
\bibitem{myst15m} I. M. Dremin, {\em Eur. Phys. J. C} {\bf 56} (2008) 81; I. M. Dremin {\em et al.}, arXiv:0809.2472[nucl-th].
\bibitem{concl1} Yu. A. Simonov, {\em JETP Lett.} {\bf 55} (1992) 605; E. L. Gubankova and Yu. A. Simonov, {\em Phys. Lett. B}
{\bf 360} (1995) 93; N. O. Agasian, {\em JETP Lett.} {\bf 57} (1993) 208.
\bibitem{concl2} Yu. A. Simonov, {\em Nucl. Phys. B} {\bf 324} (1989) 67; Yu. A. Simonov,
in "Sense of Beauty in Physics",  volume in honor of Adriano Di Giacomo, Pisa  univ. Press, p.29, 2006.
\bibitem{concl5} O. Kaczmarek and F. Zantow, {\em Phys. Rev. D} {\bf 71} (2005) 114510.
\bibitem{trusov31} Yu. A. Simonov, {\em Phys. At. Nucl.} {\bf 60} (1997) 2069; {\em Phys. Rev. D} {\bf 65} (2002) 094018.
\bibitem{lat1} F.Karsch, {\em PoSCPOD07} (2007) 026, {\em PoSLAT2007} (2007) 015 and refs. therein.
\bibitem{concl8} Z. Fodor, {\em PoSLAT2007} (2007) 011.
\bibitem{concl9} P. Buividovich, E. Lushchevskaya, M. Polikarpov, {\em Phys. Rev. D} {\bf 78} (2008) 074505.
\bibitem{concl10} F. Karsch, M. Lutgemeier, {\em Nucl. Phys. B} {\bf 550} (1999) 449;
G. Cossu {\em et.al.}, {\em Phys. Rev. D} {\bf 77} (2008) 074506; J. Engels, S. Holtmann, and T. Schulze, {\em Nucl. Phys. B}
{\bf 724} (2005) 357.
\bibitem{concl11} see {\em e.g.} H. Meyer, arXiv:0809.5202[hep-lat].
\bibitem{concl14} N. O. Agasian, {\em Phys. Lett. B} {\bf 562} (2003) 257.
\bibitem{concl**} V. P. Nair, arXiv:hep-th/0309061; {\em Nucl. Phys. B} {\bf 691} (2004) 182.
\bibitem{concl***} N. O. Agasian and S. M. Fedorov, {\em Phys. Lett. B} {\bf 663} (2008) 445;
N. O. Agasian, {\em Phys. At. Nucl.} {\bf 71} (2008) 1967.
\bibitem{concl17} A. M. Baldin, {\em Short Communications in Physics},
{\em Lebedev Institute for Physics} {\bf 1} (1971) 35; Yu. D. Bayukov {\em et al.},
{\em Izvestia Akad. Nauk SSSR, Fizika} {\bf 30} (1966) 521; {\em Yad. Fiz.} {\bf 18} (1973) 1246.
\end{thebibliography}
\end{document}